\documentclass[bibyear]{aa} 

\usepackage{longtable}
\usepackage{natbib,twoopt}
\usepackage[breaklinks=true, colorlinks=true, allcolors=blue]{hyperref}
\usepackage{cleveref}
\usepackage{graphicx}
\usepackage{adjustbox}
\usepackage{txfonts}
\usepackage{comment}
\usepackage{url}
\usepackage{threeparttablex}
\usepackage{makecell}
\usepackage{colortbl}
\usepackage{verbatim}
\usepackage{graphics,epsfig,amssymb,color,times}
\usepackage{subfig}
\usepackage{lscape}
\usepackage{arydshln}
\usepackage{graphicx}
\usepackage{pdflscape}
\usepackage{multirow,tabularx}
\usepackage{orcidlink}

\makeatletter
\newcommandtwoopt{\citeads}[3][][]{\href{http://adsabs.harvard.edu/abs/#3}%
{\def\hyper@linkstart##1##2{}%
\let\hyper@linkend\@empty\citealp[#1][#2]{#3}}}
\newcommandtwoopt{\citepads}[3][][]{\href{http://adsabs.harvard.edu/abs/#3}%
{\def\hyper@linkstart##1##2{}%
\let\hyper@linkend\@empty\citep[#1][#2]{#3}}}
\newcommandtwoopt{\citetads}[3][][]{\href{http://adsabs.harvard.edu/abs/#3}%
{\def\hyper@linkstart##1##2{}%
\let\hyper@linkend\@empty\citet[#1][#2]{#3}}}
\newcommandtwoopt{\citeyearads}[3][][]%
{\href{http://adsabs.harvard.edu/abs/#3}
{\def\hyper@linkstart##1##2{}%
\let\hyper@linkend\@empty\citeyear[#1][#2]{#3}}}
\makeatother
\newcommand{\xmm}{XMM-{\it Newton}}
\newcommand{\swift}{\textit{Swift}}
\newcommand{\xrism}{\textit{XRISM}}
\newcommand{\nustar}{\textit{NuSTAR}}
\newcommand{\hst}{\textit{HST}}

\defcitealias{petrucci18}{P18}
\defcitealias{petrucci2020}{P20}
\defcitealias{peluso2026}{SUBWAYS V}

\definecolor{modone}{RGB}{0,102,204} 
\definecolor{modthree}{RGB}{220,20,60}   
\definecolor{modtwo}{RGB}{60,179,113}    
\titlerunning{SUBWAYS VI. Soft excess}
\authorrunning{V. E. Gianolli et al.}
\begin{document}

   \title{Supermassive Black Hole Winds in X-rays -- SUBWAYS. VI. 
   Interplay between the soft X-ray excess and ultra-fast outflows}

   \author{V. E. Gianolli$^{1,2,3,4}$\orcidlink{0000-0002-9719-8740}, P-O. Petrucci$^{1}$\orcidlink{0000-0001-6061-3480}, S. Bianchi$^{3}$\orcidlink{0000-0002-4622-4240}, M. Brusa$^{5,6}$\orcidlink{0000-0002-5059-6848}, G. Lanzuisi$^{6}$\orcidlink{0000-0001-9094-0984}, G. A. Matzeu$^{7,5,6}$\orcidlink{0000-0003-1994-5322}, M. Mehdipour$^{8}$\orcidlink{0000-0002-4992-4664}, F. Ursini$^{3}$\orcidlink{0000-0001-9442-7897}, N. Torres-Alb\`a$^{9,2}$\orcidlink{0000-0003-3638-8943}, 
   E. Amenta$^{5,10}$\orcidlink{0009-0006-9487-0358}, E. Behar$^{11}$\orcidlink{0000-0001-9735-4873}, L. Borrelli$^{5,6}$, M. Cappi$^{6}$\orcidlink{0000-0001-6966-8920}, A. Comastri$^{6}$\orcidlink{0000-0003-3451-9970}, E. Costantini$^{12}$\orcidlink{0000-0001-8470-749X}, M. Dadina$^{6}$\orcidlink{0000-0002-7858-7564}, B. De Marco$^{13}$\orcidlink{0000-0003-2743-6632}, A. De Rosa$^{14}$\orcidlink{0000-0001-5668-6863}, M. Giustini$^{15}$\orcidlink{0000-0002-1329-658X}, S. Kraemer$^{16}$, G. Kriss$^{17}$\orcidlink{0000-0002-2180-8266}, Y. Krongold$^{18}$\orcidlink{0000-0001-6291-5239}, A. L. Longinotti$^{18}$\orcidlink{0000-0001-8825-3624}, S. Mathur$^{19,20}$\orcidlink{0000-0002-4822-3559}, R. Middei$^{21,22}$\orcidlink{0000-0001-9815-9092}, G. Miniutti$^{15}$\orcidlink{0000-0003-0707-4531}, E. Nardini$^{23}$\orcidlink{0000-0001-9226-8992}, S. Peluso$^{24}$\orcidlink{0009-0008-7843-2950}, E. Piconcelli$^{21}$\orcidlink{0000-0001-9095-2782}, G. Ponti$^{25,26,27}$\orcidlink{0000-0003-0293-3608}, R. Serafinelli$^{28}$\orcidlink{0000-0003-1200-5071}, F. Tombesi$^{29,21,30}$\orcidlink{0000-0002-6562-8654}, A. Tortosa$^{6}$\orcidlink{0000-0003-3450-6483}, C. Vignali$^{5,6}$\orcidlink{0000-0002-8853-9611}, I. Cox$^{2}$\orcidlink{0000-0003-2287-0325}, S. Marchesi$^{5,2,6}$\orcidlink{0000-0001-5544-0749}, M. Parra$^{31}$\orcidlink{0009-0003-8610-853X}, M. Ajello$^{2}$\orcidlink{0000-0001-5544-0749}, T. Mukherjee$^{2}$\orcidlink{0009-0004-0580-8823}, I. Pal$^{2}$\orcidlink{0000-0002-7825-1526}, P. Patel$^{2}$, A. Pizzetti$^{32}$\orcidlink{0000-0001-6412-2312}, D. Sengupta$^{33}$\orcidlink{0009-0002-6991-1534}, R. Silver$^{34,35,36}$\orcidlink{0000-0001-6564-0517}, and X. Zhao$^{37}$\orcidlink{0000-0002-7791-3671}}
   \institute{Affiliations are shown after the references}

   \date{Received XXX; accepted YYY}
 
  \abstract{ We present the first systematic broadband and multiwavelength investigation of the interplay between the soft X-ray excess (SE), Ultra-Fast Outflows (UFOs), and Warm Absorbers (WAs) in 21 Active Galactic Nuclei (AGN; 0.1 $\leq$ z $\leq$ 0.4) from the SUBWAYS project. We combine simultaneous or quasi-simultaneous UV, optical, and X-ray photometry and spectroscopy of XMM-{\it Newton} (EPIC pn and OM), \nustar, and Hubble Space Telescope. We investigate the presence and physical origin of the SE by modeling the data within the two-coronae paradigm, where a warm, optically thick corona (WC) produces the optical/UV--soft X-ray emission and a hot, optically thin corona generates the hard X-ray continuum. We also test a hybrid model consisting of a relativistic reflection component in addition to the two-coronae emission. The optical/UV coverage allows us to directly constrain the seed photon temperature of the Comptonization process, with values spanning 0.1--5.5 eV.
  The two-coronae model provides a consistent description of the broadband emission, with the warm corona electron temperatures of kT$_{\rm e} \simeq$ 0.1–0.9 keV and photon indices $\Gamma \geq$ 2, consistent with previous studies. However, 9/21 AGN require the hybrid configuration, indicating that the SE arises from a combination of processes whose relative contribution varies among sources. This is particularly evident in AGN hosting UFOs, where 4/7 observations favor the hybrid model. 
  Compared with no-UFO sources, UFO hosts show lower WC electron temperatures, harder hot corona spectra, and positive hardness ratios. However, these trends should be interpreted cautiously given the limited sample size and their dependence on spectral parameters and instrumental response. AGN with and without WAs show no significant differences.
  Finally, our correlation analysis reveals several links between WC, SE, AGN, and UFO properties, including correlations of the amplification factor with WC temperature, of WC optical depth with hardness ratio, and of WC slope and optical depth with UV luminosity, together with a tentative anticorrelation between WC optical depth and UFO outflow velocity.}



   \maketitle
%
\section{Introduction}\label{sec:intro}

The so-called soft excess \citep[SE;][]{turner89} is a strong and featureless emission below 2 keV. Firstly observed in the Seyfert galaxy Mrk~841 by \citet{arnaud85}, the SE is now identified as a nearly ubiquitous component in the X-ray spectra of Type-1 Active Galactic Nuclei \citep[AGN; e.g.,][and references therein]{halpern1984,garcia14,turner09,Boissay2016,porquet2018,petrucci18,gliozzi2020,nandi2023,vaia24}. Despite decades of study, its physical origin remains debated. The SE cannot be reproduced by the primary X-ray emission alone and was historically modeled with a blackbody component with a temperature of kT$\sim 0.1-0.2$ keV \citep[e.g.,][]{vasudevan14}. Although possibly associated with the higher energy component of thermal emission from the accretion disk \citep[AD; e.g.,][]{walter1993,dewangan07}, the observed invariance of the blackbody temperature across wide ranges of luminosity and black-hole (BH) mass scales challenges this simple interpretation \citep[e.g.,][]{Gierlinski2004,Piconcelli2005,Crummy2006,bianchi09}.
Alternative explanations invoke relativistically blurred ionized reflection \citep[e.g.,][]{ballantyne01,done07,garcia19}.
In this scenario, coronal X-ray photons irradiate the inner AD, where they are absorbed in the upper layers of the AD and reprocessed, producing a reflection spectrum characterized by absorption features and fluorescent emission lines.
Meanwhile, the relativistic motion of the inner accretion flow would blur the sharp atomic features into the smooth profile characterizing the soft excess. Nevertheless, the presence of this component is also observed in objects lacking a broad iron line, and the soft and hard X-rays do not always correlate, as would be expected in the reflection scenario.
Other hypotheses consider smeared absorption \citep{gierlinski06} or partial covering ionized absorption. 
However, in the first case, the required blurring is too large to be feasible \citep{schurch07}, while in the latter, the soft excess cannot explain all observations \citep[e.g.,][]{miller08,turner09}.

An additional interpretation is provided by the so-called ``two-coronae model''. This framework invokes the presence of two distinct Comptonizing media: the traditional hot X-ray corona, consisting of an optically thin and geometrically thick hot \citep[with a typical temperature of kT$_e \simeq$ 10--100 keV; e.g.,][and references therein]{fabian15,fabian17,tortosa18,Serafinelli2024} electron plasma \citep[e.g.,][]{haardt93}, and a warm optically thick and geometrically thin corona.
The accretion flow geometry considered for this model is driven by the properties of the two Comptonizing media \citep[see e.g.,][]{petrucci13}.
The hot X-ray corona (``HC'') is located in the innermost region of the accretion flow \citep[R $<$ R$_{\mathrm{in}}$, where R$_{\rm in}$ = 10--20 R$_{\rm g}$ is a transition radius; see][]{petrucci13}. 
This inner HC produces the hard ($>$2 keV) primary X-ray emission through inverse Compton scattering of UV photons from the disk, and it illuminates the outer AD.
Beyond the radius $R_{\mathrm{in}}$, the AD extends outward with a vertically stratified structure, consisting of a cold, optically thick body and a warm surface, the so-called warm corona (``WC''). The WC (whose emission is also modeled as thermal Comptonization) has a typical temperature of 0.1--1 keV and an optical depth in the $\tau \sim 10-40$ range \citep[e.g.,][]{magdziarz98,czerny03,done12,petrucci18}. Given the large $\tau$ observed, the warm corona is indeed expected to extend radially over a passive disk \citep[where all the accretion power is released in the warm plasma, see also][]{janiuk01,czerny03,rozanska15}. The WC heats the deeper layers of the disk, Comptonizing its optical–UV emission. In turn, part of this reprocessed radiation enters and cools the hot corona. 
In this configuration, the two coronae are thus highly interconnected and jointly produce the observed optical–UV, soft X-ray, and hard X-ray emission of AGN \citep[e.g.,][and references therein]{done12,petrucci13,Boissay2016,kubota18,petrucci18,porquet2018,zoghbi23,Ballantyne2024}.
Moreover, in this scenario, the correlation between optical/UV and soft X-rays is naturally explained \citep[e.g.,][]{petrucci13}.

Studies of individual sources suggest that the SE may not have a single origin across the AGN population. In some cases, the excess is well explained by a warm corona (e.g., 3C~382, \citealp{Ursini2018}; Mrk~359, \citealp{middei20}; Mrk~509, \citealp{noda2011,petrucci13}; Mrk~590, \citealt{Mathur2018}; Mrk~841, \citealp{Mehdipour2023}; NGC~5548, \citealp{ursini15,Mathur2017}; NGC~7469, \citealp{Middei2018}; RBS~1055, \citealp{Marinucci2022}; Ton~S180, \citealp{Matzeu2020}; some of these works also include UV data), while in others, relativistically blurred reflection provides a better description (e.g., Mrk~478, \citealp{Waddell2019}; Mrk~766, \citealp{Mason2003}; Mrk~1044, \citealp{Mallick2018}, PG~1211+143, \citealp{demarco2011}; PG 1448+273, \citealp{ding21}). Some sources even show hybrid properties, suggesting that an unidentified factor may determine which process dominates the soft X-ray emission (e.g., Ark~120, \citealp{porquet2018}; Mrk~110, \citealp{Porquet2021}; HE~1143-1820 and NGC~4593, \citealp{Xiang2022}; analyses of Type-1 AGN samples, \citealp{Ballantyne2024,VCeballos2026}). These results highlight the need for systematic tests of the two-coronae scenario on large, homogeneous samples.

The picture is even more complicated as AGN often exhibit sub-parsec scale outflows that may interact with the inner accretion flow \citep[see][for reviews]{fabian12,laha21}. The most powerful of these nuclear winds are the so-called Ultra-Fast Outflows (UFOs). Detected in approximately 35--40\% of nearby and intermediate-z AGN \citep[z $\leq$ 0.4;][]{Tombesi2010,Gofford2013,Matzeu2023,gianolli24}, UFOs manifest as blueshifted absorption lines of highly ionized iron (\ion{Fe}{xxv} and \ion{Fe}{xxvi}) in the 7–9 keV energy range, with column densities N$_{\mathrm{H}}$ $\sim$ 10$^{22}$–10$^{24}$ cm$^{-2}$ and ionization parameters $\log(\xi / \mathrm{erg\,cm\,s^{-1}}) \simeq 4-6$. UFOs extreme nature is further highlighted by their sub-relativistic outflow velocities, v$_{\rm out}$ = 0.03-0.6c \citep[e.g.,][]{Cappi2006,Tombesi2010,Gofford2013,Chartas2021,Borrelli2026}. 
Although their launching mechanism remains uncertain, with both radiative \citep[e.g.,][]{Matzeu2017} and magnetohydrodynamic \citep[MHD; e.g.,][]{Fukumura2010,Fukumura2015} processes likely playing a role \citep{gianolli24}, their high velocities suggest they originate close to the SMBH, at small radii within the AD.

Recent high-resolution X-ray spectroscopy is now refining this picture. Early results from \xrism\ indicate that Fe K absorption features previously modeled as single, broad components in CCD spectra can instead be resolved into multiple discrete velocity and ionization layers, as observed in PDS~456 \citep{xrismpds2025, xu2025}, NGC~4151 \citep{xiang2025,Miller2026}, and PG~1211+143 \citep{Mizumoto2026, Reeves2026}, revealing that these winds are intrinsically stratified and clumpy. This emerging complexity suggests that UFOs are not launched from a single characteristic radius, but instead span a range of disk radii and may consist of multiple coexisting phases. This structure further supports scenarios in which MHD processes play a significant role, either alone \citep[e.g.,][]{xiang2025} or in combination with radiation pressure \citep[e.g.,][]{xrismpds2025, xu2025}. Interestingly, \xrism\ results also suggest a connection between some UFO components, coronal activity, and magnetic reconnection near the inner accretion flow \citep{Gu2025,Vardi2026}.

In addition, the soft X-ray band reveals both fast \citep{Gupta2013,Gupta2015,Reeves2016,Longinotti2019,Serafinelli2019} and more moderate winds, the latter commonly referred to as Warm Absorbers \citep[WAs;][]{reynolds97,george98,blustin05,McKernan07,winter10,tombesi13}. WAs imprint a rich array of absorption lines and edges, primarily from O, Ne, Mg, Si and Fe ions \citep[e.g.,][]{blustin05,McKernan07, tombesi13, laha14}. These outflows have an ionization parameter that spans $\log(\xi / \mathrm{erg\,cm\,s^{-1}}) \sim 0-4$, and present column densities of N$_{\mathrm{H}} \sim 10^{20}$–10$^{23}$ cm$^{-2}$. The outflow velocities typically range from a few hundred up to several thousand km s$^{-1}$, suggesting that they trace a stratified, multi-phase wind emerging from different regions of the central engine \citep[e.g.,][]{blustin05,tombesi13,kaastra14,Zhao2026}. The observed variability in both ionization state and column density on timescales of days to months indicates that at least some of these absorbers reside at sub-parsec scales, possibly within or near the broad-line region \citep[e.g.,][]{krongold07,mehdipour17,Serafinelli2025}. 

The limited number of simultaneous and contemporaneous broad-band observations, needed to separate the different components of the AGN spectra, has limited the study of the soft excess and X-ray winds. As a result, their interplay has not yet been  explored in a large AGN sample. 
However, a link is hinted in single source studies. PG~1448+273, a Narrow Line Seyfert 1 (NLSy1) galaxy with a strong soft X-ray excess, presents in its X-ray spectrum both an UFO and WA \citep{kosec20}. Interestingly, \cite{ding21} observed that the variable UFO and SE appear together during the low flux state (a factor $\sim 5$ lower than the high state) of PG~1448+273. Another interesting NLSy1 is 1H~0707--495: \citet{parker21} studied its fractional excess variance spectra to investigate how X-ray variability changes with energy and flux state. They observed that the SE is present during both high- and low-flux states. However, in the 2008 and 2010 high-flux observations, the X-ray spectra also show a variance spike near 8 keV from the \ion{Fe}{XXV/XXVI} UFO absorption. They concluded that while the UFO variability follows the same frequency dependence as the continuum, indicating an ionization response to intrinsic X-ray flux changes, the soft excess remains relatively stable. Similar behavior is also present in IRAS~13224--3809. In this AGN, the variance spectra reveal an even less variable soft excess and strong UFO feature, both responding coherently to the X-ray continuum \citep{alston2020,parker2020}. However, the interpretation of the UFO variability in this source remains debated: while \citet{Parker2017} reported an equivalent width--flux anticorrelation and an approximately constant UFO velocity, \citet{Chartas2018} showed that the treatment of the background can significantly affect these conclusions.
Timing analysis also supports the simultaneous presence of UFOs and SE. \citet{xu2024} investigated the X-ray time lags in three AGN (PG~1448+273, IRAS~13224–3809, and PG~1211+143) and found that UFOs can appear alongside soft excess. However, when UFOs are detected, the low-frequency hard lags are suppressed, while the high-frequency soft lags remain unaffected, suggesting that these winds can alter the inner accretion flow and variability patterns without impacting the soft excess emission.

The SUpermassive Black hole Winds in the x-rAYS \citep[SUBWAYS;][]{brusa2022} sample is particularly suited to this study because it combines broad-band optical/UV/X-ray coverage from XMM-\textit{Newton}/EPIC-pn and OM, \textit{NuSTAR}, and Hubble Space Telescope (HST), with a mixed population of quasars with and without detected UFOs and WAs. Thus, enabling a homogeneous comparison of their soft excess properties as a function of wind detection. 
We compare our results with those of \citet[][``P18'' hereafter]{petrucci18}, who analyzed 22 AGN (0.01 $\leq$ z $\leq$ 0.57; of which five AGN host UFOs and three sources are shared with the SUBWAYS sample) with broadband optical/UV and X-ray coverage. By combining the two analyses, we examine one of the largest AGN samples with simultaneous UV, optical, and X-ray coverage for which the SE has been studied within the two-coronae framework, extending previous larger X-ray--only studies \citep[e.g.,][]{Boissay2016,gliozzi2020}.

The paper is organized as follows. In Sect.~\ref{data}, we describe the sample selection, the data reduction procedures for XMM-{\it Newton}, {\it NuSTAR}, and HST, and the methodology adopted in the analysis. 
Sect.~\ref{analysis} presents the results of the spectral analysis. In Sect.~\ref{SE_strength}, we discuss the proxies for the SE strength. In Sect.~\ref{discussion}, we compare our results with previous works and explore the possible connection between the SE and X-ray wind. We also carry out a correlation analysis between WC, SE, UFOs, and AGN properties. Sect.~\ref{conclusions} summarizes our findings. We assume as cosmological constants H$_{\rm 0}$ = 70 km s$^{-1}$ Mpc$^{-1}$, $\Omega_{\Lambda _{\rm 0}}$ = 0.73, and $\Omega _{\rm M}$ = 0.27, and errors are quoted at the 68\% (1$\sigma$) confidence level for one single parameter, unless otherwise stated. In this paper, we consider a probability of 0.05 (roughly corresponding to $\sim$2$\sigma$ for a Gaussian distribution) as a statistically significant threshold for the null hypothesis probability (NHP; i.e., log(NHP) $\leq-1.30$).

\section{Data and methodology} \label{data}
\subsection{Sample selection}\label{sampl_sel}

The SUBWAYS multiwavelength program targets 22 quasars (0.09 $\leq$ z $\leq$ 0.36) to study the various manifestations of UFOs \citep{brusa2022}. 
\citet[][SUBWAYS paper I]{Matzeu2023} present an X-ray spectroscopic analysis of the XMM-{\it Newton} data for this sample: UFOs and WAs have been detected in seven and 13 AGN, respectively, while six sources do not present either UFO or WA, allowing a consistent comparison between AGN hosting UFOs and/or WAs and those without such features. \citet[][SUBWAYS paper II]{mehdipour23} analyze the ionized outflows in the UV band using the HST. \citet[][SUBWAYS paper III]{gianolli24} extend the study to a larger population of radio-quiet quasars and Seyfert galaxies (a total of 77 AGN with 0.002 $\leq$ z $\leq$ 3.91) to investigate the physical mechanisms driving UFOs, their incidence, and properties. \citet[][SUBWAYS paper IV]{amenta2026} focus on the radio properties of the sample. \citet[][SUBWAYS paper V]{peluso2026} present a combined fit of XMM and \nustar\ data to constrain the hot corona properties of the SUBWAYS sources, and compare them with a large sample ($>$100 sources) spanning both lower and higher redshift.

To test the two-coronae model, we require bright AGN with simultaneous or contemporaneous optical/UV and X-ray observations, given the strong intrinsic variability of these sources across all wavelengths \citep[e.g.,][]{ponti12}. When simultaneous data are unavailable, we select observations obtained within two years. For larger temporal separations we check for the presence of spectral variability before adopting the data.

We analyze 21 of the 22 AGN in the SUBWAYS sample. 17 sources were observed with \xmm\ as part of an \xmm\ Large Program carried out between 2019 and 2020 (PI: M. Brusa), while for the remaining five AGN we use archival XMM observations. 
We exclude the archival source PG~1416--129 because its XMM observation lacks OM coverage, and the available HST spectrum was obtained eight and 25 years before the XMM and \nustar\ observations, respectively.
To assess whether HST data could be used, we examine the source variability considering archival \swift/UVOT observations. The multi-filter photometric light curve\footnote{\label{ref_uvot_lc}The \swift/UVOT light curve is extracted using the \swift\ Analysis Pipeline for Lightcurve Extraction (SAPLE), \url{https://github.com/leamarcotulli/saple}, see \citet{Marcotulli2026} for details.} (Fig.~\ref{fig:uv_var}, left panel) shows significant UV variability, confirming that the HST spectrum is not suitable for our analysis. 

Three AGN in the analyzed sample, PG~0804+761, PG~0947+396, and PG~1114+445, have multiple XMM observations contemporaneous with the available multiwavelength data. For PG~0804+761, two XMM observations separated by two days are fitted jointly, since no significant X-ray variability is found during the spectral analysis (see also Sect.~\ref{red}). This source also has a third XMM observation with a UFO detection, and is therefore included in the UFO sub-sample in the original SUBWAYS X-ray classification \citep{Matzeu2023}. However, this observation cannot be used in our analysis because it lacks simultaneous or contemporaneous optical/UV coverage: no OM filters are available, and the HST spectrum, obtained ten years earlier, is not representative given the significant UV variability shown by the archival \swift/UVOT light curve (Fig.~\ref{fig:uv_var}, lower left panel). We therefore analyze only the two XMM observations without a UFO detection, together with the corresponding optical/UV data, and classify PG~0804+761 as part of the no-UFO sub-sample for the purposes of this work (see below for more details).
PG~0947+396 and PG~1114+445 are special cases. The former AGN has two XMM observations, one with a UFO detection, while the latter source has three XMM observations, two of which show UFOs (obsID 0651330101,  ``101'' hereafter, and obsID 0651330301 ,``301'' hereafter). We therefore include the relevant observations of these AGN in both the UFO and no-UFO sub-samples. Our final working sample thus consists of 21 distinct AGN and 24 observation entries. The list of AGN and available data is reported in Table~\ref{tab:obs_info}.

Following \citet{gianolli24}, we divide the SUBWAYS sample into a ``UFO sub-sample'' and a ``no-UFO sub-sample'', referring to AGN with and without a detected ultra-fast outflow in the X-ray spectra. In \citet{Matzeu2023} analysis, a detection was considered significant when the absorption feature satisfied the F-test probability significance, P$_{\rm F} > 99\%$, and the Monte Carlo approach significance, P$_{\rm MC} > 95\%$.
Based on these criteria, the original UFO sub-sample consists of seven AGN. In our paper, however, PG~0804+761 is assigned to the no-UFO sub-sample, see above. Consequently, our UFO sub-sample contains six distinct AGN, corresponding to seven observation entries.

We note that the detection confidence is not uniform across the sample: two AGN (2MASS~J16531505+2349427 and LBQS~1338$-$0038) are highly significant with P$_{\rm MC} > 99.99\%$, while the others have P$_{\rm MC} = 95.0-97.2\%$. We also stress that the absence of a UFO detection in the X-ray spectra does not necessarily imply the absence of the wind. Detectability depends on signal-to-noise ratio, source state, ionization, inclination, and observing epoch. Therefore, throughout this paper, the UFO/no-UFO division should be interpreted as an observational classification based on the published detections, rather than as implying either identical detection robustness among UFO sources or a strict physical dichotomy between AGN with and without winds.

For each AGN in the SUBWAYS sample, Table~\ref{tab:sources_list} reports the redshift, BH mass, and X-ray flux in the 2-10 keV energy band \citep[for details on the properties, their derivation, and relevant references, see Table~A.1 in][]{gianolli24}, and notes the presence of X-ray WAs and of UFOs.

\subsection{Data reduction}\label{red}
\paragraph*{\bf X-ray data.}

The 21 AGN analyzed in this work have deep XMM-{\it Newton} data, providing both high-quality X-ray spectra with the EPIC-pn camera \citep{struder01} and optical/UV data with the optical monitor \citep[OM;][]{mason01}. The OM filters used in this work span the near-optical/UV range, from UVW2 to V. The adopted filters are reported in Table~\ref{tab:obs_info}. For the four archival AGN (PG~1402+261, PG~0804+761, PG 1114+445, and HB89 1257+286), we select the XMM observations that best satisfy our simultaneity criteria and maximize multiwavelength coverage. For PG~1114+445, we retain the two observations with detected UFOs as well as the no-UFO observation, as discussed in Sect.~\ref{sampl_sel}. We refer to Table \ref{tab:obs_info} for the list of adopted observations.
Given that the XMM/EPIC pn spectra used in this work are taken from \citet{Matzeu2023}, we refer to them for details on the reduction process.

All sources are also observed with {\it NuSTAR} \citep{harrison13}, which provides hard X-ray coverage (3--30 keV) and allows us to derive the high-energy cutoff \citep[see also][]{peluso2026}.
To produce and calibrate cleaned event files, we use \texttt{nupipeline} task and calibration files available in the data base (CALDB 20221229). 
The background is extracted using a circular region with a radius of $\sim$75\arcsec\ and the source radius is selected to maximize the signal-to-noise ratio (SNR; the radii vary within the 24--80\arcsec\ range, see Table~\ref{tab:obs_info}). We then binned the spectra using \texttt{ftgrouppha} with the \texttt{optsnmin} option, implementing the optimal scheme of \citet{kaastra16}, ensuring an SNR $>$ 3 in each bin. Table~\ref{tab:obs_info} lists the effective exposure times and observation dates for each adopted spectrum. 

For AGN where XMM-{\it Newton} and \nustar\ observations are separated by more than two years, we investigate possible X-ray variability to verify whether the spectra can be analyzed together. 
To this end, we consider all the available XMM data, and for each \nustar\ observation, we analyze the simultaneous \swift-XRT spectrum\footnote{Each \nustar\ observation of the SUBWAYS sample has a simultaneous \swift-XRT observation to cover the soft X-ray band. To obtain the \swift-XRT spectra, we use the {\it Build Swift-XRT products} tool: \url{https://www.swift.ac.uk/user_objects/}.}. For sources with limited XMM coverage, such as PG~0804+761, we only rely on \swift-XRT data. The spectra are then modeled with a phenomenological absorbed power-law to track flux and spectral changes across epochs.
Significant X-ray variability is observed in 3/4 archival AGN, PG~0804+761, PG~1114+445, and HB89~1257+286 (see contour plots of the photon index versus normalization for the primary emission in Fig.~\ref{fig:xray_var}) and thus, the hard X-ray spectra are not included in our analysis (see Sect.~\ref{comp} for details on the methodology adopted during the modeling for these sources).

\paragraph*{\bf Optical/UV data.}
\hst\ spectra are available for 16 targets and were obtained with the Cosmic Origins Spectrograph \citep[COS;][]{green2012}, the Space Telescope Imaging Spectrograph \citep[STIS;][]{Woodgate1998}, and the Faint Object Spectrograph \citep[FOS;][]{Harms1991}. They cover the rest-frame far-UV band from the \ion{Ly}{$\beta$}/\ion{O}{vi} region to \ion{C}{iv}, partially overlapping with the bluest XMM-OM bands while extending the continuum coverage to shorter wavelengths and improving the determination of the AD emission. 
Three AGN (2MASS~J10514425+3539306, 2MASS~J16531505+2349427, and WISE~J053756.30$-$024513.1) of the SUBWAYS program were not observed due to significant reddening. PG~1402+261 presents UV variability (see the multi-filter light curve\footref{ref_uvot_lc} in Fig.~\ref{fig:uv_var}, right panel) and thus, HST data cannot be used during the analysis. 
All available \swift\ observations of HB89~1257+286 are highly offset ($>$14 arcmin), preventing the derivation of a multi-filter light curve and thus assessing possible UV variability. Consequently, the HST data for this source are excluded from our analysis.
The fully calibrated and merged HST spectra used here are those presented by \citet{mehdipour23}, and we refer to that work for details of the data processing and calibration.

To constrain the disk emission, we extract up to four photometric points from each HST spectrum after excluding major absorption/emission lines to map solely the continuum. 
The spectra were divided into four wavelength intervals defined relative to the observed Ly$\alpha$ wavelength, $\lambda_{\rm {Ly \alpha,obs}}$ = 1216  $\times$ (1 + $z$) \AA\ (where $z$ is the AGN redshift): 1) $\lambda \leq$ 1300\AA; 2) 1300 $< \lambda \leq \lambda_{\rm {Ly \alpha,obs}}$; 3) $\lambda_{\rm {Ly \alpha,obs}} <  \lambda \leq 1750$\AA; 4) $\lambda > 1750$\AA. 
For each interval, the continuum flux (in units of erg s$^{-1}$ cm$^{-2}$ \AA$^{-1}$) was computed as the mean flux within line-free regions. The corresponding uncertainties were estimated using inverse-variance weighting to account for the signal-to-noise distribution across each interval. In Table \ref{tab:obs_info}, we report the observation dates and the number of photometric points derived from each spectrum.
\smallskip

In summary, the final sample (see Table \ref{tab:obs_info}) consists of 21 AGN with XMM–{\it Newton} EPIC-pn and OM coverage, of which 16 also have HST data and 18 have {\it NuSTAR} observations. A total of 14 sources have complete UV-to–hard X-ray coverage, while the remaining seven AGN lack either HST or \nustar\ observations (or both), resulting in more limited spectral coverage.

\subsection{Methodology}\label{comp}

The main assumption of the two-coronae model is that optical/UV seed photons from an optically thick multicolor AD are Comptonized in two distinct media: the warm corona, producing the optical/UV/soft X-rays, and the standard hot corona, generating the high-energy continuum emission. We follow the methodology adopted by \citetalias{petrucci18}. We perform the spectral analysis using \texttt{XSPEC}, the model components and fitting procedure are described below:

\begin{enumerate}
    \item {\bf fit optical/UV data.} The OM and HST photometric points are modeled with a disk blackbody, where the temperature and normalization are free to vary. The fraction of photons scattered into a power-law component is monitored with the \texttt{simpl} component. The reddening effects are incorporated (\texttt{redden} component) and calculated based on Galactic extinction as described by \citet{guver09} for each AGN. The reddening parameter remains fixed throughout the fitting procedure (the value for each source is reported in Table~\ref{tab:sources_list}). From the fit, we derive the disk temperature, which serves as the initial value in the joint optical/UV and X-ray fitting (see step 2 below). In \texttt{XSPEC} notation, the model is \texttt{simpl*diskbb*redden}.

    \item {\bf mod1: fit of the full broad-band data (from optical/UV to hard X-rays).}
    We model the warm and hot corona emissions with two \texttt{nthComp}\footnote{We opt for \texttt{nthComp} instead of \texttt{CompTT} as the former present less parameter degeneracy (between photon index and high-energy cut-off) compared to the latter, which shows strong degeneracy between the electron temperature and optical depth.}. The free parameters are the electron temperatures of the hot corona (kT$_{\rm e}^{HC}$) and of the warm corona (kT$_{\rm e}^{WC}$), the seed photon temperature (kT$_{\rm bb}$, which is linked between the two coronae), and the X-ray photon indices of the two coronae ($\Gamma ^{HC}$ and $\Gamma ^{WC}$).
    Given the absence of simultaneous hard X-ray coverage for PG~0804+761, PG~1114+445, and HB89~1257+28, we follow \citetalias{petrucci18} and fix kT$_{\rm e}^{HC} =100$ keV. 
    
    The disk temperature obtained from the optical/UV fit (step 1) is initially frozen when the X-ray data are included, preventing the optical/UV data from being poorly fitted. Once a new best fit is found for the parameters describing the primary and reflected emission, the disk temperature is then allowed to vary freely again. For four sources (2MASS~J10514425+3539306, PG~0804+761, PG~1435$-$067, and WISE~J053756.30$-$024513.1), the disk temperature in the broadband fit is fixed to the value obtained from the optical/UV, as it is either unconstrained or reached unrealistically high values. See Table~\hyperlink{fit-table}{\ref*{tab:fit}} for the best-fit values. 
    
    The distant reflector (e.g., the torus) is modeled with \texttt{xillverCp} \citep{garcia10}, with only the normalization left free to vary. The photon index and the electron temperature are linked to those of the hot corona. As in \citetalias{petrucci18}, the iron abundance, the inclination, and the logarithm of the ionization parameter are set to 1, 30$^\circ$ and 0, respectively. The density of the AD is fixed to 10$^{15}$ cm$^{-3}$.
    The resulting two-coronae configuration is hereafter referred to as ``{\bf mod1}'' and is expressed in \texttt{XSPEC} notation as: 
    
    \begin{equation*}
    \begin{split}
       Mod1 = TBabs*redden*WA*UFO*(nthComp\_WC +\\
         nthComp\_HC + xillverCp);
    \end{split}
    \end{equation*}

    where \texttt{TBabs} accounts for Galactic absorption, with the column density tailored to the sky coordinates of each source using \citet{HI4PI2016} and kept fixed throughout the fitting process. For 11 AGN of the sample (i.e., 2MASS~J02201457$-$0728591, HB89~1529+050, LBQS~1338$-$0038, PG~0052+251, PG~0953+414, PG~1114+445, PG~1202+281, PG~1216+069, PG~1307+085, PG~1425+267, and PG~1626+554; see \citealt{Matzeu2023} for details), we add a warm absorber component (\texttt{WA} in the \texttt{XSPEC} notation) described by a \textsc{cloudy} table \citep{Gunasekera2025}. This table is computed assuming an NGC~5548-like AGN spectral energy distribution \citep[SED; see][]{Mehdipour2015,cappi16}, with both the column density and ionization parameter left free to vary\footnote{Although NGC~5548 is a Seyfert galaxy and the SUBWAYS sample consists of quasars, its SED provides a standard AGN ionizing continuum. This choice is also consistent with the first paper in the SUBWAYS series, which used an \texttt{XABS} table based on NGC~5548 SED \citep{Matzeu2023}. We verify that our \textsc{CLOUDY} table yields results consistent, within the uncertainties, with those of \citet{Matzeu2023}; therefore, the adopted SED does not affect our conclusions.}.

    For the six AGN with a detected UFO (i.e., 2MASS~J10514425+3539306, 2MASS~J16531505+2349427, LBQS~1338$-$0038, PG~0947+396, PG~1114+445, PG~1202+281; see Sect.~\ref{data}), we model the spectral imprint of the wind using two separate \texttt{XSTAR} photoionization grid tables (\texttt{UFO} in the \texttt{XSPEC} notation above): a multiplicative table for line-of-sight wind absorption and an additive table for the UFO reprocessed emission\footnote{The \texttt{XSTAR} absorption and emission tables are constructed by \citet{Matzeu2023} using the SED of each AGN. For each source, the adopted turbulent velocities are listed in Table~3 of \citet{Matzeu2023}. The tables adopt \texttt{CFRAC}=1, corresponding to \texttt{XSTAR} closed-sphere radiative-transfer prescription.}.
    We note that 2MASS~J10514425+3539306 and 2MASS~J16531505+2349427 require additional components with respect to the baseline mod1, we follow the source-specific modeling adopted in \citet{Matzeu2023}\footnote{For 2MASS~J10514425+3539306, we include intrinsic neutral absorption and two soft X-ray emission components associated with collisionally ionized gas. For 2MASS~J16531505+2349427, classified as a Seyfert~2, we account for the strong intrinsic neutral absorption (N$_{\rm H}\sim2.2\times10^{23}$~cm$^{-2}$), together with a distant scattered continuum and two collisionally ionized plasma components.}.

    Most AGN show residuals around the Fe band, suggesting that additional reflection-related complexity is present.

    \item {\bf mod1*: Gaussian broadening of the Fe K emission line.} To investigate the origin of the iron-band residuals, we perform a diagnostic fit of the X-ray spectrum above 3 keV to test whether the residuals are consistent with line broadening\footnote{For this diagnostic test, we fit the $>$3 keV spectrum with \texttt{TBabs*(nthComp + zgauss)} to characterize the Fe K$\alpha$ line profile. We first assume an intrinsically narrow line, with intrinsic width $\sigma=0$, and allow the line energy to vary. Once the uncertainties on the line energy are derived, $\sigma$ is left free to vary.}.
    When evidence for broadening is found, we additionally explore a phenomenological description in which a Gaussian smoothing component (\texttt{gsmooth}) is applied to the \texttt{xillverCp} reflection component in the broad-band fit. Since this does not alter the underlying physical emission model, but only approximates line broadening effects, we refer to this configuration as ``\textbf{mod1*}''.
    In \texttt{XSPEC} notation:
    
    \begin{equation*}
    \begin{split}
       Mod1* = TBabs*redden*WA*UFO*(nthComp\_WC +\\
         nthComp\_HC + gsmooth*xillverCp).
    \end{split}
    \end{equation*}

    An example of the best-fit result obtained with mod1* is presented in Fig.~\ref{fig:hybrid_models_fit}, middle panel.

    We note that the \texttt{gsmooth} Gaussian widths are generally $\gtrsim 50$ eV (see Table~\hyperlink{fit-table}{\ref*{tab:fit}}), suggesting a hybrid model scenario where disk reflection is significantly contributing to the emission. We therefore explore, in parallel, a physically motivated description of the iron-band residuals using the relativistic reflection component \texttt{relxillCp}. This model is tested for all sources showing significant residuals in mod1, independently of whether a phenomenological broadening component is required.

    \item {\bf mod2: evaluate relativistic reflection from the accretion disk.} When \texttt{relxillCp} is introduced, its photon index and electron temperature are tied to those of the hot corona. As in \citetalias{petrucci18}, the iron abundance and spin are set to 1 and 0.998, respectively. The outer disk radius is fixed at 400 $r_{\rm g}$, while the inner radius $R_{\rm in}$ and the ionization parameter $\log\xi$ are free to vary. In the following, we will refer to this model as ``{\bf mod2}'', written in \texttt{XSPEC} as:  
    \begin{equation*}
    \begin{split}
       Mod2 = TBabs*redden*WA*UFO*(nthComp\_WC +\\
         nthComp\_HC + relxillCp + xillverCp).
    \end{split}
    \end{equation*}

    An example of the best-fit result obtained with mod2 is presented in the lower panel of Fig.~\ref{fig:hybrid_models_fit}.
\end{enumerate}

\begin{figure}[]
\centering
\includegraphics[width=.49\textwidth, trim=15 17 20 10, clip]{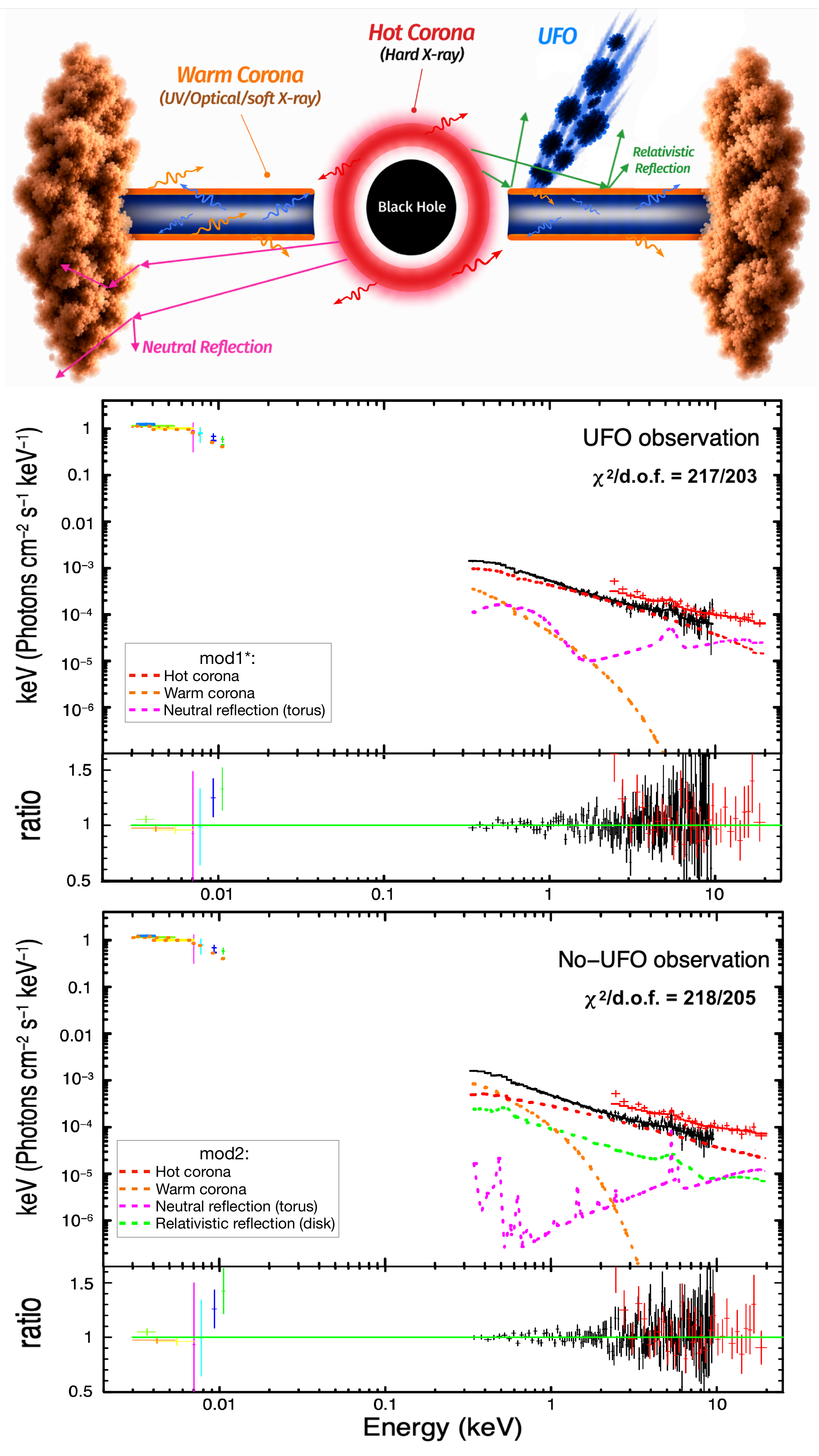}
    \caption[]
    {{{\it Upper panel:} Sketch of the hybrid model. A hot (kT $\sim$ 10--100 keV), optically thin ($\tau \lesssim 1$) corona (red) is located in the inner regions of the accretion flow near the BH, producing the hard X-ray emission via Comptonization of the soft photons originating from the accretion disk (AD). The AD is composed of cold, optically thick material in the deeper layers (white to blue), characterized by a temperature T$_{\rm bb}$, and warmer, optically thick upper layers (orange) representing a warm corona (kT $\sim$ 0.1--1 keV, $\tau \sim 10-40$). A stratified, clumpy ultra-fast outflow, launched from the inner disk regions, is shown in blue. Green and pink arrows indicate X-ray photons from the hot corona that are scattered by near (disk and/or UFO) and distant (torus) material, producing relativistic and neutral reflection components, respectively. {\it Middle and lower panels:} Best-fit models and corresponding data-to-model ratios for two observations of PG~0947+396. The middle panel shows the observation with a UFO, modeled using a pure two-coronae framework (mod1*). The lower panel shows the observation without a UFO, fitted with the hybrid model (mod2). The colors of the model components match those of the sketch. The spectra are shown in black for XMM/EPIC-pn, red for grouped \nustar\ FPMA/B, blue, light blue, pink, and yellow for the HST photometric points, and orange, lime green, and spring green for the XMM OM M2, W1, and V2 filters, respectively.}}
    \label{fig:hybrid_models_fit}
\end{figure}

\noindent We refer to mod1 and mod1* as pure two-coronae models, and to mod2 as hybrid model. We note that the models differ in the number of free parameters: mod1 has one additional degree of freedom (d.o.f) compared to mod1*, mod2 has two fewer d.o.f, the same number of d.o.f as mod1* when both $R_{\rm in}$ and $\log\xi$ are fixed, or one fewer d.o.f when only one of them is fixed.
A schematic illustration of the AGN regions and components considered in this work, as discussed above, is shown in the upper panel of Fig.~\ref{fig:hybrid_models_fit}.\\ 

\section{Spectral analysis results}\label{analysis}

In this section, we present the results of the broadband optical/UV/X-ray spectral analysis of the SUBWAYS sample, adopting the three physically motivated models (i.e., mod1, mod1*, and mod2; see Sec.~\ref{comp} for more details). The values obtained for each model are reported in Table~\hyperlink{fit-table}{\ref*{tab:fit}} and the best-fit model is highlighted in bold. A list of the best-fit models is reported in Table~\ref{tab:color_bestfit_model}. In Appendix~\ref{app_c}, we report the fitted spectra and model-to-data ratios for each AGN analyzed.

\subsection{Best-fit model selection}
\label{bestfit_selection}

We select the best-fit model on the basis of the goodness of fit and on the requirement that the derived parameters are physically acceptable.
Model comparison is performed using the Akaike Information Criterion (AIC), which provides a statistically robust framework for balancing goodness of fit against model complexity without relying on hypothesis testing with boundary constraints. Models with lower AIC values are preferred.
In cases where the AIC differences are small, the preferred solution is chosen based on the physical plausibility of the inferred spectral parameters, rejecting solutions that yield unphysical or clearly unrealistic values.
The selected models provide an adequate description of the data with the reduced $\chi^2$ values in the range 0.99–1.3 for all spectra, indicating statistically acceptable fits within the assumptions of the adopted framework.

We note that for AGN for which mod2 is the
best-fitting model, the improvement provided relative to mod1 is mainly mainly associated with broad and/or asymmetric Fe K-band structure and, where \nustar\ coverage is available, related curvature in the harder X-ray continuum.
The free parameters of the relativistic reflection component (i.e., the inner disk radius $R_{\rm in}$ and/or the ionization parameter $\xi$), are constrained only for LBQS~1338$-$0038 and PG~1202+281. Meanwhile, for six AGN (2MASS~J10514425+3539306, PG~0804+761, PG~0947+396 and PG~1114+445 the observations without UFO, PG~1216+069, and PG~1352+183), they remain unconstrained during the fitting procedure. We therefore use the F-test to compare fits with these parameters free and fixed at their best-fit values. Since the free-parameter fits are not statistically preferred, we retain the fixed-parameter solutions.
We compare the seed photon temperatures inferred from the optical/UV data with those obtained from the combined optical/UV/X-ray modeling. This comparison does not show any systematic bias that affects the best-fit model selection or the conclusions of the analysis (see Appendix~\ref{comp_kt_bb} for more details).

Overall, nine (including the observation with detected UFO of PG~1114+445, obsID 301) AGN are better described by mod1, six by mod1* (including the observation with detected UFO of PG~0947+396), and nine by mod2 (including the observation with detected UFO of PG~1114+445, obsID 101); see Table~\ref{tab:color_bestfit_model}.
\begin{table}[]
\centering
\caption{Best-fit models for each AGN.}\label{tab:color_bestfit_model}
\begin{tabular}{lcc}
\hline
\multicolumn{1}{c}{\textbf{Source}} & \multicolumn{2}{c}{\textbf{Models}} \\
 & Two-coronae & Hybrid \\
\hline\hline
\multicolumn{3}{c}{\texttt{UFO sample}} \\
2MASS~J10514425+3539306 & & \cellcolor{modtwo} mod2 \\
2MASS~J16531505+2349427 & \cellcolor{modthree} \textcolor{white}{mod1} & \\
LBQS~1338$-$0038 & & \cellcolor{modtwo} mod2 \\
PG~0947+396 & \cellcolor{modone} \textcolor{white}{mod1*} & \\
PG~1114+445 (101) & & \cellcolor{modtwo} mod2 \\
PG~1114+445 (301) & \cellcolor{modthree} \textcolor{white}{mod1} \\
PG~1202+281 & & \cellcolor{modtwo} mod2 \\
\hline
\multicolumn{3}{c}{\texttt{No-UFO sample}} \\
2MASS~J02201457$-$0728591 & \cellcolor{modthree} \textcolor{white}{mod1} \\
HB89~1257+286 & \cellcolor{modthree} \textcolor{white}{mod1} & \\
HB89~1529+050 & \cellcolor{modthree} \textcolor{white}{mod1} & \\
PG~0052+251 & \cellcolor{modone} \textcolor{white}{mod1*} & \\
PG~0804+761 & & \cellcolor{modtwo} mod2 \\
PG~0947+396 & & \cellcolor{modtwo} mod2 \\
PG~0953+414 & \cellcolor{modthree} \textcolor{white}{mod1} & \\
PG~1114+445 & & \cellcolor{modtwo} mod2 \\
PG~1216+069 & & \cellcolor{modtwo} mod2 \\
PG~1307+085 & \cellcolor{modthree} \textcolor{white}{mod1} & \\
PG~1352+183 & & \cellcolor{modtwo} mod2 \\
PG~1402+261 & \cellcolor{modone} \textcolor{white}{mod1*} \\
PG~1425+267 & \cellcolor{modthree} \textcolor{white}{mod1} & \\
PG~1435$-$067 & \cellcolor{modthree} \textcolor{white}{mod1} & \\
PG~1626+554 & \cellcolor{modone} \textcolor{white}{mod1*} & \\
SDSS~J144414.66+063306.7 & \cellcolor{modone} \textcolor{white}{mod1*} & \\
WISE~J053756.30$-$024513.1 & \cellcolor{modone} \textcolor{white}{mod1*} & \\
\hline
\end{tabular}
\begin{flushleft}
\footnotesize  \textbf{Notes.} In red and blue the two-coronae models, without and with the \texttt{gsmooth} component, respectively. In green the hybrid model, mod2.
The best-fitting values are reported in Table~\hyperlink{fit-table}{\ref*{tab:fit}}.
\end{flushleft}
\end{table}
The pure two-coronae (mod1 and mod1*) configurations are therefore the most frequent best-fit models. We test whether the warm corona emission component (\texttt{nthComp\_WC}) is statistically required. Removing \texttt{nthComp\_WC} always results in significant residuals in the soft X-ray band and in a poorer fit of the optical/UV data, confirming that warm Comptonization is generally needed. The only exception is 2MASS~J10514425+3539306, for which the WC component accounts only for the disk emission rather than producing a genuine soft X-ray excess (see also Sec.~\ref{SE_strength}, where we discuss the absence of a significant SE component in this AGN).

At the same time, mod2 is also preferred in a substantial fraction of the sample, indicating that relativistic reflection from the AD plays an important role in shaping the X-ray spectra. We note that for these AGN, the inclusion of the \texttt{relxillCp} component always yields a statistically significant improvement of the fit with respect to models without relativistic reflection.
Together, these results suggest that the soft X-ray excess may not arise from a single physical mechanism across the AGN population. Instead, both warm Comptonization and relativistic reflection appear to play important roles, with their relative contributions varying from source to source. This interpretation is consistent with previous studies of individual AGN \citep[e.g.,][]{petrucci13,ursini15,Mallick2018,Waddell2019,middei20,ding21,Mehdipour2023}.

\subsection{Best-fit models in the UFO and WA sub-samples}
\label{ufo_wa_results}
Focusing on the UFO sub-sample: two AGN are best described by mod1, one by mod1*, and four by mod2 (see Table~\ref{tab:color_bestfit_model}). Thus, the hybrid model is generally favored. This may suggest that, in the majority of UFO sources, the inner AD is both strongly illuminated and coronally active. However, this interpretation does not establish that the entire reflection-like component originates in the AD.
Across the seven observations, the modeled UFO component contributes $\sim$0.4\%–30\% of the total 0.5–10 keV flux. Thus, a contribution from UFO-related emission, such as ionized Fe emission, cannot be ruled out.

The AGN with multiple observations are an interesting case. In PG~0947+396, the observation with a detected UFO is best described by a pure two-coronae model, mod1*, while the observation without UFO is best fitted by the hybrid model mod2 (the fitted spectra and data-to-model ratios for two observations are shown in Fig.~\ref{fig:hybrid_models_fit}). In PG~1114+445: the two observations with detected UFOs are best fitted by mod1 and mod2, whereas the observation without a UFO is best described by mod2. 
Overall, the detection of UFOs appears to be compatible with different continuum scenarios. However, given the small sample size and the fact that UFO detectability depends on several factors (as discussed in the next section), we refrain from implying a direct link between UFO detection and a specific continuum configuration or from attributing the reflection emission uniquely to the AD.

Considering the sample of AGN with an X-ray WA, five objects are best fitted by mod1, three by mod1*, and five by mod2. The distribution thus shows a mild preference for the two-coronae scenario over the hybrid configuration. However, this is broadly consistent with the full sample, suggesting that the presence of a WA does not introduce a clear preference.

\subsection{Impact of multiwavelength data coverage and UFO detection}
\label{data_coverage_results}

Not all sources have complete \hst+XMM+\nustar\ coverage (see Sec.~\ref{data} for details), which may affect the model selection, as different spectral components are not equally constrained in the absence of full multiwavelength data. Among the four AGN lacking simultaneous \hst\ coverage (2MASS~J10514425+3539306, 2MASS~J16531505+2349427, PG~1402+261, and WISE~J053756.30$-$024513.1), the preferred models span mod1, mod1*, and mod2, with two sources best fitted by mod1*. This may indicate a mild tendency toward two-coronae solutions; however, given the small number of AGN and that two-coronae models are also common in the full sample, no firm conclusion can be drawn. A similar caveat applies to the AGN lacking simultaneous \nustar\ coverage. For PG~0804+761 and PG~1114+445, four observations are considered in total, and the preferred models are distributed across mod1, mod1*, and mod2. Overall, although we find no clear systematic dependence of the preferred model on data availability, we caution that, for individual AGN with incomplete spectral coverage, the model preference may partly reflect observational limitations rather than intrinsic physical differences.

As discussed in Sects.~\ref{data} and \ref{ufo_wa_results}, the UFO/no-UFO division introduces a possible observational bias. 
Although useful for comparison, this classification reflects whether a UFO is significantly detected in the available X-ray spectra, rather than the intrinsic presence or absence of an outflow. 
Sources in the no-UFO sub-sample should therefore not necessarily be interpreted as lacking winds altogether, since UFO detectability can be affected by observational, geometrical, and physical factors. 
Given the small sample size, this uncertainty might contribute to the scatter in the UFO/no-UFO comparison and reduce the sensitivity to intrinsic differences between the two groups.
\begin{figure*}[h!]
\centering
\includegraphics[width=0.8\textwidth, trim=5 5 5 5, clip]{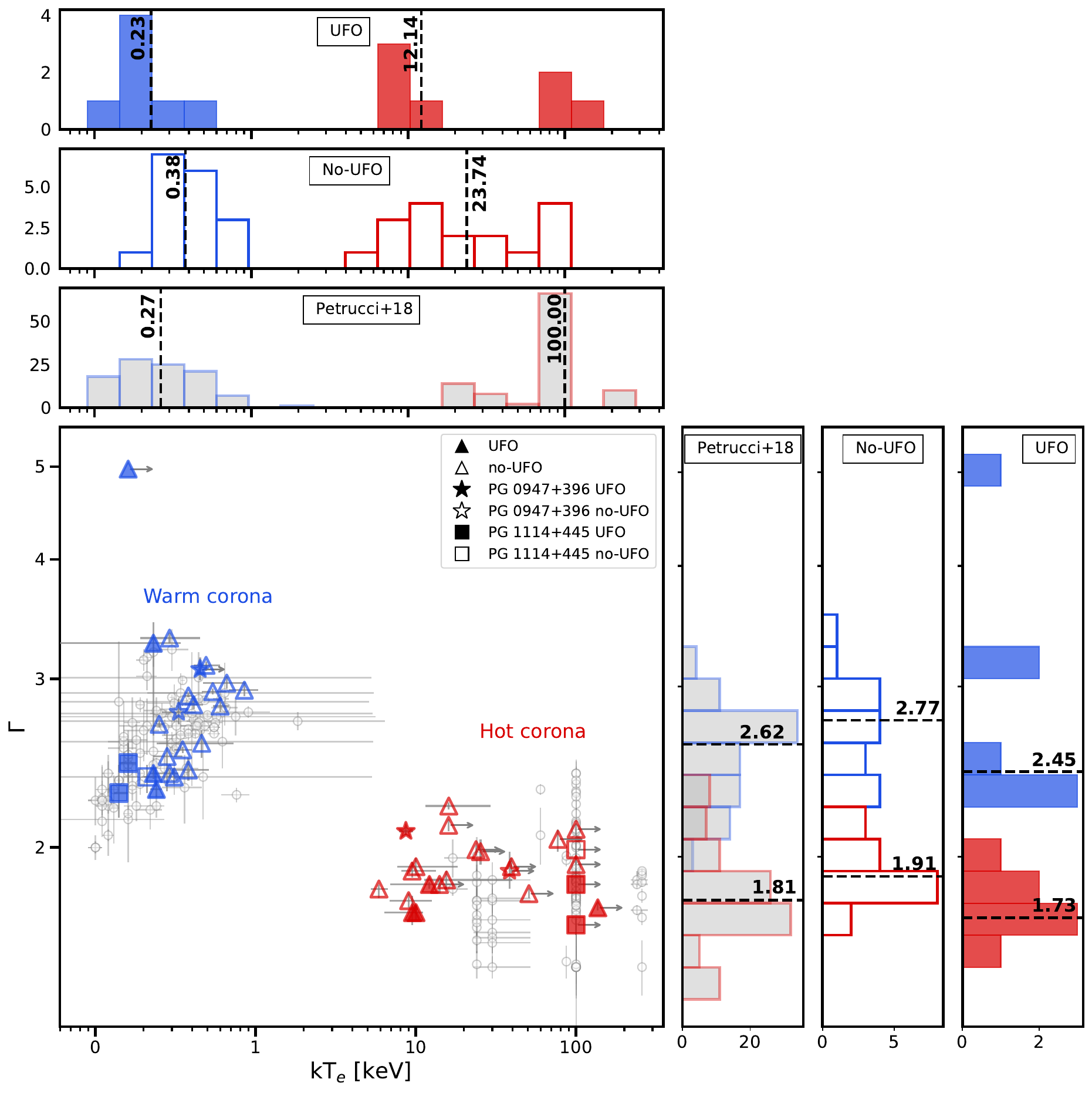}
    \caption[]
    {{{\it Main panel:} scatter plot of the warm and hot corona physical parameters. \citetalias{petrucci18} comparison sample is shown in gray points. In \citetalias{petrucci18}, the HC electron temperatures are fixed at kT$_{\rm e}=100$~keV, producing the vertical alignment in the plot. For 7/22 AGN of their sample, literature estimates of the cut-off energy are available \citep{laha2014,fabian15,jiang2019,tortosa23,vaia24}. For visualization purposes, we estimate their coronal temperatures as kT$_{\rm e}=E_{\rm cut}/2$ and show them in the plot; we stress that these estimates are not included in the analysis.
    {\it Upper and side histograms:} distributions for UFO AGN (outer), no-UFO AGN (middle), and the comparison sample (inner), with dashed lines and labels indicating the respective medians.}}
    \label{fig:2coronae_val}
\end{figure*}

\subsection{Warm and hot corona parameters}
\label{coronal_parameters}
The best-fit warm and hot coronal parameters span a broad range across the SUBWAYS sample for both two-coronae only (mod1/mod1*) and hybrid (mod2) models.
The derived values and their distributions are shown in Fig.~\ref{fig:2coronae_val}.

Focusing on the WC, the derived physical parameters (temperatures kT$_{\rm e} \simeq 0.1$–0.9 keV and photon indices $\Gamma \geq 2$) are consistent with previous studies (e.g., \citealp{magdziarz98,czerny03,done12}; \citetalias{petrucci18}). 
In particular, the photon index spans a broad range in the pure two-coronae models, with $\Gamma^{\rm WC} \sim 2.37-4.05$. For the hybrid model, the upper end of the range is driven by 2MASS~J10514425+3539306, for which only a very steep lower limit is obtained ($>$4.97). Excluding this AGN, the hybrid-model values cluster in the narrower range $\Gamma^{\rm WC} \sim 2.28-3.10$. 
The electron temperature of the WC is well constrained, with a slightly narrower range of values for sources fitted with mod2, i.e., kT$_{\rm e}^{\rm WC} \sim$ 0.14-0.49~keV. 
For each model, the distributions of both parameters are shown in Fig.~\ref{fig:distributions}.

Regarding the HC, the photon index spans $\sim$1.66–2.21 across the 24 spectra. The median values are $\Gamma_{\rm HC} \sim 1.83$ for the hybrid model and $\sim$1.91 and $\sim$1.96 for the two-coronae configurations (Fig.~\ref{fig:distributions}, first column of panel a). 
For AGN fit with the hybrid model, the constrained electron temperatures are in the kT$_{\rm e}^{\rm HC} \sim 9.5$-12.1~keV range, with four sources providing only lower limits, and three cases (PG~0804+761 and two observations, no-UFO and 101, of PG~1114+445) fixed at 100~keV due to lack of constraints (see Sect.~\ref{data}). 
In sources favoring mod1/mod1*, the constrained temperatures span kT$_{\rm e}^{\rm HC} \sim 5.9$–16.1 keV, while five AGN provide only lower limits and two (HB89~1257+286 and PG~1114+445, obsID 301) are fixed at 100~keV (see Fig.~\ref{fig:distributions}, second column of panel a).
Overall, the results for the HC are broadly consistent with those reported by \citet{peluso2026} (hereafter ``SUBWAYS V''). We refer to Sect.~\ref{subwaysv} for a detailed comparison between their and our results.

To assess whether differences between models are statistically significant, we perform the two-sample Kolmogorov-Smirnov (KS) test comparisons on the distributions of $\Gamma_{\rm WC}$, kT$_{\rm e}^{\rm WC}$, $\Gamma_{\rm HC}$, and kT$_{\rm e}^{\rm HC}$. We note that the relatively small sample sizes may limit the sensitivity of the test, thus the resulting significance levels should be interpreted with caution and together with the overall trends. Since mod1* is a variant of mod1, we first compare the combined mod1+mod1* sample and the mod2. Considering only constrained values, i.e., excluding fixed parameters and lower limits, we find no statistically significant differences; including all values leads to the same conclusion.

We then carry out pairwise comparisons among the individual model distributions. In almost all cases, the p-values are well above the 0.05 threshold, indicating statistically consistent distributions.
The only exceptions are kT$_{\rm e}^{\rm WC}$ for mod1 versus mod1* and mod1* versus mod2 when all values (including fixed parameters and lower limits) are considered, with Log(NHP)=$-$1.55 in both cases.
Since these differences arise only when fixed values and lower limits are included, they should be interpreted cautiously. Overall, despite small differences in the median values, the coronal properties inferred from the different model configurations are statistically consistent within the current sample. We refer to Sect.~\ref{comp_pop18} for a detailed discussion between UFO and no-UFO samples, and the comparison with \citetalias{petrucci18} work.

\section{Soft excess strength}\label{SE_strength}
In this section, we investigate the soft excess strength of the AGN in the SUBWAYS sample. 

\subsection{Model-independent and luminosity based diagnostics}
\label{SE_HR_Q}

We first compute the hardness ratio (HR) from the background-subtracted counts in the soft (0.5--2 keV) and hard (2--10 keV) bands, as a qualitative, model-independent indication of the spectral shape, reflecting the relative contribution of the soft and hard X-ray emissions. In particular, the HR is defined as:
\begin{equation}
HR = \frac{H - S}{H + S},
\end{equation}
where $S$ and $H$ are the counts in the soft and hard bands, respectively. As discussed by \citet{palit2024}, the absolute value of the HR depends on the instrumental response (e.g., XMM-{\it Newton}, similarly to other X-ray focusing facilities working below 10 keV, has a higher effective area in the soft band and thus, for similar flux levels in the two energy ranges, the HR will be actually negative). Moreover, the HR can be significantly affected by line-of-sight absorption. Therefore, the HR does not provide a direct measure of the SE strength.
However, extreme HR values can be reliably interpreted as indicating genuinely soft or hard spectra. The HR values are reported in Table~\ref{tab:se_par}.

18/21 AGN exhibit negative HR values in all reported observations, indicating that their spectra are generally dominated by soft X-ray emission. 2MASS~J10514425+3539306 and 2MASS~J16531505+2349427 show positive HR values, while PG~1114+445 displays both positive (UFO observations) and negative values (no-UFO observation).

To quantify the SE strength, a commonly adopted approach in the literature is to use a band-limited ratio between the soft excess emission and the primary hard X-ray continuum \citep[e.g.,][]{Gierlinski2004,Piconcelli2005,Crummy2006,Boissay2016,palit2024}. In this framework, the SE strength is estimated from the ratio of the soft-band flux (or luminosity) attributed to the excess component to that of the underlying continuum, measured over specific energy ranges. We note that these estimates are inherently model-dependent, as they rely on the spectral decomposition adopted to separate the soft excess and primary continuum components. A closely related formulation was introduced by \citet{Vasudevan2013} and recently revisited by \citet{jana2026}, who define a dimensionless parameter $Q$ to quantify the relative strength of the SE compared to the primary X-ray continuum (PC):
\begin{equation}
Q = \log_{10} \left( \frac{L_{\text{SE, 0.5--2.0 keV}}}{L_{\text{PC, 2.0--10 keV}}} \right),
\end{equation}
where $L_{\rm{SE, 0.5-2.0\,keV}}$ is the luminosity of the soft excess component in the 0.5--2.0~keV energy band, and $L_{\rm{PC, 2.0-10\,keV}}$ is the luminosity of the primary intrinsic power-law continuum in the 2.0--10~keV range.
For consistency with the adopted definition and to provide a homogeneous comparison across the whole sample, in both the two-coronae and hybrid scenarios $L_{\text{SE}}$ is computed considering only the \texttt{nthComp\_WC} component. We stress that, in the hybrid model, this may not capture the full soft excess emission, as blurred reflection can also contribute significantly below 2 keV.
In our sample (excluding 2MASS~J10514425+3539306 for which no SE is detected), the WC and disk reflection components contribute 34--54\% and 2--18\%, respectively, to the total 0.5--2 keV flux. The corresponding values for individual AGN are listed in Table~\ref{tab:hybrid_soft_band_contributions}. Including the disk reflection contribution would increase the inferred SE luminosity by a factor of $\sim$1.05--1.5, corresponding to a change in $Q$ of only $\Delta Q \simeq 0.02$--0.18 dex. Therefore, although $Q$ formally traces only the WC contribution and may underestimate the total SE strength in hybrid sources, this effect is modest for the SUBWAYS sample.

We derive the $Q$ parameters for the SUBWAYS sample and report the values in Table~\ref{tab:se_par}.
The parameter spans a broad range with $-5.1\leq  Q \leq -1.6$, and most sources (19 out of 21 AGN) show $Q \geq -3$, suggesting that the soft excess luminosity relative to the primary continuum is broadly similar across the majority of the sample. Only two AGN, 2MASS~J10514425+3539306 and WISE~J053756.30$-$024513.1, exhibit significantly lower values ($Q \leq -3.5$), indicative of a very weak or absent soft excess.
In WISE~J053756.30$-$024513.1, modeled with mod1*, the \texttt{nthComp\_WC} component is statistically required ($\Delta\chi^2 = 247.5$ for 3 additional free parameters) by the data and together with $Q = -3.85^{+1.13}_{-0.29}$, indicates a weak SE. In contrast, in 2MASS~J10514425+3539306 ($Q=-5.07 \pm 1.37$ and modeled with mod2), the \texttt{nthComp\_WC} component accounts only for the optical/UV emission associated with the AD, implying that the SE is absent.

\subsection{Warm corona physical diagnostic}
\label{SE_tau_A}
As a second diagnostic, we investigate the radiative balance of the warm Comptonizing medium following \citet[][``P20'' hereafter]{petrucci2020}.
In an accretion disk--warm corona system, the spectral shape of the Comptonized emission is governed by the competition between the power dissipated within the corona and the flux of soft seed photons supplied by the AD. The former heats the electrons, while the latter cools them through inverse Compton scattering. 

This balance is described through the Compton amplification factor, $A$, with larger values indicating a greater contribution from coronal heating relative to the seed photon supply, and values closer to unity indicating more efficient cooling by disk photons. To estimate $A$, we first derive the Thomson optical depth of the WC, $\tau^{\rm WC}$. Its value depends on the adopted prescription and geometrical assumptions \citep[e.g.,][]{Middei2019,peluso2026}, with larger uncertainties in the high-optical depth regime characteristic of the WC. We therefore tested both \citet{Beloborodov1999} and \citet{Longair2011} prescriptions; the full comparison for $\tau^{\rm WC}$, $A$, and $\chi$ is provided as online material (link in Appendix~\ref{app_c}). For consistency with \citetalias{petrucci2020}, we adopt the Beloborodov formulation throughout the main analysis\footnote{\label{fn:tau} We estimate the WC optical depth following \citetalias{petrucci2020}, using our best-fit spectral parameters ($\Gamma^{WC}$ and kT$_{\rm e}^{WC}$) together with Eq.~(13) of \citet{Beloborodov1999}, i.e., $\Gamma \simeq \frac{9}{4} \, y^{-2/9}$ with the Compton parameter defined as $y = 4\left[\frac{kT_e}{m_e c^2} + 4\left(\frac{kT_e}{m_e c^2}\right)^2\right]\tau(\tau + 1)$.}.
Following Appendix~A of \citetalias{petrucci2020}, we can define $A$ as: 
\begin{equation}\label{eq3}
A = 1+\frac{F_{\rm cor}}{F_{\rm tot,disk}},
\end{equation}
where $F_{\rm cor}$ is the power generated through internal dissipation in the WC and $F_{\rm tot,disk}$ is the total
seed photon flux emitted by the AD and entering WC before Comptonization. 
Thus, $A-1$ measures the coronal heating relative to the AD seed photon flux available for Compton cooling. We estimate $F_{\rm tot,disk}$ by reconstructing the un-Comptonized seed photon spectrum from our best-fits, forcing the \texttt{nthComp\_WC} component to a very steep spectral slope ($\Gamma=10$) while preserving the seed photon normalization. To derive $F_{\rm cor}$, we first consider the flux emerging from an optically thick, plane-parallel WC, $F_{\rm out}$, which is derived from the unabsorbed best-fitting \texttt{nthComp\_WC} component. Then, rearranging Eq.~(A.3) of \citetalias{petrucci2020} to solve for the coronal flux, we obtain: 
\begin{equation}
F_{\rm cor} =
2\left[
F_{\rm out}
-
\frac{F_{\rm tot,disk}}
     {1+\frac{3}{4}\tau^{\rm WC}}
\right],
\end{equation}
and the amplification factor can be derived directly from Eq~\ref{eq3}.

For a WC fully covering a non-dissipative AD, Eq.~\ref{eq3} is:
\begin{equation}
A= 1+
\frac{2}
     {1+\frac{3}{4}\tau^{\rm WC}},
\end{equation}
Within this limit, an increase in optical depth enhances the radiative feedback between AD and WC: disk photons undergo many scatterings and provide an increasingly strong cooling field relative to the additional coronal heating. Consequently, $A-1$ becomes small
and $A$ approaches unity.

Lastly, we characterize the WC--AD radiative balance by defining the coronal dissipation fraction $\chi$ (see Eq.~(6) of \citetalias{petrucci2020} for the full derivation) as:
\begin{equation}
\chi \equiv \frac{F_{\rm cor}}{F_{\rm out}}.
\end{equation}
Within the geometry considered by \citetalias{petrucci2020} and adopted here, $\chi = 1$ represents radiative equilibrium between the upward and downward fluxes at the base of the WC, as expected for a fully covering WC above a non-dissipative AD. 
Values $0\leq\chi<1$ imply a net upward flux entering the WC from below and are therefore consistent with a slab-like or extended WC above an intrinsically dissipative AD. For $1<\chi<2$, the upward flux entering the WC at its base is smaller than the downward one. A configuration compatible with a patchy WC above a weakly dissipative or non-dissipative AD.

The optical depth, amplification factor, and coronal dissipation fraction for the SUBWAYS sample are reported in Table~\ref{tab:se_par}. For both two-coronae and hybrid models, these quantities are computed considering only the \texttt{nthComp\_WC} component. Since no significant soft X-ray excess is detected in 2MASS~J10514425+3539306 (see Sect.~\ref{SE_HR_Q}), we do not derive $A$ and $\chi$ for this source.

\begin{figure}[]
\centering
\hspace*{-0.7cm}
    \includegraphics[width=0.53\textwidth, trim=0 5 5 0,clip]{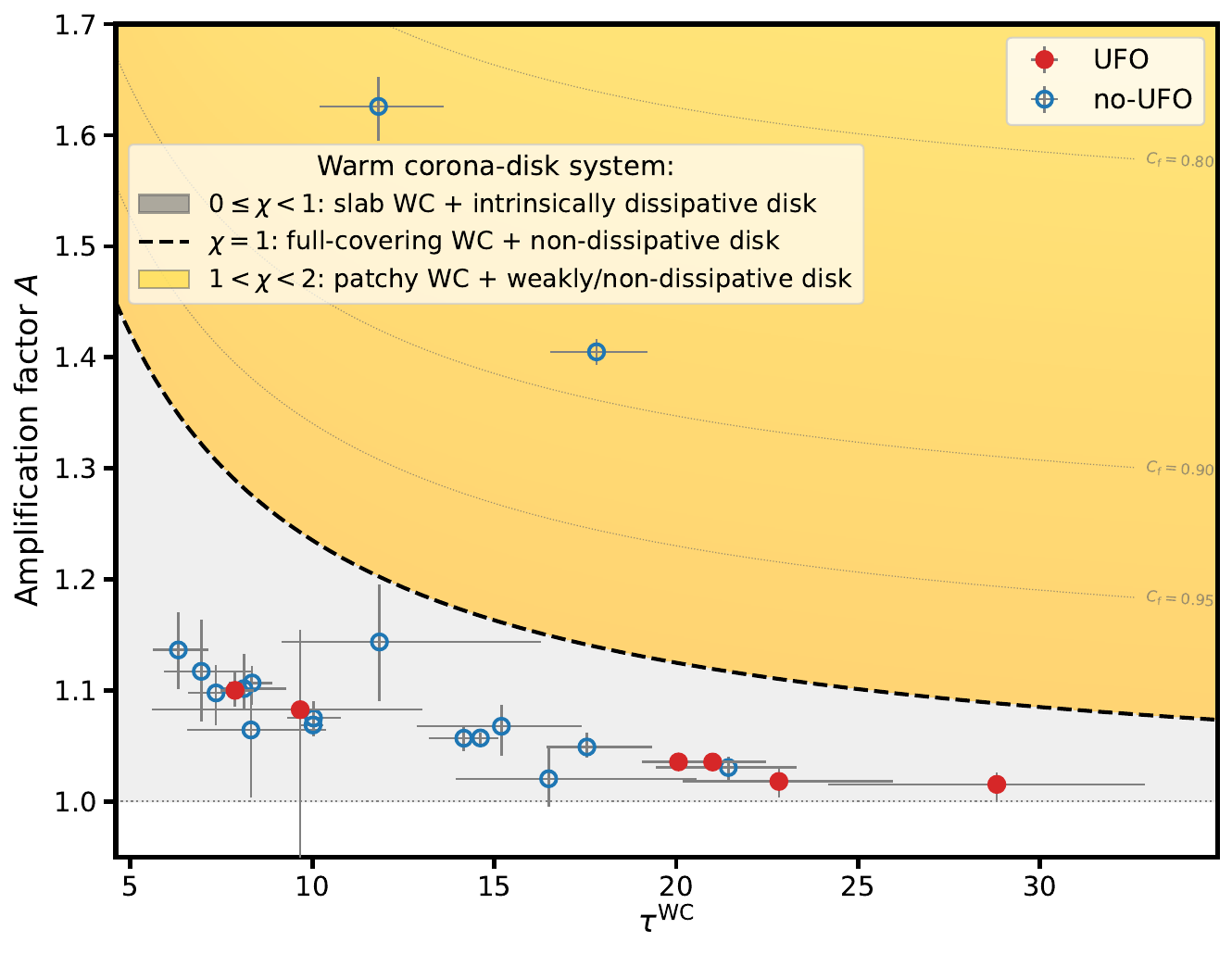}
    \caption[]
    {{$\tau^{\rm WC}$--$A$ plane. The dashed black curve marks $\chi=1$, corresponding to a WC fully covering a non-dissipative AD. The gray region between $A=1$ and the $\chi=1$ curve is consistent with an extended or slab-like WC above an intrinsically dissipative AD. The region above the dashed curve, i.e. $1<\chi<2$, is compatible with a patchy WC above a weakly or non-dissipative AD. In this region, the color scale shows the fraction of the disk surface, C$_{\rm f}$, covered by the warm Comptonizing layer, assuming a non-dissipative AD; lower values correspond to increasingly patchy configurations. Given that C$_{\rm f}$ and the intrinsic disk dissipation are degenerate, the color scale should be regarded as indicative rather than as a measurement of the geometry. The region below $A=1$ is reported in white as it implies $F_{\rm cor}<0$.
    Filled red and open blue circles indicate UFO and no-UFO sources, respectively. 2MASS~J10514425+3539306 is not shown as no significant SE is detected.
    $\tau^{\rm WC}$ is derived adopting \citet{Beloborodov1999} prescription.}}
    \label{fig:A_tau_geometry}
\end{figure}

Fig.~\ref{fig:A_tau_geometry} shows the $\tau^{\rm WC}$--$A$ plane. For 15 AGN of the SUBWAYS sample (17 observations, including PG~0947+396 no-UFO and all observations of PG~1114+445), $\tau^{\rm WC}$ falls within the expected interval for an optically thick WC (i.e., $\sim$10--40; \citealp{magdziarz98,czerny03,done12}; \citetalias{petrucci18}). Meanwhile, six sources (including PG~0947+396 UFO spectra) show lower values $\tau \simeq 6-9$. For 2MASS~J10514425+3539306, only a lower limit is obtained ($\tau > 4.3$), preventing a firm classification.
The amplification factor generally spans $A\simeq1.0$--1.1, indicating efficient WC cooling. HB89~1257+286 and PG~0947+396 (no-UFO observatio) have $A \gtrsim 1.4$, implying a larger coronal heating contribution.

Focusing on $\chi$, 18 sources (including PG~0947+396 UFO observation) are consistent with an extended or slab-like warm corona above an intrinsically dissipative disk ($\chi < 1$). WISE~J053756.30$-$024513.1 has $\chi=0.83\pm0.30$, consistent with $\chi=1$ within uncertainties.
HB89~1257+286 and PG~0947+396 (the no-UFO observation), which populate the upper part of Fig.~\ref{fig:A_tau_geometry}, have $\chi \sim 1.5$, suggesting a patchy WC covering a weakly or non-dissipative disk.

Finally, we test whether the inferred WC properties depend on the adopted best-fit model. The two-coronae and hybrid sub-samples span strongly overlapping ranges, with no statistically significant differences between their distributions.

\subsection{SE diagnostics in sub-samples with and without UFOs and WAs}
\label{SE_UFO_model}
\begin{figure*}[]
\centering
    \includegraphics[width=1\textwidth, trim=320 310 320 160,clip]{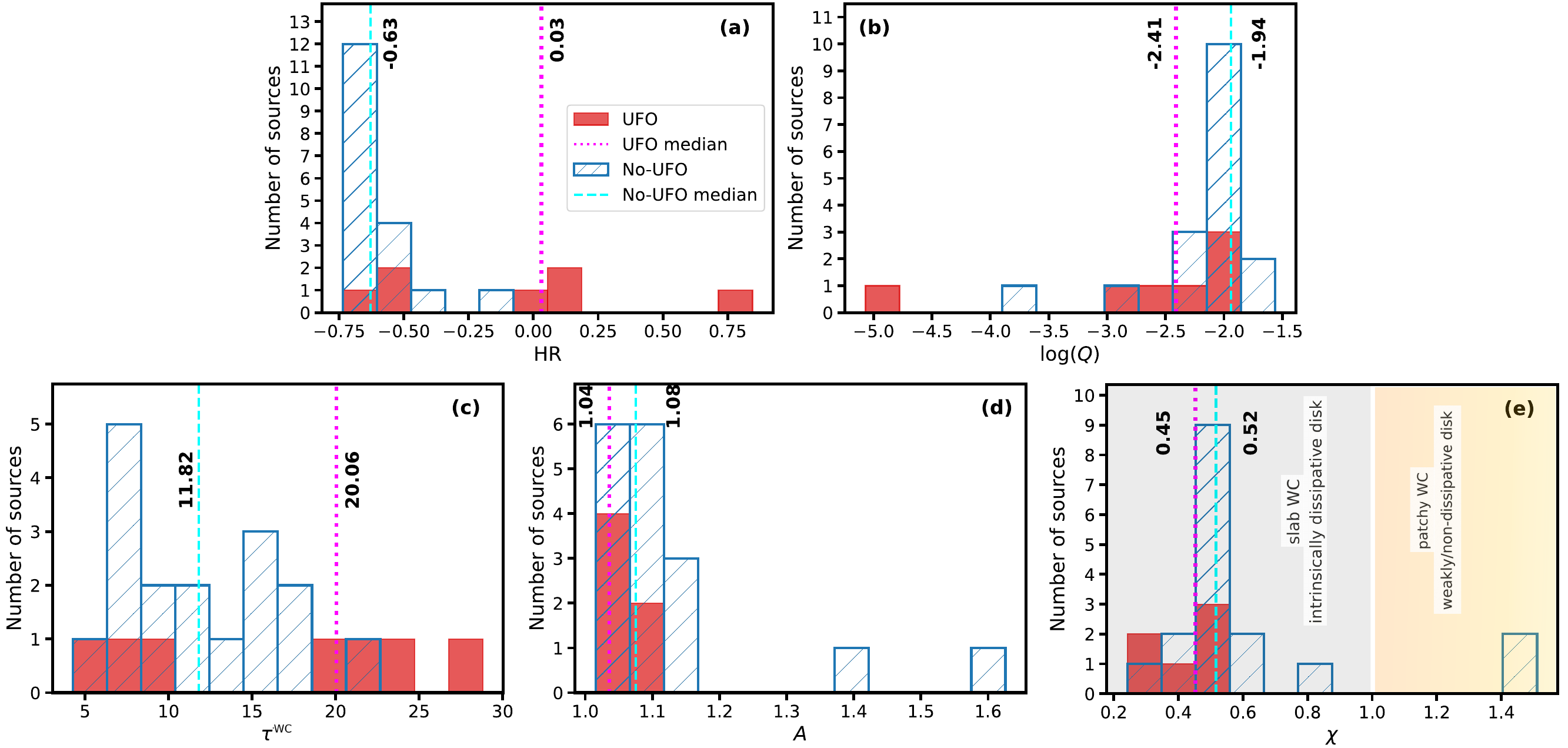}
    \caption[]
    {{Distributions of the SE related parameters for the UFO and no-UFO sub-samples. {\it Panel a and b:} hardness ratio and $Q$ parameter. {\it Panel c and d:} warm corona optical depth and amplification factor $A$. {\it Panel e:} coronal dissipation fraction $\chi$. The range consistent with a slab WC covering a dissipative AD is shown in gray, while the patchy WC over a weakly or non-dissipative disk is shown in yellow.
    The magenta and cyan dashed lines show the median values for the UFO and no-UFO sub-samples distributions. The KS tests indicate a statistically significant difference between the HR distributions of the two sub-samples.}}
    \label{fig:distr_se_par}
\end{figure*}

Focusing the UFO and no-UFO sub-samples, we compare the distributions of the SE related parameters (HR, $\log(Q)$, $\tau^{\rm WC}$, $A$, and $\chi$) in Fig.~\ref{fig:distr_se_par}. The KS tests indicate differences in the HR (Log(NHP)=$-$1.70) distribution, with the UFO sub-sample generally showing positive HR. We note that, given the limited sample size, this result should be regarded with caution.

Although no significant differences emerge, we report the results of PG~0947+396 and PG~1114+445, for which both UFO and no-UFO observations are available. In PG~0947+396, the HR and $Q$ values are consistent between the two observations (within the uncertainties) and are also compatible with the range spanned by the rest of the sample. The main difference between the two epochs is found in the WC optical depth, $A$, and $\chi$. In particular, the UFO observation shows lower values than the no-UFO observation (see Table~\ref{tab:se_par}).
For PG~1114+445, the three observations show  consistent values of both $A$ and $\chi$, and high optical depths of the WC. However, the two UFO observations are characterized by positive HR values, whereas the no-UFO observation presents a negative HR.

We also repeat the same analysis by dividing the sample into AGN with and without WA, finding no statistically significant differences in the distributions of the SE-related parameters.

\section{Discussion}\label{discussion}

In the following sections, we compare the results obtained from modeling the (quasi)simultaneous optical/UV/X-ray emission of the SUBWAYS sample with previous X-ray studies of AGN coronal properties and SE. In particular, we consider the recent analysis of the SUBWAYS sample presented in \citet{peluso2026} and the multi-epoch study of \citetalias{petrucci18}. These comparisons allow us to assess the robustness of our findings and highlight the impact of different observational and methodological approaches.
In addition, we perform a correlation analysis among the WC parameters, SE strengths, AGN and UFO properties, to identify any statistically significant relation.

\subsection{Comparison with \citetalias{peluso2026}}\label{subwaysv}

Our work extends the study of AGN coronal properties presented in \citetalias{peluso2026} by incorporating a broader multiwavelength context.
While they focus on the hot corona, and therefore the X-ray band, assuming a two-coronae scenario for the SE, here we focus on the warm corona and take advantage of simultaneous optical/UV and X-ray data to explore both a pure two-coronae configuration and hybrid model that include relativistic reflection (see Sect.~\ref{comp} for further details on the methodology and models). This approach allows us to assess the relative contribution of different physical mechanisms to the SE and directly constrain the seed photon population of the Comptonization process, thereby coupling the disk and coronal emission in a physically self-consistent framework. In particular, while \citetalias{peluso2026} fix kT$_{\rm BB} = 10$ eV, we constrain it from the fits and find generally lower values (with median 2.57$^{+1.99}_{-1.52}$ eV), reducing the dependence of the inferred coronal parameters on this assumption.
A further difference concerns the data selection. We prioritize optical/UV and X-ray observations that are as simultaneous as possible to minimizes the impact of source variability across wavelengths and within the same energy band. As a result, our selection of \nustar\ observations differs from that of \citetalias{peluso2026} (see Sect.~\ref{data} and Table~\ref{tab:obs_info} notes).

We compare our WC optical depths with those obtained by \citetalias{peluso2026} from their \texttt{compTT} fits. The values are broadly consistent, although source-to-source differences remain. In 10 observations, the Longair slab prescription gives values closer to the \texttt{compTT} estimates than the Beloborodov relation, while in the remaining cases the agreement is comparable. This further highlights the systematic uncertainty in deriving $\tau^{\rm WC}$. A detailed comparison of the two prescriptions adopted here is provided as online material (see Appendix~\ref{app_c}).

For the remaining comparison, we consider the two-coronae model of \citetalias{peluso2026} in which the WC is described with \texttt{nthComp} (their ``Model 2''). The WC temperatures are broadly consistent between the two studies. \citetalias{peluso2026} find a relatively uniform distribution with median kT$_{\rm e}^{\rm WC}\sim0.40$ keV, consistent with our two-coronae configurations (mod1 and mod1*), for which the median is 0.32 keV and the 16th--84th percentile range is 0.19--0.40 keV.
We also find a dependence on the adopted spectral model, with the hybrid configuration yielding lower WC temperatures, with a median of $\sim0.24$ keV. Thus, although a warm, optically thick corona is commonly present, the SE may arise from a combination of warm Comptonization and reflection whose relative contributions vary among sources and depend on the modeling assumptions (see Sect.~\ref{analysis}). This points toward a more complex physical picture, in which the geometry and radiative processes of the inner accretion flow vary from source to source.

We detect the SE in a larger fraction of AGN than \citetalias{peluso2026}, who found that six sources did not require this component. We confirm its absence only in 2MASS~J10514425+3539306 (see Sect.~\ref{SE_HR_Q}), while WISE~J053756.30$-$024513.1 shows only a very weak SE. This may reflect the greater sensitivity provided by the optical/UV data and the fact that, in our fits, the \texttt{nthComp\_WC} component also accounts for disk emission. 

The HC appears more sensitive to both model choice and data selection. For 14 AGN, both studies provide either a constrained HC temperature or a lower limit. Restricting our comparison to mod1 and mod1*, and including both constrained values and lower limits, we obtain a median kT${\rm e}^{\rm HC}=22.6^{+16.0}{-6.4}$ keV\footnote{The median is estimated using a log-normal likelihood that accounts for both constrained values and lower limits. A constrained measurement contributes through the log-normal probability density evaluated at its best-fit value, and are propagated directly through bootstrap resampling. A lower limit is treated as indicating that the true temperature lies above the quoted value. Measurements fixed at 100~keV are excluded. The reported interval corresponds to a 68\% c.l. on the median.}, consistent (within uncertainties) with the value reported by \citetalias{peluso2026}, kT$_{\rm e}^{\rm HC}\simeq15$ keV with a 68\% confidence interval of 14--17 keV. 

Considering the UFO and no-UFO sub-samples (as further discussed in Sect.~\ref{comp_pop18}), we find a statistically significant harder $\Gamma^{\rm HC}$ distribution in UFO sources. Although this may appear in tension with \citetalias{peluso2026}, who associate UFOs with lower high-energy cut-offs and hence cooler HC, the $\Gamma$--kT${\rm e}$ behavior of the SUBWAYS sample shows that flatter spectra are preferentially associated with cooler, optically thicker coronae, implying that the spectral slope is largely driven by changes in optical depth rather than temperature alone \citepalias[see][]{peluso2026}. The two results are therefore physically consistent, both pointing to cooler and more optically thick HC in UFO hosts. The median kT${\rm e}^{\rm HC}$ is also lower for the UFO sub-sample, although this difference is not statistically significant and should be interpreted cautiously because many HC temperatures are constrained only by lower limits.
In addition, the relatively low HC temperatures obtained in some fits should be interpreted with caution. In the SUBWAYS flux and redshift regime, the models lose sensitivity to high coronal temperatures, potentially biasing the estimates toward lower values. We refer to \citetalias{peluso2026} for a dedicated discussion of this caveat and of the HC properties of the sample.

\subsection{Comparison with \citetalias{petrucci18}}\label{comp_pop18}

Here, we compare the SUBWAYS results with those of \citetalias{petrucci18}, who apply the two-coronae model to broadband optical/UV/X-ray data for 22 AGN, analyzing 100 XMM-\textit{Newton} observations with simultaneous EPIC-pn and OM coverage over timescales from a few days to $\sim$15 years.  Their study represents one of the most extensive multi-epoch and multi-band applications of the two-coronae model to date, and therefore provides a valuable benchmark for our analysis.

The \citetalias{petrucci18} sample spans a broader range in redshift, bolometric luminosity, and Eddington ratio than SUBWAYS: $0.01\leq z\leq0.57$ versus $0.09\leq z\leq0.36$, (Log(L$_{\rm bol}/{\rm erg\,s^{-1}}) = 43.2$--46.3 versus 44.8--46.2, and $L_{\rm bol}/L_{\rm Edd}=0.009$--1.89 versus 0.02--0.99. The BH mass distributions are broadly comparable, although SUBWAYS extends to slightly higher masses, up to Log(M$_{\rm BH}/M_\odot) = 9.2$ compared with 9.0. Together, the two datasets provide one of the largest collections of AGN in which the soft excess has been studied with contemporaneous optical/UV and X-ray coverage, enabling a comprehensive comparison of physical properties across different sources, epochs, and regions of parameter space.
A relevant difference between the two analyses concerns the electron temperature of the hot corona. For the SUBWAYS sample, kT$_{\rm e}^{\rm HC}$ is derived self-consistently from the spectral analysis (see Sect.~\ref{comp}). On the contrary, \citetalias{petrucci18} fixed this parameter to 100~keV in the absence of simultaneous \nustar\ observations that would allow constraints on the high-energy cut-off. In Fig.~\ref{fig:2coronae_val}, we show the distributions (and the medians) of the electron temperature and photon index for both WC and HC in the two samples. 

To assess the statistical differences between our sample and the comparison sample, we perform the two-sample KS tests on the distributions (i.e., considering kT$_{\rm e}^{WC}$, $\Gamma^{WC}$, and $\Gamma^{HC}$). We find no statistically significant differences (see upper section of Table~\ref{tab:ks_results_gamma_kte}). Given the different methodologies adopted, a direct comparison in terms of kT$_{\rm e}^{HC}$ between the two works is not attempted.
\begin{table}[h]
\centering
\renewcommand{\arraystretch}{1.2}
\caption{Comparison between the UFO, No-UFO, and \citetalias{petrucci18} samples.}\label{tab:ks_results_gamma_kte}
\begin{tabular}{c| c c c c}
Samples & kT$_{\rm e}^{WC}$ & $\Gamma_{WC}$  & $\Gamma_{HC}$ \\
\hline\hline
SUBWAYS vs. P18 & {\bf x} & {\bf x} & {\bf x} \\ \hline
UFO vs. No-UFO & $-$2.90 & {\bf x} & $-$1.58 \\
UFO vs. P18   & {\bf x} & {\bf x} & {\bf x} \\
No-UFO vs. P18 & $-$2.23 & {\bf x} & $-$2.00 \\
\hline
UFO vs. P18 (UFO)\\
All obs. & {\bf x} & {\bf x} & {\bf x} \\
Medians  & {\bf x} & {\bf x} & {\bf x} \\
Monte Carlo & {\bf x} & {\bf x} & {\bf x} \\
\end{tabular}
\begin{flushleft}
\footnotesize  \textbf{Notes.} The lower the Log(NHP) values, the more statistically different the compared samples become. We mark ``\textbf{x}'' when the difference between two samples is below the adopted significance threshold (Log(NHP) $> -1.30$, i.e., the compared samples are statistically consistent). 
For the \citetalias{petrucci18} UFO sub-sample, ``All obs.'', ``Medians'', and ``Monte Carlo'' refer to the three different ways of treating the multiple observations per source.
\end{flushleft}
\end{table}

We then repeat the KS analysis by considering the SUBWAYS UFO and no-UFO sub-samples, and compare each of them independently with the \citetalias{petrucci18} sample. The results are reported in the middle section of Table~\ref{tab:ks_results_gamma_kte}. 
Firstly, focusing on the SUBWAYS sample, kT${\rm e}^{WC}$ and $\Gamma^{HC}$ differ significantly between the two groups, with AGN hosting UFOs showing lower values in both cases. 
Since two AGN (PG~0947+396 and PG~1114+445) are observed in both UFO and no-UFO states, we repeat the KS tests at the source level to assess the impact of repeated observations\footnote{We construct two alternative source-level samples. In the first, only the UFO-state observations are retained, excluding the corresponding no-UFO epochs; when more than one UFO observation is available for the same source, we use the median value. In the second, only the no-UFO observation is retained and the UFO epochs are excluded. In this way, each AGN contributes only once to a given KS test comparison.}. The differences in both $\Gamma^{\rm HC}$ and kT${\rm e}^{WC}$ remain statistically significant.
Physically, lower kT$_{\rm e}^{WC}$ values in UFO sources may indicate that the thermal state of the warm Comptonizing region is connected to the presence of fast winds. This could point to a more efficient cooling of the WC when UFOs are present, to a different vertical structure of the inner accretion flow, or to part of the accretion power being redistributed into launching and/or accelerating the wind rather than heating the warm corona.
For $\Gamma^{\rm HC}$, a harder spectrum can in general reflect changes in the coronal temperature, optical depth, heating-to-cooling balance, or geometry of the X-ray emitting region. In the specific parameter regime probed by SUBWAYS, however, the $\Gamma$--$kT_{\rm e}$ relation favors an interpretation in terms of cooler and optically thicker hot coronae for the harder-spectrum sources, as discussed in Sect.~\ref{subwaysv}. We caution that these results suggest an interplay between UFOs and coronal properties in the current sample, rather than establishing whether the wind and corona are causally connected or driven by a common underlying parameter. 
Secondly, the SUBWAYS UFO sub-sample is statistically consistent with the full \citetalias{petrucci18} sample. By contrast, the no-UFO sub-sample differs from \citetalias{petrucci18} in kT$_{\rm e}^{WC}$ (Log(NHP) = $-$2.23) and $\Gamma^{HC}$ (Log(NHP) = $-$2.00).

We also repeated the KS tests separating the sources according to the presence or absence of warm absorbers, and found no statistically significant differences in either the warm or hot corona parameters.

Five AGN in \citetalias{petrucci18} host UFOs (1H~0419-577, Mrk~509, NGC~4593, PG~0804+761, and PG~1114+445). We therefore assess whether the best-fit corona parameters derived for these sources are statistically comparable to those found in our UFO sub-sample.
We address this via a multi-step statistical approach.
First, we perform the KS test by considering all individual observations \citepalias[for a total of 45 observations for the five sources, see][]{petrucci18} as independent data points. Second, we re-derive the KS test using only the median values for each of the five AGN to ensure the results were not biased by sources with more frequent monitoring.
We adopt a Monte Carlo approach to further validate these findings\footnote{Our approach considers $5 \times 10^3$ iterations where the values from \citetalias{petrucci18} are resampled from Gaussian distributions. The uncertainties reported in \citetalias{petrucci18} are quoted at the 90\% c.l. and, therefore, we convert them to approximate 1$\sigma$ equivalents by dividing by 1.645.}.
In all cases, the p-values remain above the significance threshold for the tested parameters (i.e., kT$_{\rm e}^{WC}$, $\Gamma^{WC}$, and $\Gamma^{HC}$; lower section of Table~\ref{tab:ks_results_gamma_kte}). These results confirm that the corona properties of the AGN hosting UFO in the \citetalias{petrucci18} sample are statistically consistent with our own best-fit results, demonstrating the robustness of our findings against spectral variability and observational uncertainties.

Additionally, we compare the seed photon temperatures (derived from optical/UV/X-ray data) obtained for our AGN sample with those reported by \citetalias{petrucci18}. The two distributions are statistically consistent according to a KS test (Fig.~\ref{fig:ktbb_comp}). A small sub-set of sources in both samples (PG~1435$-$067 in SUBWAYS and 1H~0419-577, IRASF~12397+3333, and RE~1034+396 in \citetalias{petrucci18}) shows high temperatures (kT$_{\rm bb} \gtrsim 10$ eV). We investigate whether these AGN share common physical or spectral properties, but found no statistically significant differences with respect to the rest of the population. Details of this analysis are reported in Appendix~\ref{comp_kt_bb}.

\subsection{Correlations analysis}\label{corr}

We perform a Spearman correlation analysis (accounting for asymmetric uncertainties\footnote{To properly account for asymmetric uncertainties, we adopt the perturbation method by \citet{gianolli24}, which also estimate the linear regressions and corresponding uncertainty between parameter pairs.}) to investigate possible relations among the best-fit spectral parameters of the warm corona (i.e., seed photon temperature, X-ray photon index, electron temperature, and optical depth), SE diagnostics (i.e., hardness ratio, $Q$ parameter, amplification factor, and coronal dissipation fraction), AGN \citep[i.e., redshift, BH mass, full width at half maximum of the broad H$\beta$ emission line, monochromatic UV luminosity at 2500 \AA, bolometric and ionizing luminosities, Eddington ratio, X-ray/UV ratio, and X-ray weakness; see][for the derivation and a complete list of values]{gianolli24}, and UFO observed \citep[ionization parameter, column density, and outflow velocity; the values are taken from][]{Matzeu2023} and derived \citep[distance between the wind and the central source, mass outflow rate, mechanical power, and outflow momentum rate; the values are taken from][]{gianolli24} parameters. The significant correlation plots are provided as online material, with links listed Appendix~\ref{app_c}. For correlations involving the hot corona properties, we refer to \citetalias{peluso2026}.

We find a tentative anticorrelation between $\tau^{\rm WC}$ and the UFO outflow velocity, with Log(NHP)=$-$1.44. Given the small number of UFO-hosting AGN, this result should be interpreted cautiously. No other significant correlations are found between WC and UFO properties.

Among the SE diagnostics, the amplification factor $A$ positively correlates with $kT_{\rm e}^{\rm WC}$ (Log(NHP)=$-$3.39), considering only AGN with a slab WC above a dissipative disk (see Sect.~\ref{SE_tau_A}). We also find a weaker positive correlation between $\tau^{\rm WC}$ and HR.

Among the WC parameters, $\tau^{\rm WC}$ anticorrelates with both $\Gamma_{\rm WC}$ and $kT_{\rm e}^{\rm WC}$, with Log(NHP)=$-$7.63 and $-$2.47, respectively. These trends are expected because $\tau^{\rm WC}$ is derived from the photon index and electron temperature through the adopted Comptonization relation\textsuperscript{\ref{fn:tau}}.

We identify four significant correlations between AGN and WC properties. $\Gamma_{\rm WC}$ positively correlates with $L_{2500,\AA}$, with Log(NHP)=$-$3.19. Although the trend is visible across the full range of $\Gamma_{\rm WC}$, it is characterized by substantial intrinsic scatter ($0.39\pm0.01$ dex), with a broad dispersion in $L_{2500 \AA}$ at similar $\Gamma_{\rm WC}$.
Moreover, $\tau^{\rm WC}$ anticorrelates with $L_{2500,\AA}$, with Log(NHP)=$-$2.47 and it is likely related to the previous trend through the dependence of $\tau^{\rm WC}$ on $\Gamma_{\rm WC}$. We also find weaker positive trends between kT$_{\rm e}^{\rm WC}$ and the bolometric luminosity. A weak correlation with the Eddington ratio is formally present, but should be treated cautiously: most AGN cluster at $\lambda_{\rm Edd}\lesssim0.2$ over a relatively narrow range of temperature, while the apparent trend is influenced by a small number of high-$\lambda_{\rm Edd}$ objects and is accompanied by a high intrinsic scatter of $0.26\pm0.06$ dex.

\section{Summary and Conclusions}\label{conclusions}

In this paper, we present the first systematic broadband and multiwavelength study of the possible interplay between the soft X-ray excess and ionized winds in AGN. To this aim, we analyze 21 quasars (0.1 $\leq$ $z$ $\leq$ 0.4) from the SUBWAYS sample, leveraging simultaneous or quasi-simultaneous multiwavelength observations from XMM-{\it Newton}, \nustar, and HST (see Sect.~\ref{data} for the sample selection and Table~\ref{tab:obs_info} for the list of adopted observations for each instrument). This broadband coverage enables self-consistent modeling of the disk–corona system and provides a unique opportunity to explore the connection between the inner accretion flow and outflowing material in a sample where six out of 21 sources host ultra-fast outflows and 9 quasars show X-ray warm absorbers (see Table~\ref{tab:sources_list} for the list of AGN hosting UFOs and WAs).

We model the optical/UV/X-ray data within the two-coronae framework, which includes emission from a warm and a hot Comptonizing medium together with neutral reflection from distant material (e.g., the torus). We systematically test whether this baseline model is sufficient to reproduce the data, or whether an additional relativistic reflection component (from the accretion disk) is required (see Sect.~\ref{data} for the methodology and models description). This procedure provides a homogeneous set of coronal properties and SE diagnostics for the full SUBWAYS sample, allowing us to perform consistent comparisons and correlation analyses.

Our main results can be summarized as follows:
\begin{itemize}
\item \textbf{The soft X-ray excess (SE) explained by the two-coronae and the hybrid models.}
    A warm, optically thick corona represents an essential component of the AGN emission in the SUBWAYS sample (20 out of 21 quasars, 2MASS~J10514425+3539306 is the only AGN without SE). The warm corona (WC) physical properties, electron temperatures kT$_{\rm e} \simeq$ 0.1--0.9 keV and photon indices $\Gamma \geq 2$, are consistent with previous studies (see Sect.~\ref{analysis} for details and Table~\hyperlink{fit-table}{\ref*{tab:fit}} for the spectral modeling results).
    In nine out of 21 sources, the inclusion of relativistic reflection significantly improves the fit. This indicates that the soft excess is frequently best described by a combination of warm Comptonization and disk reflection, rather than by a single universal mechanism.

    \item \textbf{The SE diagnostics across the sample.}
    The soft excess shows broadly consistent properties across the adopted diagnostics: hardness ratio HR, $Q$ parameter, optical depth $\tau^{\rm WC}$, amplification factor $A$, and coronal dissipation fraction $\chi$. 
    Most AGN cluster within the expected ranges, although a few outliers are present (see Sect.~\ref{SE_strength} for a complete discussion). The parameters derived for each AGN are reported in Table~\ref{tab:se_par} and their distributions are shown in Fig.~\ref{fig:distr_se_par}. We find no clear trends with the adopted spectral model (pure two-coronae or hybrid), suggesting that the inferred SE properties are robust against model assumptions.
    These findings indicate that the underlying physical properties of the WC are broadly similar across the population and are consistent with a WC extending over an intrinsically dissipative disk.
    However, the few sources exhibiting more extreme properties (e.g., weak or absent SE emission, or high $A$ and $\chi$ values), point to more complex disk--corona configurations. These could involve enhanced coronal heating, patchy geometries, or a non-negligible contribution from intrinsic disk emission.

    \item \textbf{The connection between the SE and UFOs.}
    AGN hosting UFOs do not favor a unique description of the SE (see Sect.~\ref{ufo_wa_results}). Both pure two-coronae and hybrid scenarios provide the best-fit, with the latter being slightly more frequent, accounting for four of the seven analyzed observations. These AGN shows statistically significant differences in the coronal properties compared with no-UFO sources, with lower WC electron temperatures and steeper HC photon indices (Log(NHP) = $-$2.90 and $-$1.58, respectively; see Sect.~\ref{comp_pop18} and Fig.~\ref{fig:2coronae_val}).
    
    While the presence of UFOs appears to be associated primarily with a difference in hardness ratio, caution is warranted given the limited sample size and the dependence of the HR on the instrumental response. No statistically significant evidence are found for the SE strength, WC optical depth, or coronal dissipation fraction (see Sect.~\ref{SE_UFO_model}). The distributions are shown in Fig.~\ref{fig:distr_se_par}.
    
    The multi-epoch cases of PG~0947+396 and PG~1114+445 further illustrate that UFO detection is not associated with a unique spectral configuration or with systematic variations in the SE diagnostics (see Sect.~\ref{SE_UFO_model}).

    Finally, we find a tentative anticorrelation between the WC optical depth and the UFO outflow velocity, although this result should be interpreted cautiously given the limited sample size (see Sect.~\ref{corr} and Appendix~\ref{app_c} for the plot).

    \item \textbf{Comparison with the \citetalias{petrucci18} sample.}
    No statistically significant differences are found between the SUBWAYS sample and the \citetalias{petrucci18} sample. Moreover, five AGN in the \citetalias{petrucci18} sample host UFOs. Their coronal parameters (i.e., kT$_{\rm e}^{WC}$, $\Gamma^{WC}$, and $\Gamma^{HC}$) are statistically consistent with those of the SUBWAYS UFO sub-sample. See Sect.~\ref{comp_pop18} for more details and Table~\ref{tab:ks_results_gamma_kte} for the KS test results.

    \item \textbf{Interplay between the SE component and warm absorbers.}
    When the SUBWAYS sample is divided according to the presence or absence of WAs, we find no statistically significant differences either in the soft excess diagnostics or in the physical parameters of the warm and hot coronae (see Sects.~\ref{ufo_wa_results}, \ref{SE_UFO_model}, and \ref{comp_pop18}).

    \item \textbf{Correlation analysis between WC, SE proxies, and AGN properties.} We find several significant correlations involving the WC parameters. The expected anticorrelations of WC optical depth with photon index and electron temperature are recovered. Among the SE diagnostics, the amplification factor correlates positively with the WC electron temperature, while the hardness ratio correlates with the WC optical depth. The WC photon index and optical depth correlate with the UV luminosity, while the electron temperature shows a weaker trend with the bolometric luminosity (see Sect.~\ref{corr} and Appendix~\ref{app_c} for the plots).
    
\end{itemize}

In conclusion, this work supports a scenario in which the soft X-ray excess is a common component of AGN emission, primarily arising from a warm Comptonizing corona extended over a intrinsically dissipative disk and, in many cases, accompanied by relativistic reflection. At the same time, the presence of sources with a weak or absent SE indicates that the relative importance of these processes can vary significantly across the AGN population. Our results point to differences in the hardness ratio and physical properties of both the warm and hot coronae between AGN with and without UFOs. This suggests that, the conditions in the inner accretion flow may play a role in enabling or regulating the launching of these outflows. Hence, the disk–corona structure emerges as a key element linking accretion and ejection processes. 
This work constitutes the first systematic broadband and multiwavelength investigation of the connection between the SE and ionized outflows in a sample of AGN, establishing a benchmark framework for future studies. In particular, future observations with higher spectral resolution and larger, statistically robust samples will be crucial for disentangling the physical mechanisms involved and better understanding the coupling between the inner accretion flow and AGN winds.

\begin{acknowledgements}
ChatGPT (OpenAI) was used for language editing, code debugging, and the design of the upper panel of Fig.~\ref{fig:hybrid_models_fit}. All scientific content, analysis, and conclusions were independently developed and verified by the authors. This work was supported by the "Action Thématique Phénomènes Extremes et Multimessagers " (ATPEM) of CNRS/INSU PN Astro and by the french national spatial agency CNES. This research used observations (associated with program 15890) made with the NASA/ESA Hubble Space Telescope obtained from the Space Telescope Science Institute, which is operated by the Association of Universities for Research in Astronomy, Inc., under NASA contract NAS 5–26555. VEG acknowledges funding under NASA contract 80NSSC24K1403. GM acknowledges support from grants n.~PID2023-147338NB-C21 funded by MICIU/AEI/10.13039/501100011033 and ERDF/EU. BDM acknowledges support from the Spanish MINECO grants PID2023-148661NB-I00, PID2022-136828NB-C44, and CNS2025-166335. EB is partially supported by The Israel Science Foundation (grant No. 2617/25). RS acknowledges logistical support by the Istituto Nazionale di Astrofisica (INAF) at Osservatorio Astronomico di Roma (OAR). ALL acknowledges support from DGAPA-PAPIIT IA-103625 grant.
\end{acknowledgements}
\vspace{-0.8cm}
{\scriptsize
   \bibliographystyle{aa} 
   \bibliography{biblio}} 
%

\vspace{0.2cm}

\noindent \textbf{Affiliations:} \\
\noindent
\textit{\scriptsize
$^{1}$Université Grenoble Alpes, CNRS, IPAG, 38000 Grenoble, France;\\
$^{2}$Department of Physics and Astronomy, Clemson University, Kinard Lab of Physics, Clemson, SC 29634, USA;\\
$^{3}$Dipartimento di Matematica e Fisica, Università degli Studi Roma Tre, Via della Vasca Navale 84, 00146 Roma, Italy;\\
$^{4}$INAF-Osservatorio Astronomico di Brera, Via Brera 28, 20121 Milano, Italy;\\
$^{5}$Dipartimento di Fisica e Astronomia (DIFA), Università di Bologna, Via Gobetti 93/2, I-40129 Bologna, Italy; \\
$^{6}$INAF-Osservatorio di Astrofisica e Scienza dello Spazio di Bologna, Via Gobetti, 93/3, I-40129 Bologna, Italy;\\
$^{7}$Quasar Science Resources SL for ESA, European Space Astronomy Centre (ESAC), Science Operations Department, 28692, Villanueva de la Ca\~{n}ada, Madrid, Spain;\\
$^{8}$Space Telescope Science Institute, 3700 San Martin Drive, Baltimore, MD 21218, USA;\\
$^{9}$Department of Physics, University of Virginia, Charlottesville, VA 22904, USA;\\
$^{10}$INAF -- Istituto di Radioastronomia, Via P. Gobetti 101, 40129 Bologna, Italy;\\
$^{11}$Physics Department, The Technion, 32000 Haifa, Israel;\\
$^{12}$SRON Space Research Organisation Netherlands, Niels Bohrweg 4, 2333 CA Leiden, The Netherlands;\\
$^{13}$Departament de F\'isica, EEBE, Universitat Polit\`ecnica de Catalunya, Av. Eduard Maristany 16, Barcelona, E-08019, Spain;\\
$^{14}$INAF -- Istituto di Astrofisica e Planetologia Spaziali, Via Fosso del Cavaliere, 00133 Roma, Italy;\\
$^{15}$Centro de Astrobiologia (CAB), CSIC-INTA, Camino Bajo del Castillo s/n, Campus ESAC, 28692, Villanueva de la Cañada, Madrid, Spain;\\
$^{16}$Department of Physics, Institute for Astrophysics and Computational Sciences, The Catholic University of America, Washington, DC 20064, USA;\\
$^{17}$Space Telescope Science Institute, 3700 San Martin Drive, Baltimore, MD 21218, USA;\\
$^{18}$Instituto de Astronomía, Universidad Nacional Autónoma de México, Circuito Exterior, Ciudad Universitaria, Ciudad de México 04510, México; \\
$^{19}$Department of Astronomy, The Ohio State University, 140 West 18th Avenue, Columbus, OH 43210, USA;\\
$^{20}$Center for Cosmology and Astroparticle Physics, 191 West Woodruff Avenue, Columbus, OH 43210, USA;\\
$^{21}$INAF -- Osservatorio Astronomico di Roma, Via Frascati 33, 00078 Monte Porzio Catone (Roma), Italy;\\
$^{22}$Space Science Data Center, Agenzia Spaziale Italiana, Via del Politecnico snc, 00133 Roma, Italy;\\
$^{23}$INAF -- Osservatorio Astrofisico di Arcetri, Largo Enrico Fermi 5, 50125 Firenze, Italy;\\
$^{24}$European Southern Observatory (ESO), Karl-Schwarzschild-Strasse 2, D-85748 Garching bei München, Germany;\\
$^{25}$INAF -- Osservatorio Astronomico di Brera, Via Bianchi 46, 23807 Merate (LC), Italy;\\
$^{26}$Max-Planck-Institut für extraterrestrische Physik, Giessenbachstra{\ss}e 1, 85748 Garching bei München, Germany;\\
$^{27}$Como Lake Center for Astrophysics (CLAP), DiSAT, Università
degli Studi dell’Insubria, via Valleggio 11, 22100 Como, Italy;\\
$^{28}$Instituto de Alta Investigación, Universidad de Tarapacá, Casilla 7D, Arica, Chile;\\
$^{29}$Dipartimento di Fisica, Universita of Rome `Tor Vergata', Via della Ricerca Scientifica 1, 00133 Rome, Italy;\\
$^{30}$INFN - Rome Tor Vergata, Via della Ricerca Scientifica 1, 00133 Rome, Italy;\\
$^{31}$Department of Physics, Ehime University, 2-5, Bunkyocho, Matsuyama, Ehime 790-8577, Japan;\\
$^{32}$European Southern Observatory, Alonso de Còrdova 3107, Casilla 19, Santiago, 19001, Chile;\\
$^{33}$CNRS, CEA, AIM, Université Paris-Saclay, Université Paris Cité, 91191, Gif-sur-Yvette, France;\\
$^{34}$NASA Goddard Space Flight Center, Greenbelt, MD 20771, USA;\\
$^{35}$Southeastern Universities Research Association, Washington, DC 20005, USA;\\
$^{36}$School of Physics and Astronomy, University of Minnesota, Minneapolis, MN 55455, USA;\\
$^{37}$Cahill Center for Astrophysics, California Institute of Technology, 1216 East California Boulevard, Pasadena, CA 91125, USA.}
\begin{appendix}
\onecolumn
\section{Source and observations details}

For each AGN here analyzed, we present in Table~\ref{tab:obs_info} the details of the available observations. In Table~\ref{tab:sources_list}, we report the physical properties of the analyzed AGN.

\begin{table}[h]
    \centering
    \renewcommand{\arraystretch}{2}
    \caption{List of sources with information on adopted data.}
    \begin{adjustbox}{max width=\textwidth,center}
\begin{tabular}{c|ccc|ccc|cc}
\hline
\textbf{Source} & \multicolumn{3}{c|}{\textbf{XMM-Newton}} & \multicolumn{3}{c|}{\textbf{NuSTAR}} & \multicolumn{2}{c}{\textbf{HST}} \\
\vspace{-0.2cm}
& pn [ks] & OM filters & Obs. date & FPMA/B [ks] & Radii [arcsec] & Obs. date & points & Obs. date \\
\scriptsize (1) & \scriptsize(2) & \scriptsize(3) & \scriptsize(4) & \scriptsize(5) & \scriptsize(6) & \scriptsize(7) & \scriptsize (8) & \scriptsize (9) \\
\hline\hline
2MASS~J02201457$-$0728591 & 77.09 & M2, W1, U, B, V & 2020-02-03  & 39.79 (0.43\%) & 37.0 and 37.0 & 2021-03-06 & 4 & 2020-01-03 \\ 
2MASS~J10514425+3539306${\bf *}$& 80.26 & M2, W1, U, B, V & 2020-05-26 & 36.65 (0.44\%) & 24.5 and 37.0 & 2020-06-04 & / &  / \\
2MASS~J16531505+2349427${\bf *}$& 59.83 & M2, W1, U, B, V & 2020-02-11 & 27.60 (0.16\%) & 61.5 and 61.5 & 2018-01-19 & / & /  \\
HB89~1529+050 & 75.25 & W1, U, B, V & 2019-08-22 & 30.22 (0.43\%) & 49.0 and 37.0 & 2020-09-06 & 2 & 2019-08-30 \\ 
LBQS~1338$-$0038 & 91.55 & M2, W1, U, B, V & 2020-02-01  & 51.41 (0.49\%) & 49.0 and 49.0 & 2021-01-08 & 4 & 2020-01-28 \\ 
PG~0052+251 & 34.38 & M2, W2, W1, U, V & 2019-07-15  & 24.29 (0.43\%) & 74.0 and 62.0 & 2020-08-21 & 3 & 2019-11-02 \\ 
PG~0947+396 \makecell{UFO \\ no-UFO} & \makecell{31.11 \\ 45.70} & \makecell{M2, W1, U \\ M2, W1, U} & \makecell{ 2019-11-25 \\ 2020-04-16} & 23.67 (0.51\%) & 37.0 and 37.0 & 2021-06-24 & 4 & 2020-04-25 \\ 
PG~0953+414 & 45.51 & M2, W1, U, B, V & 2020-04-14 & 67.51 (0.44\%) & 41.0 and 30.0 & 2021-11-02 & 4 & 2020-04-04 \\ 
PG~1202+281 & 46.82 & M2, W1, U, B, V & 2020-06-29  & 21.09 (0.45\%) & 61.0 and 61.5 & 2020-05-23 & 3 & 2019-12-14 \\ 
PG~1216+069 & 63.44 & M2, W1, U, B, V & 2020-06-05  & 31.82 (0.41\%) & 49.0 and 25.0 & 2020-06-03 & 4 & 2019-12-26 \\ 
PG~1307+085 & 19.50 & M2, U, V & 2020-01-22  & 34.95 (0.50\%) & 49.0 and 30.0 & 2021-06-15 & 3 & 2019-11-30 \\
PG~1352+183 & 74.35 & M2, W1, U, B, V & 2020-01-26  & 32.39 (0.43\%) & 60.0 and 74.0 & 2020-12-02 & 4 & 2020-01-12 \\ 
PG~1425+267 & 73.28 & M2, W1, U, B, V & 2020-02-05 & 34.26 (0.44\%) & 49.0 and 49.0 & 2020-07-21 & 4 & 2020-01-25 \\ 
PG~1435$-$067 & 42.28 & M2, W1, U, B, V & 2019-07-24 & 21.02 (0.43\%) & 49.0 and 74.0 & 2020-12-26 & 4 & 2020-01-19 \\ 
PG~1626+554 & 51.41 & M2, W1, U, B, V & 2019-05-27 & 21.40 (0.42\%) & 61.5 and 49.0 & 2020-07-28 & 4 & 2020-07-13 \\ 
SDSS~J144414.66+063306.7 & 66.30 & M2, W1, U, B, V & 2019-07-28  & 27.27 (0.40\%) & 30.0 and 37.0 & 2020-09-04 & 4 & 2019-12-22 \\ 
WISE~J053756.30$-$024513.1 ${\bf *}$ & 69.82 & M2, W1, U, B, V & 2020-03-02 & 32.19 (0.45\%) & 24.5 and 37.0 & 2021-03-27 & / & / \\
\cdashline{1-9}
HB89~1257+286 ${\bf \#}$ & 47.50 & M2, W2, U & 2006-06-06 & / & / & / & / & / \\
PG~0804+761 ${\bf \#}$ & \makecell{29.21 \\ 25.21} & \makecell{M2, W1, U, V \\ M2, W1, U, V} & \makecell{2010-03-10 \\ 2010-03-12} & / & / & / & 4 & 2010-06-12 \\
PG~1114+445 ${\bf \#}$ \makecell{ no-UFO \\ UFO (101) \\ UFO (301) } & \makecell{ 37.04 \\ 24.5 \\ 30.3 } & \makecell{M2, W2, W1, U, B, V \\ M2, W2, W1\\ M2, W2, W1} & \makecell{ 2002-05-14 \\ 2010-05-19 \\ 2010-05-23 }  & / & / & / & \makecell{ 3 \\ / \\ /} & \makecell{ 2004-05-28 \\ / \\ /} \\
PG~1402+261 ${\bf +}$ & 67.24 & M2, W2, W1, U & 2018-12-17 & 67.73 (0.24\%) & 49.0 and 74.0 & 2018-12-17 & / & / \\
\hline \hline
\end{tabular}
    \end{adjustbox}
    \label{tab:obs_info}
    \par
    \smallskip
    \footnotesize{
    \textit{Notes:} \\ {\it Columns:} (1) Source; (2) EPIC-pn effective exposure time after data reduction; (3) Available OM filters; (4) observation dates for XMM data; (5) \nustar\ effective exposure time after data reduction, the exposures are averaged between FPMA and FPMB, with the range of percentage variance between the two in parentheses; (6) source extraction radii for the \nustar\ FPMA and FPMB spectra; (7) observation dates for \nustar\ data; (8) Number of photometric points extracted from each HST spectrum; (9) observation dates for HST data. \\ 
    The last four AGN listed are archival targets of the SUBWAYS sample.\\
    ${\bf *}$: 2MASS~J10514425+3539306, 2MASS~J16531505+2349427, and WISE~J053756.30$-$024513.1 are too heavily reddened and could not be observed with HST. \\
    ${\bf +}$: from the \swift\ light curve, see Fig.~\ref{fig:uv_var} right panel, PG~1402+261 results to be variable in the UV band. Given that the HST data were taken in 1996 (i.e., 24 years before the XMM and \nustar\ data), we will not use them during the analysis. \\
    ${\bf \#}$: HB89~1257+286, PG~0804+761, and PG~1114+445 are variable in the X-ray band (see Fig.~\ref{fig:xray_var}). Given that \nustar\ and XMM observations were taken 14, 6, and 18 years apart (respectively), we will not use the \nustar\ spectra and the electron temperature of the hot corona is fixed to kT$_{\rm e}^{HC}$ = 100 keV during the fit \citepalias[as adopted by][]{petrucci18}. Moreover, for HB89~1257+286, the observation considered here was not analyzed by \citepalias{peluso2026}; therefore, we do not adopt the temperature derived in their work.}
\end{table}

\begin{table}[h]
\renewcommand{\arraystretch}{2.2}
\centering
\caption{List the AGN properties for the analyzed sample, and presence (or absence) of WA and UFO.}
\begin{tabular}{c|c|c|c|c|c|c|}
    \hline
    \scriptsize (1) & \scriptsize(2) & \scriptsize(3) & \scriptsize(4) & \scriptsize(5) & \scriptsize(6) &  \scriptsize(7)\\
    Source & z & Log(M$_{\rm BH}$ / M$_{\rm \odot}$) & E(B-V) & F$_{\rm 2-10\, keV}$ & WA X-ray & UFO X-ray \\
    & & & & [erg/cm$^{2}$/s] &&\\
    \hline
2MASS~J02201457$-$0728591 & 0.21343 & 8.42$\pm$0.30 & 0.02 & 1.24$^{+0.02}_{-0.04}$ & Y & / \\
2MASS~J10514425+3539306 & 0.159 & 8.40$\pm$0.30 & / & 0.61$^{+0.02}_{-0.35}$ & / & Y \\
2MASS~J16531505+2349427 & 0.103 & 6.98$\pm$0.30 & / & 2.31$^{+0.05}_{-0.07}$ & / & Y \\
HB89~1257+286 & 0.091 & 7.46$\pm$0.30 & 0.01 & 1.98$^{+0.03}_{-0.02}$ & / & / \\
HB89~1529+050 & 0.21817 & 8.75$\pm$0.30 & 0.043 & 1.16$^{+0.01}_{-0.04}$ & Y & / \\
LBQS~1338-0038 & 0.23745 & 7.77$\pm$0.03 & 0.022 & 2.07$^{+0.06}_{-0.09}$ & Y & Y \\
PG~0052+251 & 0.15445 & 8.57$\pm$0.09 & 0.041 & 6.33$^{+0.14}_{-3.47}$ & Y & / \\
PG~0804+761 & 0.100 & 8.31$\pm$0.04 & 0.031 & 9.03$^{+0.10}_{-0.12}$ & / & /{\bf *} \\
PG~0947+396 \makecell{ UFO obs \\ no-UFO obs} & 0.20553 & 8.68$^{+0.08}_{- 0.10}$ & 0.017 & \makecell{1.39$^{+0.02}_{-0.05}$ \\ 1.24$^{+0.04}_{-0.64}$} & / & \makecell{Y\\/} \\
PG~0953+414 & 0.23410 & 8.27 $^{+0.06}_{-0.09}$ & 0.011 & 2.81$^{+0.03}_{-1.70}$ & / & / \\
PG~1114+445 \makecell{ no-UFO \\ UFO (101) \\ UFO (301)} & 0.144 & 8.59$\pm$0.09 & 0.014 & \makecell{ 2.29$^{+0.61}_{-0.19}$ \\ 1.71$^{+0.05}_{-0.07}$ \\ 1.67$^{+0.07}_{-0.05}$} & Y & \makecell{ / \\ Y \\ Y } \\
PG~1202+281 & 0.16501 & 8.61$^{+0.08}_{- 0.10}$ & 0.018 & 3.58$^{+0.28}_{-1.94}$ & Y &  Y \\
PG~1216+069 & 0.33130 & 9.20$^{+0.09}_{- 0.11}$ & 0.022 & 1.82$^{+0.03}_{-0.05}$ & Y &  / \\
PG~1307+085 & 0.15384 & 8.64$\pm$0.12 & 0.03 & 3.53$^{+0.04}_{-0.09}$ & Y &  / \\
PG~1352+183 & 0.15147 & 8.42$^{+0.08}_{- 0.10}$ & 0.016 & 1.35$^{+0.01}_{-0.79}$ & / &  / \\
PG~1402+261 & 0.164 & 7.94$^{+0.08}_{- 0.10}$ & 0.014 & 1.60$^{+0.03}_{-0.84}$ & / &  / \\
PG~1425+267 & 0.36361 & 9.22$\pm$0.30 & 0.015 & 1.62$^{+0.01}_{-1.09}$ & Y &  / \\
PG~1435-067 & 0.12900 & 8.37$^{+0.08}_{- 0.10}$ & 0.05 & 1.09$^{+0.02}_{-0.66}$ & / &  / \\
PG~1626+554 & 0.13170 & 8.50$^{+0.08}_{- 0.10}$ & 0.005 & 2.69$^{+0.02}_{-0.05}$ & Y &  / \\
SDSS~J144414.66+063306.7 & 0.20768 & 8.10$\pm$0.30 & 0.03 & 2.50$^{+0.01}_{-0.25}$ & / & /  \\
WISE~J053756.30$-$024513.1 & 0.11 & 7.73$\pm$0.30 & / & 1.31$^{+0.03}_{-0.05}$ & / & /  \\
\hline
\end{tabular}
\label{tab:sources_list}
    \par
    \smallskip
    \scriptsize{
    \textit{Notes:} \\ {\it Columns:} (1) Source; (2) Redshift; (3) Black hole mass; (4) E(B-V); (5) X-ray flux in the 2-10 keV energy band, obtained from the best-fit models in units of $\times$10$^{-12}$ erg/cm$^{2}$/s; (6) detected warm absorber in the X-ray band; (7) detected ultra-fast outflow in the X-ray band. \\
    We report with ``Y`` when the feature is detected in the spectrum of the observation analyzed in this work.
    Redshift and black hole masses have been taken from \citet{gianolli24} and references therein. The E(B-V) are derived from \url{https://irsa.ipac.caltech.edu/applications/DUST/} adopting \citet{Schlafly2011} estimates. The information on the WA and UFO detections are taken from \citet{Matzeu2023} and \citet{mehdipour23}.\\
    {\bf *}:Archival source. Three XMM observations are available: two without UFO detections, which were fitted jointly owing to the lack of short-term X-ray variability, and one with a UFO detection. The latter was not included in this work as it lacks simultaneous OM coverage. Moreover, the available \hst\ and \nustar\ observations are too distant in time to be associated with this epoch, given the UV and long-term X-ray variability of the AGN. The PG~0804+761 is therefore included in the no-UFO sub-sample. See Sect.~\ref{data} for further details.\\
   }
\end{table}

\clearpage

\section{X-ray and optical/UV variability}

\begin{figure}[h!]
\centering
\includegraphics[width=0.47\textwidth,trim=10 5 5 5,clip]{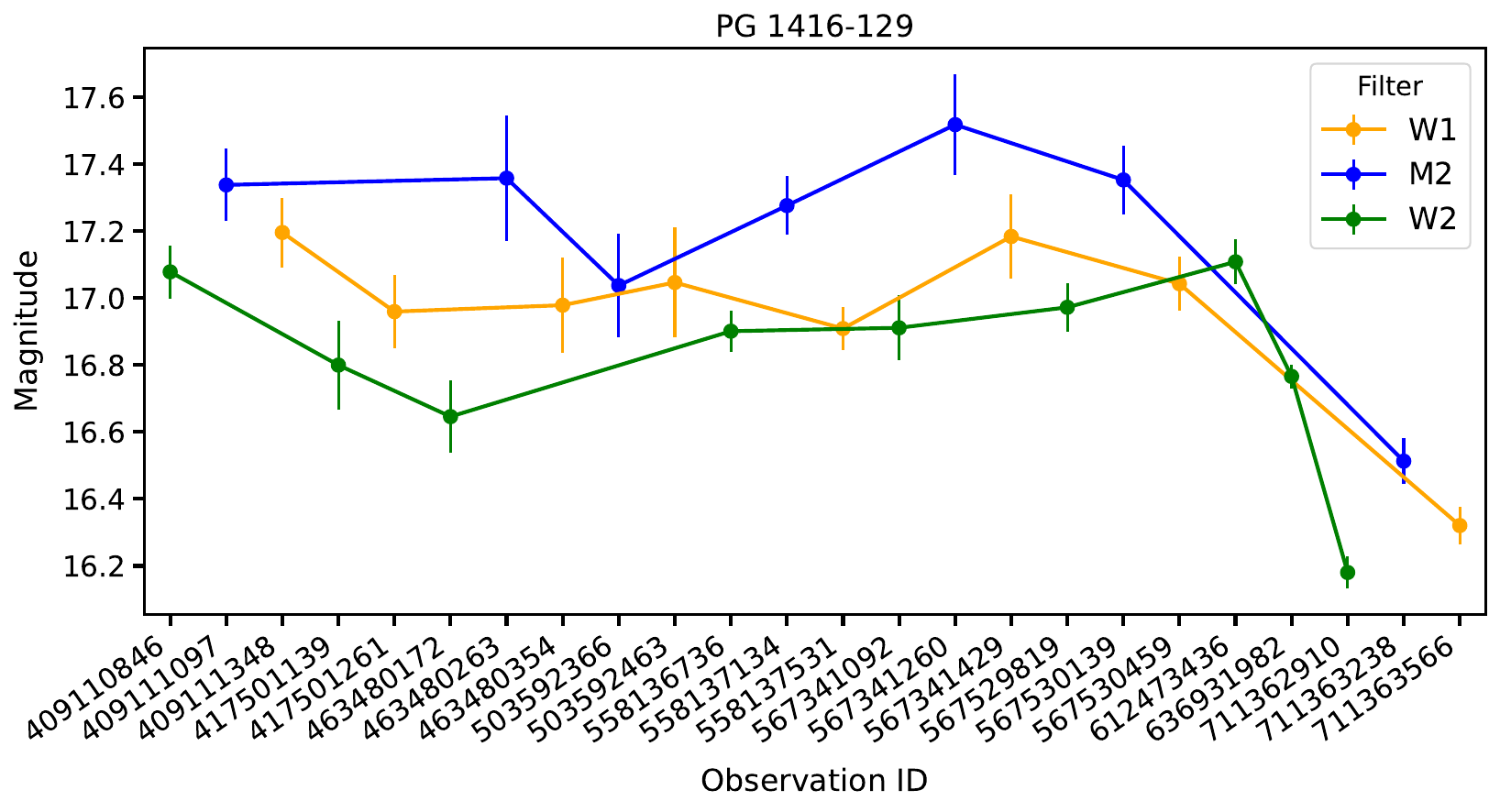}
\includegraphics[width=0.38\textwidth,trim=10 5 5 5,clip]{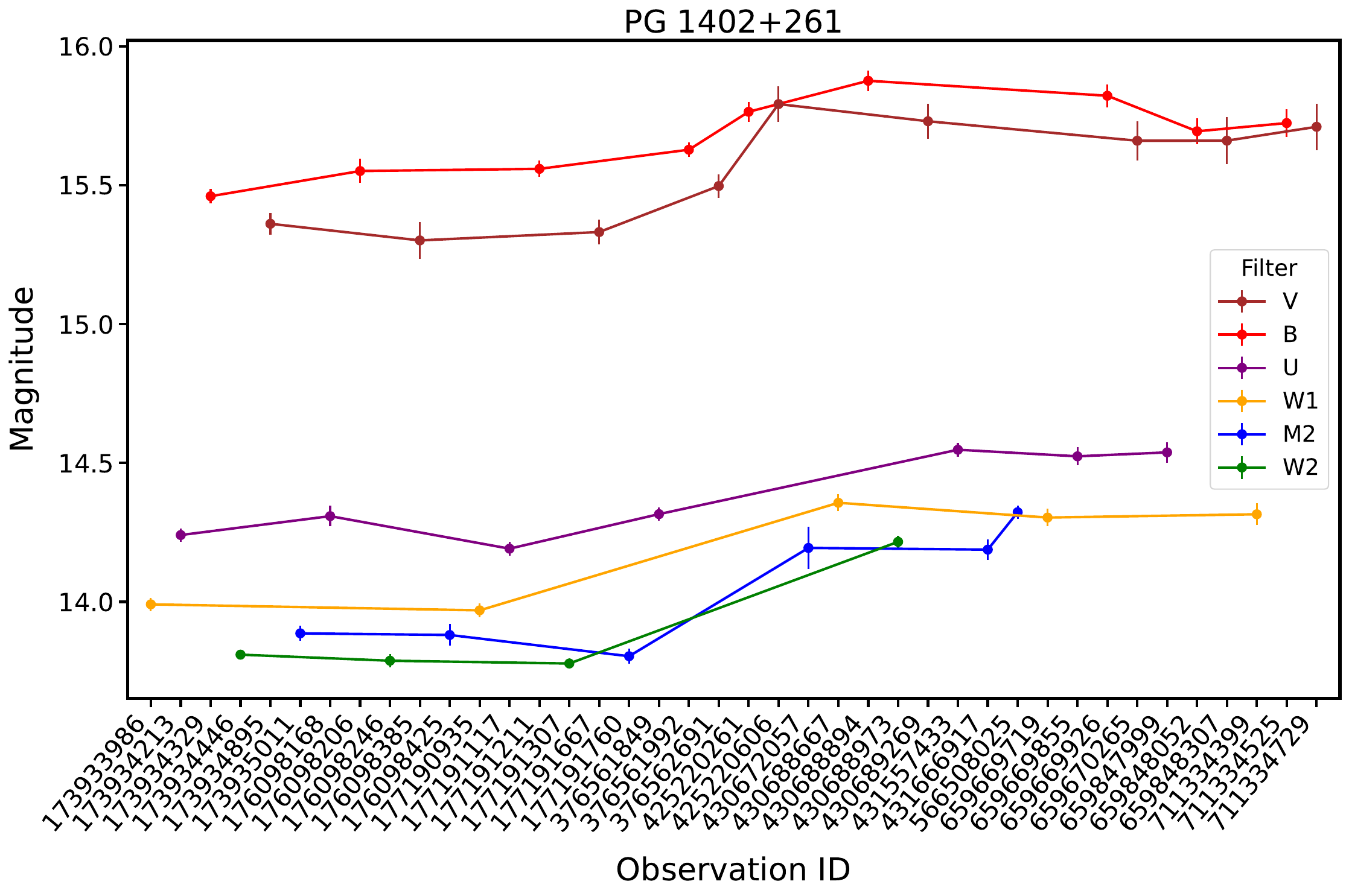}\\
\includegraphics[width=0.47\textwidth,trim=10 5 5 5,clip]{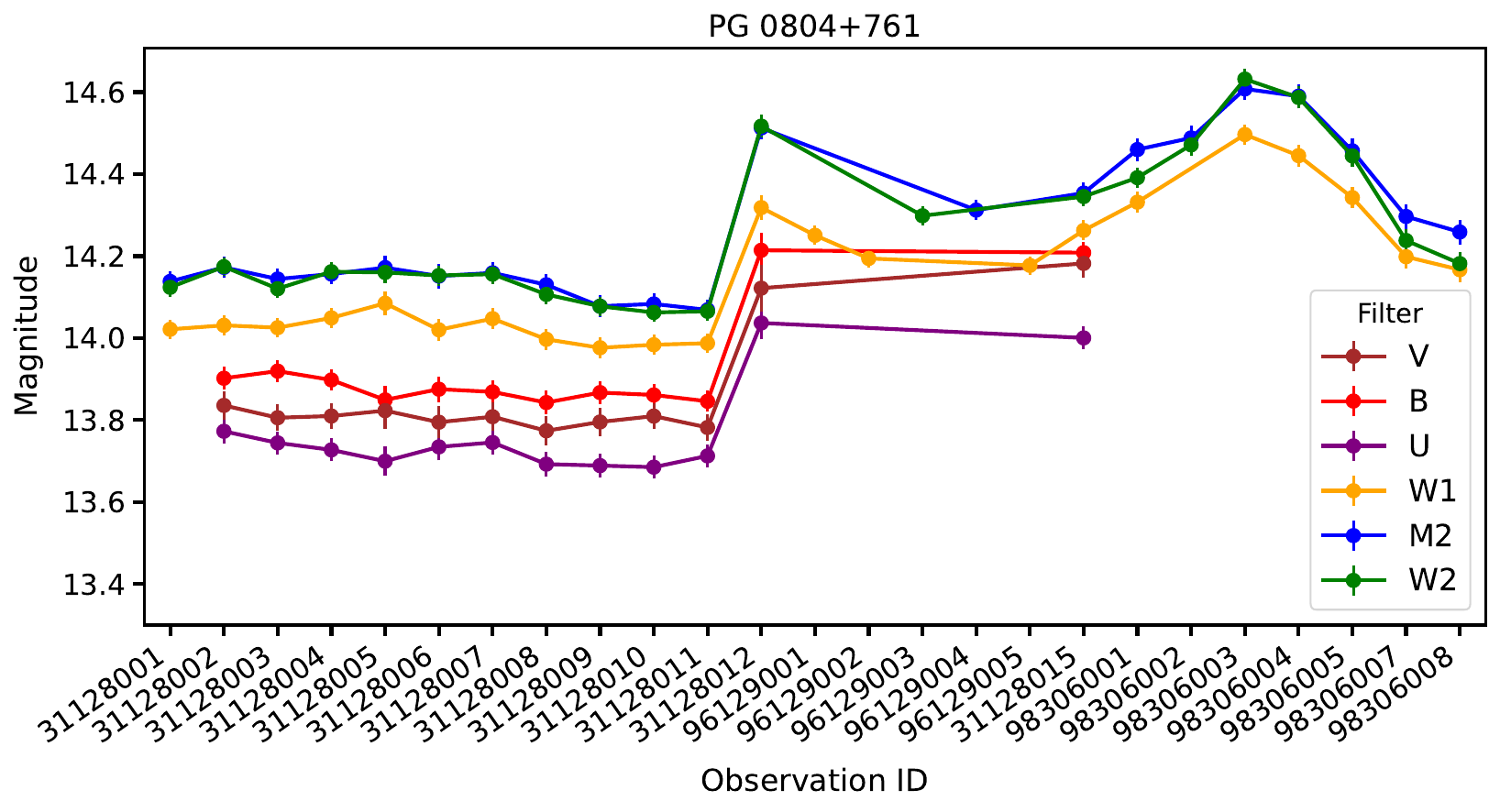}
\includegraphics[width=0.47\textwidth,trim=10 5 5 5,clip]{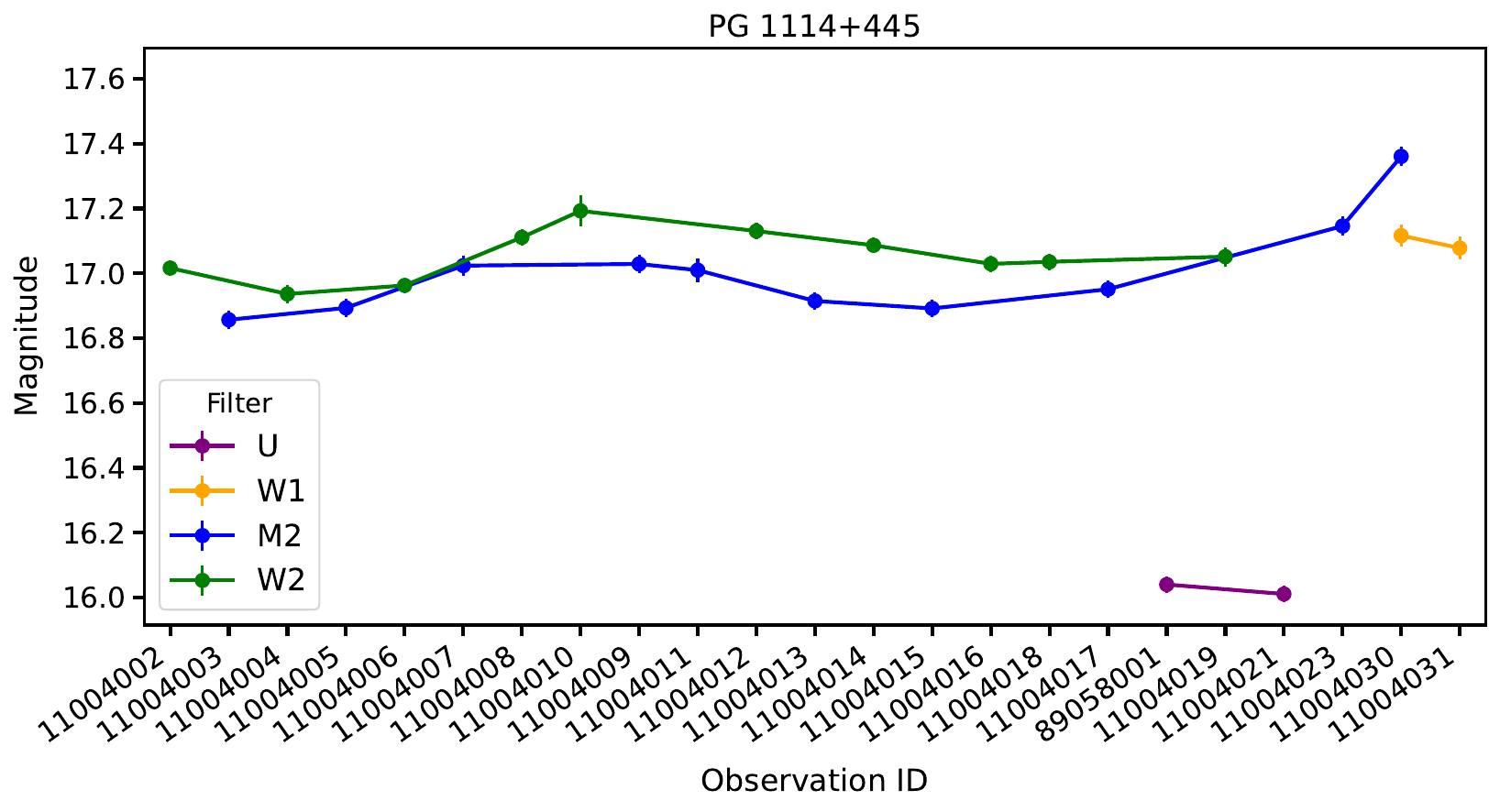}
    \caption[]
{{\swift/UVOT multi-band photometric light curves of PG~1416–129 (upper left panel), PG~1402+261 (upper right panel), PG~0804+761 (lower left panel), and PG~1114+445 (lower right panel). Each  curve represents the observed optical/UV magnitude in a different UVOT filter (W2, M2, W1, B, V, and U) as a function of observation ID. Clear UV variability is present in all AGN, preventing the use of HST data during the analysis.}}
    \label{fig:uv_var}
\end{figure}

\begin{figure*}[h!]
\centering
\includegraphics[width=.32\textwidth]{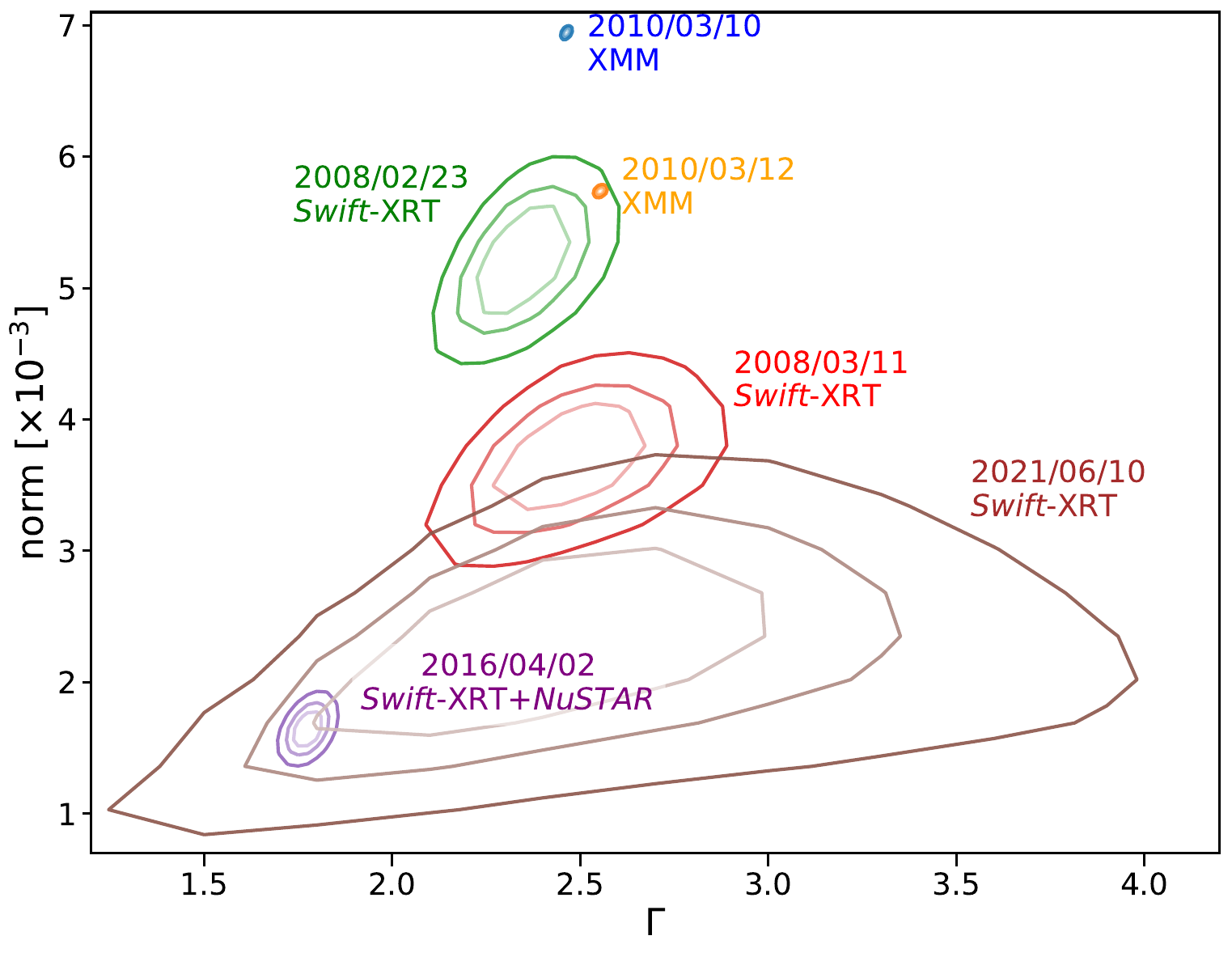}
\includegraphics[width=.32\textwidth]{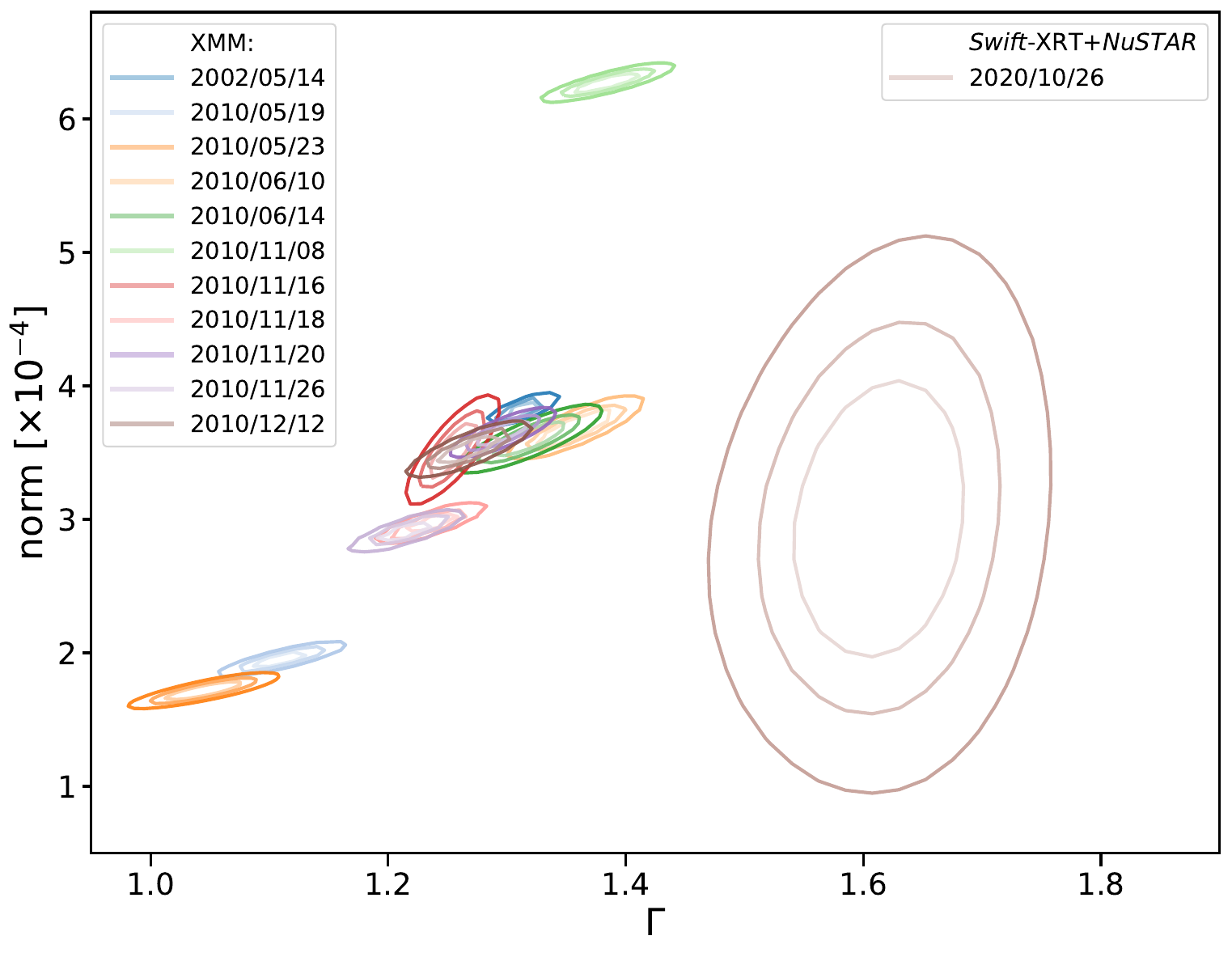}
\includegraphics[width=.32\textwidth]{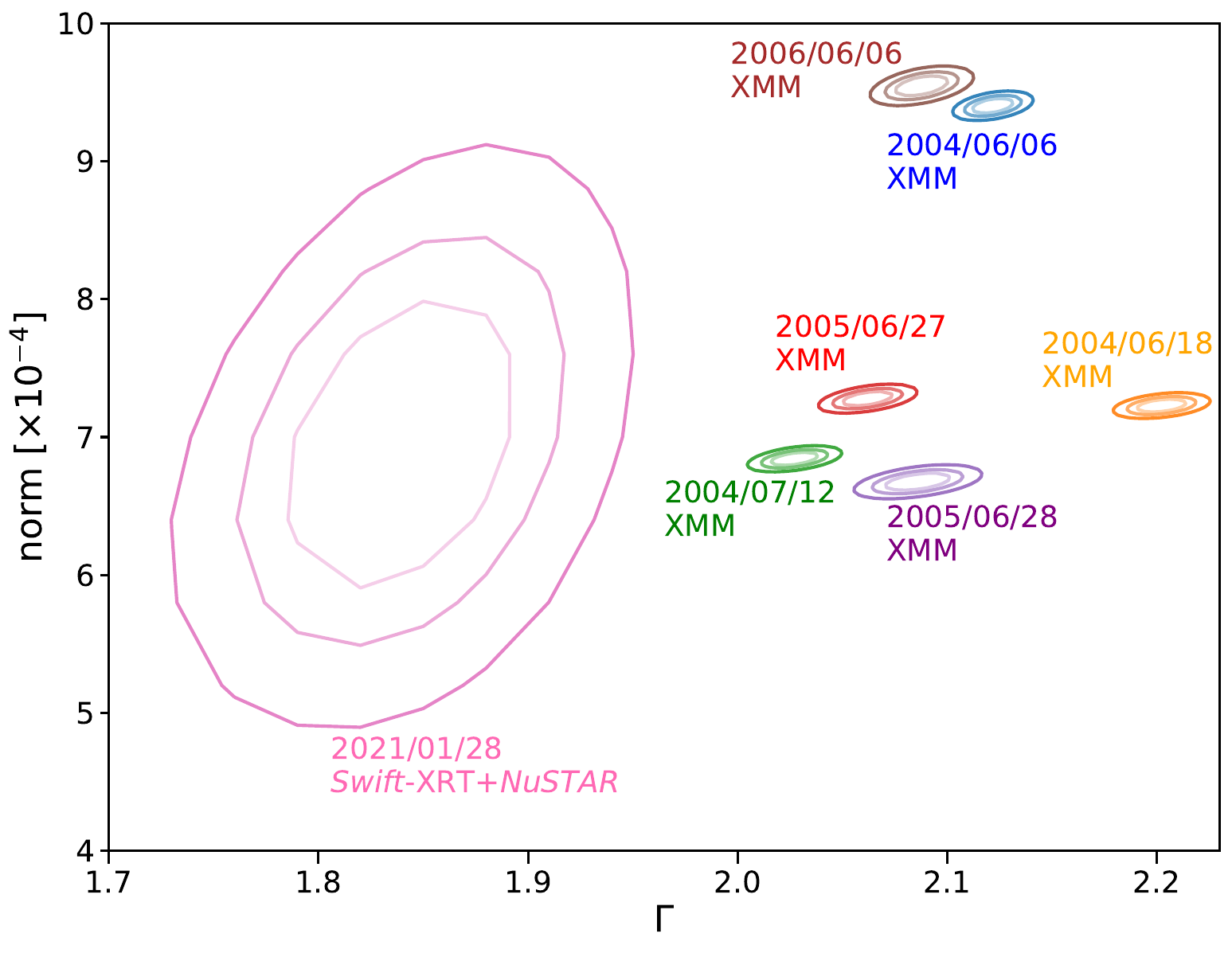} 
    \caption[]
    {{Contour plots (at 68\%, 90\%, and 99\% confidence levels) photon index versus normalization parameter for the available XMM and/or \swift\ observations and simultaneous \swift+\nustar\ observations of PG~0804+761 (left panel), PG~1114+445 (middle panel), and HB89~1257+286 (right panel). Different colors correspond to individual observations. Clear X-ray variability is present in all three AGN, preventing the use of \nustar\ data during the analysis. }}
    \label{fig:xray_var}
\end{figure*}

To determine whether the HST and \nustar\ observations taken more than two years apart from the XMM–{\it Newton} data, can be reliably used in our analysis, we examine the UV and X-ray variability. For the UV band, Fig.~\ref{fig:uv_var} shows the \swift/UVOT multi-filter light curves \citep[for more information on the analysis pipeline see][]{Marcotulli2026} of PG~1416–129, PG~1402+261, PG~0804+761, and PG~1114+445. Each curve corresponds to a different UVOT filter and is used to identify possible flux changes across epochs.
For the X-ray band, Fig.~\ref{fig:xray_var} shows contour plots of photon index versus normalization for all available XMM, \swift, and simultaneous \swift+\nustar\ observations of PG~0804+761, PG~1114+445, and HB89~1257+286. The contours, derived using an absorbed power law as a phenomenological model, illustrate whether spectral variations occur across epochs.

\clearpage
\newpage

\section{Best-fits results}\label{app_c}

\begin{figure*}
\centering
    \subfloat[][]{\includegraphics[width=1.\textwidth]{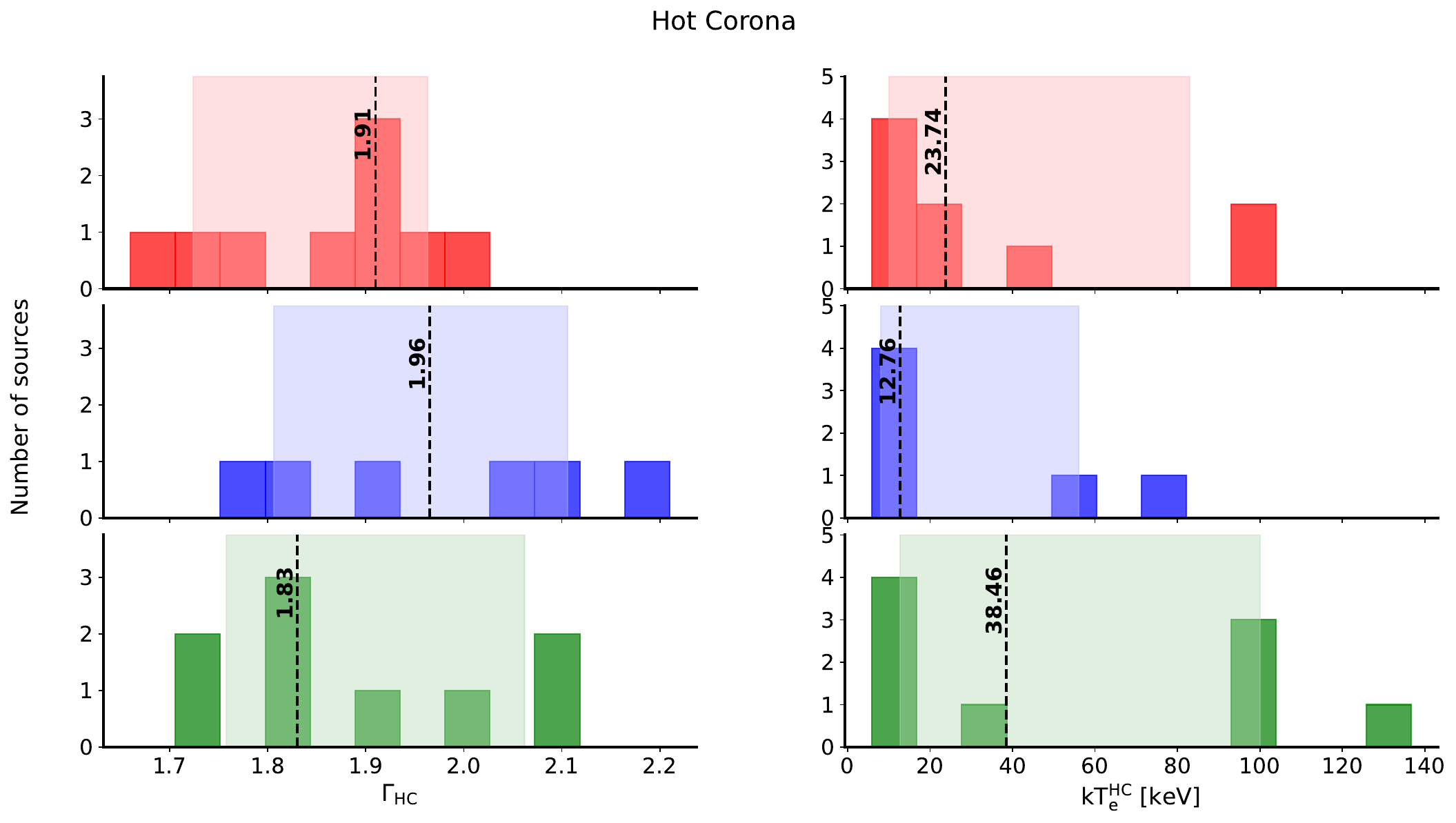}}\\
    \subfloat[][]{\includegraphics[width=1.\textwidth]{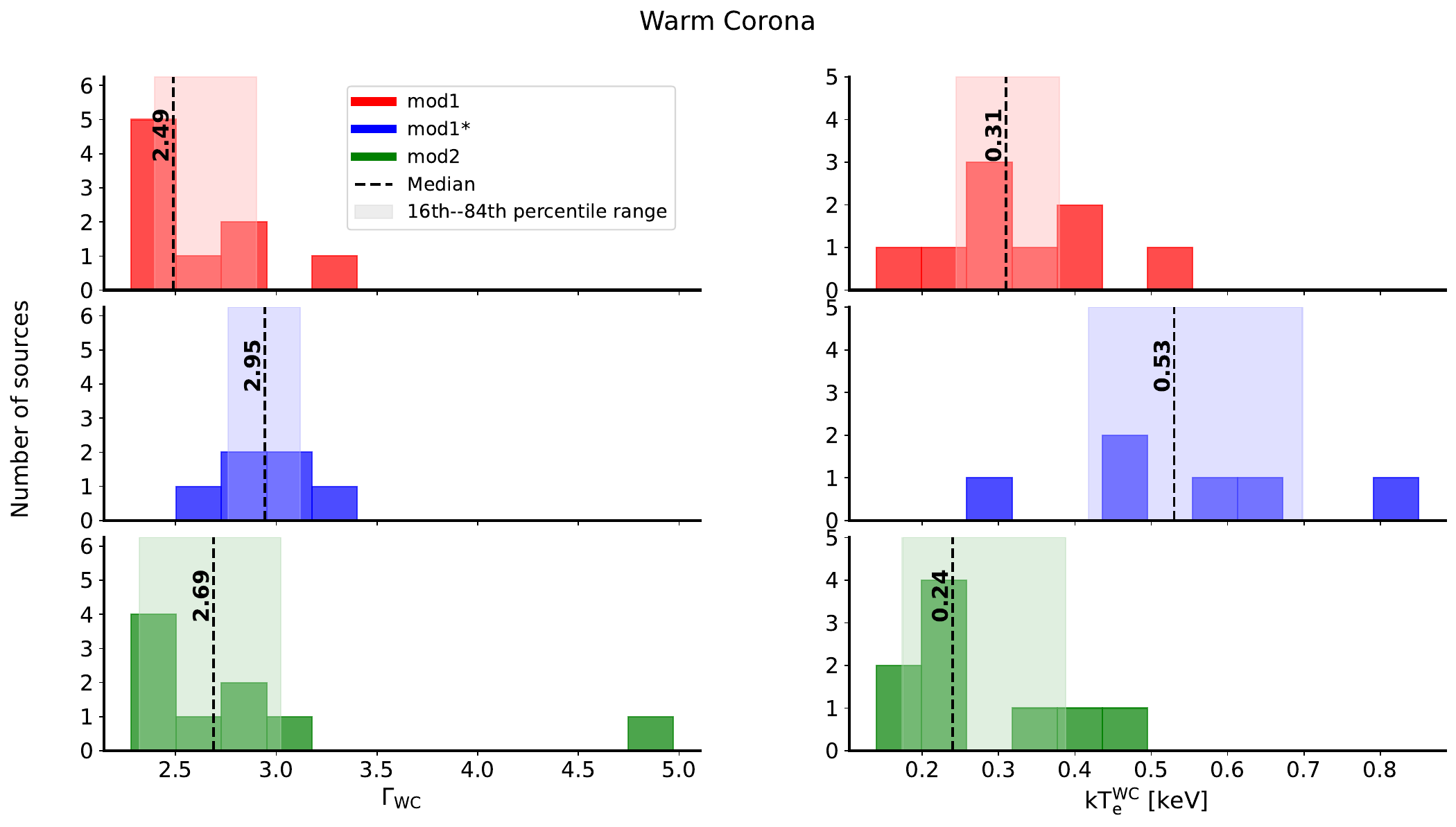}}
    \caption[]
    {{Distributions of the best-fit models photon index $\Gamma$ and electron temperature kT$_{\rm e}$ reported in Table~\hyperlink{fit-table}{\ref*{tab:fit}}. {\it Panel a:} hot corona parameters. {\it Panel b:} warm corona parameters. Mod1 in shown red, mod1* in blue, and mod2 in green. 
    The dashed lines show the median values for each distribution with the 16th–84th percentile dispersion in colored bars.}}
    \label{fig:distributions}
\end{figure*}

We present the spectral modeling results and derived physical parameters. In particular, Table~\hyperlink{fit-table}{\ref*{tab:fit}} reports the best-fit values for the tested models for each analyzed AGN and Fig.~\ref{fig:distributions} shows the distributions of key parameters for both warm and hot coronae, as well as quantities describing the soft excess in Table~\ref{tab:se_par} and Fig.~\ref{fig:distr_se_par}. In Table~\ref{tab:hybrid_soft_band_contributions}, we report the warm corona and disk reflection component contributions to the 0.5–2 keV flux for AGN best described by the hybrid model. 

The plots showing the spectra and model-to-data ratios, together with those illustrating the significant correlations identified in our analysis, and the discussion of the optical depth estimates are available only online at \href{https://zenodo.org/records/21513991?preview=1&token=eyJhbGciOiJIUzUxMiJ9.eyJpZCI6IjQyZmExNDM1LTg0YTctNDZmYi04ZjEyLTYwNTZjNDAyNGRlZCIsImRhdGEiOnt9LCJyYW5kb20iOiI4NTU1NTBiMjg5NmM0MTFmZjMzMjZmZjgzZGE1M2RkYyJ9.62_1sviPOjP7iBC_85zpIqZdE_tXXyqLHGroH6C73SsKChclM0EbGjb3wUhrhzIB0F9O98TXAPm0ea61ky-71Q}{Zenodo}.

{\centering
\renewcommand{\arraystretch}{1.4}
\setlength{\LTleft}{-1cm}
\setlength{\LTright}{\fill}
\begin{longtable}{c|c:ccccccccc}
\caption{Best-fit parameters for each tested model.\label{tab:fit}}\\
\multicolumn{10}{c}{%
  \hypertarget{fit-table}{\rule{0pt}{1pt}}%
}\\[-1.5ex]
\hline
    \scriptsize (1) & \scriptsize(2) & \scriptsize(3) & \scriptsize(4) & \scriptsize(5) & \scriptsize(6) & \scriptsize(7) & \scriptsize (8) & \scriptsize (9) & \scriptsize (10)\\
Source & Model & kT$_{\rm BB}^{\rm opt/UV}$ & kT$_{\rm BB}$ & $\Gamma^{\rm HC}$ & kT$^{\rm HC}$ & $\Gamma^{\rm WC}$ & kT$^{\rm WC}$ & $\sigma_{\texttt{gsmooth}}$ & $\chi^2$ / d.o.f \\
     & & [eV] & [eV] & & [keV] &  & [keV] & [eV] \\ \hline
     \hline
\endfirsthead
\caption{continued.}\\
\hline\hline
    \scriptsize (1) & \scriptsize(2) & \scriptsize(3) & \scriptsize(4) & \scriptsize(5) & \scriptsize(6) & \scriptsize(7) & \scriptsize (8) & \scriptsize (9) & \scriptsize (10)\\
Source & Model & kT$_{\rm BB}^{\rm opt/UV}$ & kT$_{\rm BB}$ & $\Gamma^{\rm HC}$ & kT$^{\rm HC}$ & $\Gamma^{\rm WC}$ & kT$^{\rm WC}$ & $\sigma_{\texttt{gsmooth}}$ & $\chi^2$ / d.o.f \\
     & & [eV] & [eV] & & [keV] &  & [keV] & [eV]\\ \hline
     \hline
\endhead
\endfoot
\endlastfoot
       \makecell{2MASS\\J02201457$-$0728591} & {\bf mod1} & 1.88$_{-0.15}^{+0.16}$ &1.15$^{+0.17}_{-0.50}$ & 1.76$\pm$0.03 & 9.01$_{-2.14}^{+3.57}$ & 2.41$^{+0.07}_{-0.06}$ & 0.38$^{+0.13}_{-0.08}$ & x & 223/208  \\
        \cdashline{2-10}        
        \makecell{2MASS\\J10514425+3539306} &
         mod1 & 0.30$\pm$0.03 &  0.30$^{f,\triangle}$ & 1.70$\pm$0.03 & 7.45$^{+6.44}_{-1.66}$ & $>$4.98 & $<$0.32 & x & 237/158 \\
       & {\bf mod2} &  &  & 1.71$^{+0.05}_{-0.07}$ & 9.52$^{+0.34}_{-0.38}$ & $>$4.97 & $>$0.16 & x & 210/158\\
       \cdashline{2-10}
       \makecell{2MASS\\J16531505+2349427} & {\bf mod1} & 0.26$\pm$0.14 & 0.69$_{-0.07}^{+0.06}$ & 1.71$\pm$0.03 & 10.02$_{-1.04}^{+3.67}$ & 3.27$_{-0.17}^{+0.65}$ & 0.23$_{-0.11}^{+0.21}$ & x & 265/238 \\ 
       \cdashline{2-10}
        HB89~1257+286 & {\bf mod1} & 2.23$_{-0.15}^{+0.17}$ & 1.48$_{-0.20}^{+0.22}$ & 1.92$\pm$0.05 & 100$^{f}$ & 2.39$_{-0.02}^{+0.03}$ & 0.29$\pm$0.04 & x & 148/134 \\ 
       \cdashline{2-10}
        HB89~1529+050 & {\bf mod1} & 2.29$_{-0.11}^{+0.12}$ & 2.12$\pm$0.14 & 1.91$\pm$0.04 & 10.01$_{-2.36}^{+8.26}$ & 2.37$_{-0.04}^{+0.05}$ & 0.31$_{-0.03}^{+0.05}$ & x & 217/212 \\
       \cdashline{2-10}
        \makecell{LBQS\\1338-0038 {\bf $^\$$}} & mod1 & 2.35$^{+0.11}_{-0.09}$ & 1.90 $^{+0.10}_{-0.01}$ & 1.90 $^{+0.04}_{-0.03}$ & $>$7.46 & 2.39 $^{+0.04}_{-0.10}$ & 0.29$\pm$0.04 & / & 357/245 \\
        & mod1* & & 1.90$^{+0.17}_{-0.16}$ & 1.90$^{+0.05}_{-0.07}$ & 7.42$^{+1.12}_{-0.87}$ & 2.39$^{+0.06}_{-0.18}$ & 0.29$\pm$0.07 & 380$^{+140}_{-130}$ & 331/244 \\ 
        & {\bf mod2} &  & 1.88$\pm$0.01 & 1.83$^{+0.02}_{-0.01}$ & 12.14$^{+5.20}_{-3.33}$ & 2.30$^{+0.04}_{-0.05}$ & 0.25$\pm$0.02 &/ & 322/242 \\
       \cdashline{2-10}
        PG~0052+251 & mod1 & 3.48$^{+0.16}_{-0.15}$  & 4.19$^{+0.21}_{-0.11}$ & 1.77$\pm$ 0.02 & $>$64.02 & 2.77$^{+0.03}_{-0.02}$ & 0.63$^{+0.04}_{-0.09}$ & / & 509/262\\
        &  {\bf mod1*} & & 5.13$^{+0.23}_{-0.26}$ & 1.79$\pm$ 0.02 & $>$50.78 & 2.92$^{+0.05}_{-0.04}$ & 0.85$^{+0.19}_{-0.17}$ & 350$^{+40}_{-50}$ & 341/261 \\ 
        & mod2 &  & 4.18$^{+0.15}_{-0.11}$ & 1.77$\pm$ 0.01 & $>$66.67 & 2.76$^{+0.01}_{-0.03}$ & 0.60$^{+0.08}_{-0.05}$ & / & 356/261 \\
       \cdashline{2-10}
        PG 0804+761 $\oplus$ & mod1 & 0.83$_{-0.16}^{+0.24}$ & 4.67$_{-0.10}^{+0.09}$ & 2.09$_{-0.03}^{+0.04}$ &  & 3.05$\pm$0.01 & 0.38$_{-0.04}^{+0.05}$ &/ & 456/315 \\
        & mod1* & & 4.67$_{-0.11}^{+0.09}$ & 2.08$\pm$0.03 & 100$^{f}$ & 3.05$\pm$0.01 & 0.38$_{-0.02}^{+0.03}$ & $<$110 & 436/314 \\
        &  mod2 &  & 4.79$_{-0.10}^{+0.13}$ & 2.05$\pm$0.04 &  & 3.07$_{-0.02}^{+0.01}$ & 0.44$_{-0.06}^{+0.07}$ &/ & 398/313\\
        & {\bf mod2} &  & 0.83$^{f,\triangle}$ & 2.09$\pm$0.03 & & 3.10$_{-0.04}^{+0.03}$ & 0.49$_{-0.08}^{+0.11}$ & / & 403/314\\
        \cdashline{2-10}
        PG~0947+396 & & & & & \texttt{UFO}
        & & & \\
        & mod1 & 3.22$^{+0.45}_{-0.37}$ & 2.76$^{+0.70}_{-0.55}$ & 1.96$\pm$0.07 & $>$19.85 & 2.54$^{+0.08}_{-0.06}$ & 0.24$^{+0.03}_{-0.02}$ &/ & 226/204 \\
        & {\bf mod1*} &  & 3.20$^{+0.51}_{-0.38}$ & 2.08$\pm$0.02 & 8.66$^{+0.93}_{-0.79}$ & 3.07$^{+0.09}_{-0.08}$ & $>$0.45 & 360$^{+80}_{-70}$ & 217/203 \\
        & mod2 &  & 2.89$^{+0.73}_{-0.58}$ & 1.98$^{+0.07}_{-0.12}$ & $>$12.65 & 2.59$^{+0.09}_{-0.07}$ & 0.26$^{+0.07}_{-0.03}$ &/ & 228/202 \\
        & & & & & \texttt{no-UFO} & & & \\
        & mod1 &  & 2.92$^{+0.66}_{-0.54}$ & 1.96$\pm$0.07 & $>$14.89 & 2.72$^{+0.09}_{-0.07}$ & 0.31$^{+0.17}_{-0.05}$ &/ & 222/206 \\
        &mod1* & & 2.98$^{+0.38}_{-0.34}$ & 1.92$^{+0.11}_{-0.07}$ & 6.55$^{+2.09}_{-1.43}$ & 2.79$^{+0.06}_{-0.04}$ & 0.49$^{+0.38}_{-0.13}$ & 90$^{+60}_{-50}$ & 218/205 \\
        & {\bf mod2} &  & 3.03$^{+0.62}_{-0.58}$ & 1.89$^{+0.09}_{-0.08}$ & $>$38.46 & 2.77$^{+0.09}_{-0.08}$ & 0.33$^{+0.09}_{-0.07}$ &/ & 213/204 \\
        \cdashline{2-10}
        PG~0953+414 & {\bf mod1} & 3.89$^{+0.13}_{-0.12}$ & 4.34$_{-0.12}^{+0.15}$ & 1.99$_{-0.03}^{+0.02}$ & $>$23.74 & 2.88$_{-0.01}^{+0.02}$ & 0.38$\pm$0.02 & x & 291/295 \\
       \cdashline{2-10}
       PG 1114+445 & & & & & \texttt{UFO (101)} & & & \\
       & mod1 & 2.73$_{-0.15}^{+0.17}$ & 2.56$_{-0.40}^{+0.57}$ & 1.67$_{-0.06}^{+0.05}$ & 100$^{f}$ & 2.35$_{-0.12}^{+0.20}$ & 0.15$_{-0.01}^{+0.02}$ & / & 141/108  \\
       & mod1* &  & 2.51$_{-0.41}^{+0.53}$ & 1.79$_{-0.07}^{+0.08}$ &  & 2.28$_{-0.13}^{+0.18}$ & 0.14$_{-0.01}^{+0.02}$ & 220$_{-70}^{+80}$ & 131/107 \\
       & {\bf mod2} &  & 2.50$_{-0.41}^{+0.50}$ & 1.83$\pm$0.07 & & 2.28$_{-0.13}^{+0.16}$ & 0.14$_{-0.01}^{+0.02}$ & / & 127/106 \\
       & & & & & \texttt{UFO (301)}
        & & & \\
        & {\bf mod1} &  & 2.76$_{-0.31}^{+0.36}$ & 1.66$_{-0.04}^{+0.03}$ & & 2.45$_{-0.11}^{+0.12}$ & 0.16$\pm$0.02 & x & 153/133 \\
       & & & & & \texttt{no-UFO} & & & \\
       & mod1 &  & 2.31$_{-0.17}^{+.09}$ & 1.92$_{-0.05}^{+0.07}$ &  & 2.41$_{-0.12}^{+0.07}$ & 0.22$_{-0.03}^{+0.02}$ & / & 178/145 \\
       & mod1* & & 2.27$\pm$0.12 & 1.93$_{-0.06}^{+0.05}$ &  & 2.39$_{-0.08}^{+0.09}$ & 0.22$\pm$0.02 & $<$90 & 176/144 \\
        & {\bf mod2} &  & 2.17$_{-0.10}^{+0.11}$ & 1.99$_{-0.07}^{+0.06}$ &  & 2.37$_{-0.09}^{+0.07}$ & 0.21$_{-0.01}^{+0.03}$&/ & 163/143\\
       \cdashline{2-10}
       PG~1202+281 {\bf $^\$$} & mod1 & 2.61$^{+0.15}_{-0.14}$  & 2.14$\pm$0.01 & 1.78$\pm$0.04 & $>$10.83 & 2.36$\pm$0.05 & 0.22$\pm$0.02 & x & 296/261 \\
        & {\bf mod2} &  & 2.18$\pm$0.12 & 1.73$^{+0.01}_{-0.02}$ & $>$136.67 & 2.39$\pm$0.03 & 0.23$^{+0.02}_{-0.01}$ & x & 272/259\\
       \cdashline{2-10}
       PG~1216+069 & mod1 & 2.67$^{+0.19}_{-0.17}$ & 2.60$\pm$0.16 & 1.83$\pm$0.03 & $>$19.33 & 2.78$\pm$0.02 & 0.40$\pm$0.04 &/ & 312/242\\
       & mod1* & & 2.74$^{+0.17}_{-0.15}$ & 1.86$\pm$0.04 & 16.94$^{+16.67}_{-3.63}$ & 2.78$\pm$0.03 & 0.39$\pm$0.04 & 240$^{+100}_{-80}$ &  311/241 \\ 
        & {\bf mod2} &  & 2.64$^{+0.16}_{-0.15}$ & 1.83$^{+0.03}_{-0.04}$ & $>$14.04 & 2.83$\pm$0.03 & 0.41$^{+0.06}_{-0.05}$ &/ & 306/240\\
       \cdashline{2-10}
       PG~1307+085 & {\bf mod1} & 3.55$_{-0.12}^{+0.13}$ & 3.95$_{-0.11}^{+0.15}$ & 1.85$\pm$0.04 & 15.53$_{-4.28}^{+24.60}$ & 2.91$\pm$0.02 & 0.54$_{-0.08}^{+0.12}$ & x & 272/248\\
       \cdashline{2-10}
       PG~1352+183 & mod1 & 3.72$_{-0.14}^{+0.15}$ & 3.74$_{-0.13}^{+0.10}$ & 2.10$\pm$0.03 & $>$80.78 & 2.64$_{-0.04}^{+0.03}$ & 0.22$_{-0.01}^{+0.02}$ &/ & 269/217 \\
       & mod1* & & 3.75$_{-0.17}^{+0.01}$ & 2.10$\pm$0.03 & $>$34.25 & 2.64$\pm$0.04 & 0.22$_{-0.01}^{+0.02}$ & $<$460 & 269/214 \\
        & {\bf mod2} &  & 3.75$_{-0.01}^{+0.21}$ & 2.11$\pm$0.03 & $>$16.01 & 2.69$_{-0.02}^{+0.03}$ & 0.25$_{-0.01}^{+0.02}$ &/ & 256/213 \\
       \cdashline{2-10}
       PG 1402+261 {\bf $\dagger$} & mod1 & 3.51$_{-0.39}^{+0.52}$ & 3.72$_{-0.31}^{+0.49}$ & 1.93$\pm$0.04 & $>$50.73 & 2.78$_{-0.02}^{+0.04}$ & 0.61$_{-0.05}^{+0.07}$ & / & 324/287 \\
       & {\bf mod1*} & & 3.73$_{-0.34}^{+0.50}$ & 2.04$_{-0.06}^{+0.05}$ & $>$76.98 & 2.81$_{-0.04}^{+0.05}$ & 0.60$_{-0.06}^{+0.08}$ & 300$\pm$60 & 315/286 \\
        & mod2 &  & 3.35$\pm$0.29 & 1.98$_{-0.05}^{+0.03}$ & $>$13.10 & 2.77$_{-0.04}^{+0.03}$ & 0.62$_{-0.06}^{+0.04}$ &/ & 313/285 \\
       \cdashline{2-10}
       PG~1425+267 & {\bf mod1} & 2.30$\pm$0.08 & 1.93 $^{+0.12}_{-0.13}$ & 1.91$\pm$0.04 & $>$39.49 & 2.53$\pm$0.02 & 0.35$^{+0.05}_{-0.04}$ & x & 276/231 \\
       \cdashline{2-10}
       PG~1435$-$067 & mod1 & 5.94$_{-0.36}^{+0.42}$ & 14.41$^{+1.27}_{-1.31}$ & 2.20$\pm$0.01 & 25.71$_{-12.36}^{+92.06}$ & 5.39$_{-0.37}^{+0.34}$ & 0.46$_{-0.05}^{+0.04}$ & x & 262/212 \\
        & {\bf mod1} & & 5.94$^{f,\triangle}$ & 1.98$\pm$0.06 & $>$25.34 & 2.49$_{-0.20}^{+0.14}$ & 0.28$_{-0.05}^{+0.07}$ & x & 273/214 \\
       \cdashline{2-10}
       PG~1626+554 & mod1 & 5.01$_{-0.19}^{+0.21}$ & 5.22$_{-0.14}^{+0.13}$ & 2.09$\pm$0.01 & $>$13.18 & 3.27$_{-0.02}^{+0.04}$ & 0.65$_{-0.13}^{+0.16}$ &/ & 367/234 \\
       & {\bf mod1*} & & 5.48$_{-0.16}^{+0.17}$ & 2.21$\pm$0.02 & 16.05$_{-4.61}^{+13.14}$ & 3.31$\pm$0.04 & 0.29$_{-0.10}^{+0.16}$ & 380$\pm$20 & 262/233 \\
        & mod2 &  & 5.48$_{-0.15}^{+0.13}$ & 2.05$_{-0.01}^{+0.02}$ & $>$177.60 & 3.43$_{-0.07}^{+0.04}$ & 0.74$_{-0.10}^{+0.12}$ &/ & 322/231 \\
       \cdashline{2-10}
       \makecell{SDSS\\J144414.66+063306.7} & mod1 & 3.97$_{-0.18}^{+0.20}$ & 4.66$_{-0.22}^{+0.21}$ & 1.85$\pm$0.03 & 11.82$_{-2.19}^{+6.06}$ & 2.88$\pm$0.03 & 0.60$_{-0.08}^{+0.14}$ & / & 314/238 \\
       & {\bf mod1*} & & 5.03$_{-0.24}^{+0.27}$ & 1.89$_{-0.03}^{+0.04}$ & 9.47$_{-1.35}^{+3.93}$ & 2.97$_{-0.04}^{+0.05}$ & 0.66$_{-0.19}^{+0.21}$ & 420$^{+80}_{-70}$ & 310/237\\
       & mod2 &  & 4.77$_{-0.14}^{+0.21}$ & 1.86$\pm$0.03 & 13.31$_{-3.50}^{+6.99}$ & 2.93$\pm$0.05 & 0.63$_{-0.13}^{+0.19}$ & / & 311/235 \\
       \cdashline{2-10}
       \makecell{WISE\\J053756.30$-$024513.1} & mod1 & 0.08$^{+0.02}_{-0.03}$ & 0.08$^{f,\triangle}$ & 1.72$\pm$0.03 & 6.28$^{+1.00}_{-2.64}$ & 2.40$^{+0.03}_{-0.04}$ & 0.45$^{+0.07}_{-0.05}$ &/ & 324/215 \\
       & {\bf mod1*} & & & 1.81$^{+0.02}_{-0.04}$ & 5.88$^{+0.80}_{-0.64}$ & 2.57$^{+0.08}_{-0.09}$ & 0.46$^{+0.27}_{-0.22}$ & 540$^{+40}_{-30}$ & 282/214 \\
       & mod2 &  & & 1.72$\pm$0.03 & 6.27$^{+2.66}_{-0.88}$ & 2.40$^{+0.04}_{-0.03}$ & 0.49$^{+0.07}_{-0.06}$ &/ & 324/214 \\ \hline
\end{longtable}\par
    \smallskip
    \footnotesize{
    \textbf{Notes:} \\
    \textit{Columns}: (1) Source name; (2) Model; (3) seed photon temperature obtained by modeling opt/UV data with \texttt{diskBB}; (4) seed photon temperature obtained by modeling opt/UV/X-ray data with \texttt{nthComp}; (5) photon index of the hot corona; (6) electron temperature for the hot corona derived by modeling opt/UV/X-ray data with \texttt{nthComp}; (7) photon index of the warm corona; (8) electron temperature for the warm corona obtained obtained by modeling opt/UV/X-ray data with \texttt{nthComp}; (9) width of the \texttt{gsmooth} component; (10) $\chi^2$ over d.o.f for each fit.\\
   {\it Mod1} refers to the 2corona model (\texttt{nthComp$_{WC}$+nthComp$_{HC}$+xillverCp}). {\it Mod1*} adds to mod1 the \texttt{gsmooth} component, accounting for the broadening of the Fe line. {\it Mod2} refers to the two-coronae model with the addition of \texttt{relxillCp} to account for the disk reflection. \\
   We report in bold the best-fit model. All parameter values discussed in the paper refer to the results obtained from this best-fit model. \\
   {\bf x}: the \texttt{gsmooth} component is not needed in these observations. \\
   $\oplus$: PG~0804+761 has two XMM-pn and -OM observations. No variability is detected and thus, their parameters are linked during the fit. \\
   {\bf $^\$$}: Both ionization parameter and inner radius of the \texttt{relxillCp} component are well constrained for LBQS 1338-003 (log($\xi$) = 2.86$_{-0.31}^{+0.42}$ R$_{\rm in}$ = $-$8.41$_{-13.49}^{+5.28}$), and 
   PG~1202+281 (log($\xi$) = 3.00$_{-0.21}^{+0.30}$ and R$_{\rm in}$ = $-$12.38$_{-20.98}^{+6.95}$).\\ 
   {\bf $\triangle$}: we fix kT$_{\rm BB}$ to kT$_{\rm BB}^{\rm opt/UV}$, given that only upper limits were found during spectral fitting for WISE~J053756.30$-$024513.1 and 2MASS~J10514425+3539306. For PG~1435$-$067, we fix the kT$_{\rm BB}$ to the opt/UV best-fit value ($\Delta \chi^2$ = +11 with only one d.o.f. more) as, if allowed to vary freely, it reaches $\sim$15 eV. 
   For PG~0804+761, kT$_{\rm BB}$ differs strongly from the opt/UV–only value, we thus fix it to the opt/UV best-fit ($\Delta \chi^2$ = + 5 with only one d.o.f. more). The only parameters affected are those of the WA, which become less constrained, so we fix the WA to the values from mod2. Given the small statistical change, we adopt this as the best-fit for the analysis.\\
   {\bf $\dagger$}: the AIC test indicates no statistically significant preference between mod1* and mod2. However, mod2 is rejected because it resulted in unphysical parameter values; therefore, mod1* is adopted.}
}

\begin{table}[h]
\renewcommand{\arraystretch}{1.6}
\centering
\caption{Hardness ratio, soft excess strength proxies, optical depth and $\chi$ parameter of the warm corona.}\label{tab:se_par}
\begin{tabular}{c|ccccc}
    \hline
    Source & HR & $Q$ & $\tau^{\rm WC}$ & $A$ & $\chi$\\
    \scriptsize (1) & \scriptsize(2) & \scriptsize(3) & \scriptsize(4) & \scriptsize(5) & \scriptsize(6) \\
    \hline
2MASS~J02201457$-$0728591
& $-0.518\pm0.003$
& $-1.84^{+0.20}_{-0.22}$
& $15.19^{+2.20}_{-2.34}$
& $1.07^{+0.02}_{-0.03}$
& $0.59^{+0.17}_{-0.23}$ \\
2MASS~J10514425+3539306
& $0.031\pm0.006$
& $-5.07 \pm 1.37$
& $>4.27$
& --
& -- \\
2MASS~J16531505+2349427
& $0.846\pm0.007$
& $-2.91^{+0.17}_{-0.30}$
& $9.67^{+3.36}_{-4.07}$ & $1.08^{+0.07}_{-0.14}$& $0.51^{+0.38}_{-0.88}$ \\
HB89~1257+286
& $-0.630\pm0.002$
& $-1.876^{+0.005}_{-0.004}$
& $17.81^{+1.40}_{-1.28}$
& $1.40\pm0.01$
& $1.49\pm0.04$ \\
HB89~1529+050
& $-0.598\pm0.002$
& $-1.56^{+0.45}_{-0.20}$
& $17.55^{+1.79}_{-1.12}$
& $1.05\pm0.01$
& $0.52^{+0.13}_{-0.10}$ \\
LBQS~1338$-$0038
& $-0.582\pm0.001$
& $-2.09^{+0.09}_{-0.24}$
& $21.00^{+1.47}_{-1.12}$
& $1.04\pm0.01$
& $0.46^{+0.09}_{-0.08}$ \\
PG~0052+251
& $-0.592\pm0.001$
& $-2.29^{+0.16}_{-0.29}$
& $6.32^{+0.83}_{-0.70}$
& $1.14\pm0.04$
& $0.56\pm0.14$ \\
PG~0804+761 \makecell{(obs 1)\\(obs 2)}
& \makecell{
$-0.707\pm0.002$\\
$-0.736\pm0.001$}
& $-2.190^{+0.005}_{-0.004}$
& $7.35^{+0.79}_{-0.76}$
& $1.10\pm0.03$
& $0.48^{+0.13}_{-0.14}$ \\
PG~0947+396 \makecell{(UFO)\\(No-UFO)}
& \makecell{
$-0.637\pm0.004$\\
$-0.659\pm0.002$}
& \makecell{
$-2.00^{+0.24}_{-0.23}$\\
$-2.00\pm0.06$}
& \makecell{
$7.87^{+0.51}_{-0.41}$\\
$11.82^{+1.80}_{-1.61}$}
& \makecell{
$1.10\pm0.02$\\
$1.63\pm0.03$}
& \makecell{
$0.51^{+0.08}_{-0.07}$\\
$1.51^{+0.06}_{-0.07}$} \\
PG~0953+414
& $-0.717\pm0.001$
& $-1.86^{+0.03}_{-0.02}$
& $10.02^{+0.29}_{-0.32}$
& $1.07\pm0.01$
& $0.45^{+0.04}_{-0.05}$ \\
PG~1114+445 \makecell{(101 -- UFO)\\(301 -- UFO)\\(No-UFO)}
& \makecell{
$0.150^{+0.088}_{-0.008}$\\
$0.162\pm0.010$\\
$-0.154\pm0.003$}
& \makecell{
$-2.41^{+0.12}_{-0.06}$\\
$-2.50^{+0.20}_{-0.11}$\\
$-1.92^{+0.08}_{-0.03}$}
& \makecell{
$28.81^{+4.08}_{-4.62}$\\
$22.82^{+3.13}_{-2.65}$\\
$21.43^{+1.87}_{-1.99}$}
& \makecell{
$1.02^{+0.01}_{-0.02}$\\
$1.02\pm0.01$\\
$1.03\pm0.01$}
& \makecell{
$0.30^{+0.21}_{-0.31}$\\
$0.28^{+0.20}_{-0.22}$\\
$0.41^{+0.13}_{-0.16}$} \\
PG~1202+281
& $-0.567\pm0.002$
& $-2.09^{+0.21}_{-0.28}$
& $20.06^{+0.73}_{-1.02}$
& $1.04\pm0.01$
& $0.45^{+0.05}_{-0.08}$ \\
PG~1216+069
& $-0.616\pm0.002$
& $-1.937^{+0.006}_{-0.004}$
& $10.03^{+0.76}_{-0.72}$
& $1.08\pm0.02$
& $0.49\pm0.10$ \\
PG~1307+085
& $-0.670\pm0.001$
& $-2.019\pm0.001$
& $8.12^{+1.17}_{-0.59}$
& $1.10^{+0.03}_{-0.02}$
& $0.53^{+0.16}_{-0.10}$ \\
PG~1352+183
& $-0.677\pm0.001$
& $-1.94\pm0.01$
& $14.62^{+0.39}_{-0.70}$
& $1.06\pm0.01$
& $0.51^{+0.04}_{-0.07}$ \\
PG~1402+261
& $-0.714\pm0.001$
& $-2.96\pm0.01$
& $8.34^{+0.55}_{-0.63}$
& $1.11\pm0.02$
& $0.56^{+0.08}_{-0.10}$ \\
PG~1425+267
& $-0.520\pm0.002$
& $-1.94\pm0.01$
& $14.16\pm0.96$
& $1.06\pm0.01$
& $0.50^{+0.09}_{-0.10}$ \\
PG~1435$-$067
& $-0.615\pm0.003$
& $-2.40^{+0.13}_{-0.30}$
& $16.49^{+4.06}_{-2.56}$
& $1.02\pm0.03$
& $0.24^{+0.33}_{-0.30}$ \\
PG~1626+554
& $-0.682\pm0.001$
& $-1.86^{+0.15}_{-0.04}$
& $8.31^{+2.08}_{-1.75}$
& $1.06^{+0.05}_{-0.06}$
& $0.38^{+0.29}_{-0.36}$ \\
SDSS~J144414.66+063306.7
& $-0.627\pm0.001$
& $-2.09^{+0.20}_{-0.30}$
& $6.95^{+1.40}_{-1.01}$
& $1.12^{+0.05}_{-0.04}$
& $0.53\pm0.21$ \\
WISE~J053756.30$-$024513.1
& $-0.401\pm0.004$
& $-3.85^{+1.13}_{-0.29}$
& $11.84^{+4.44}_{-2.70}$
& $1.14\pm0.05$
& $0.83\pm0.30$ \\
\hline
\end{tabular}
    \par
    \smallskip
    {\footnotesize
    \textit{Notes:} \\ {\it Columns:} (1) Source; (2) Hardness ratio HR = $\frac{H - S}{H + S}$ between the soft (S, 0.5--2 keV) and hard (H, 2--10 keV) energy bands; (3) Q = L$_{\rm 0.5-2 \, keV}^{\rm SE}$/L$_{\rm 2-10 \, keV}^{\rm PL}$, defined by \citet{jana2026}; (4) optical depth of the wamr corona; (5) amplification factor A = 1 + F$_{\rm cor}$/F$_{\rm tot,disk}$ defined in \citet{petrucci2020}; (6) coronal dissipation fraction $\chi$. $\tau^{\rm WC}$, $A$, and $\chi$ are obtained using the formula from \citet{Beloborodov1999}, whereas the corresponding quantities derived following \citet{Longair2011} are available on Zenodo.}
\end{table}

\begin{table}
\centering
\caption{Warm corona and disk reflection component contributions to the 0.5–2 keV flux.}\label{tab:hybrid_soft_band_contributions}
\begin{tabular}{lcc}
\hline
Source & Warm corona & Disk reflection \\
       & (\%)         & (\%)                \\
\hline
\hline
2MASS~J10514425+3539306 {\bf *}  & 1  & 98 \\
LBQS~1338$-$0038 & 34 & 10 \\
PG~0804+761 & 44 & 2  \\
PG~0947+396 (no-UFO) & 36 & 18 \\
PG 1114+445 (UFO - 101) & 40 & 2  \\
PG 1114+445 (no-UFO) & 54 & 4  \\
PG~1202+281 & 34 & 10 \\
PG~1216+069 & 40 & 8  \\
PG~1352+183 & 34 & 2  \\
\hline
\end{tabular}
    \par
    \smallskip
    {\footnotesize
\textit{Note.} The reported fractions are 
not included in the computation of $Q$, which considers only the warm corona \texttt{nthComp} component.\\
{\bf *}: In 2MASS~J10514425+3539306, no soft excess is detected and the warm corona \texttt{nthComp} component accounts only for the accretion disk emission.}
\end{table}

\twocolumn

\section{Comparison seed photon temperatures from optical/UV/X-ray data}\label{comp_kt_bb}

\begin{figure}[h]
\centering
\includegraphics[width=.4\textwidth, trim=10 10 10 10, clip]{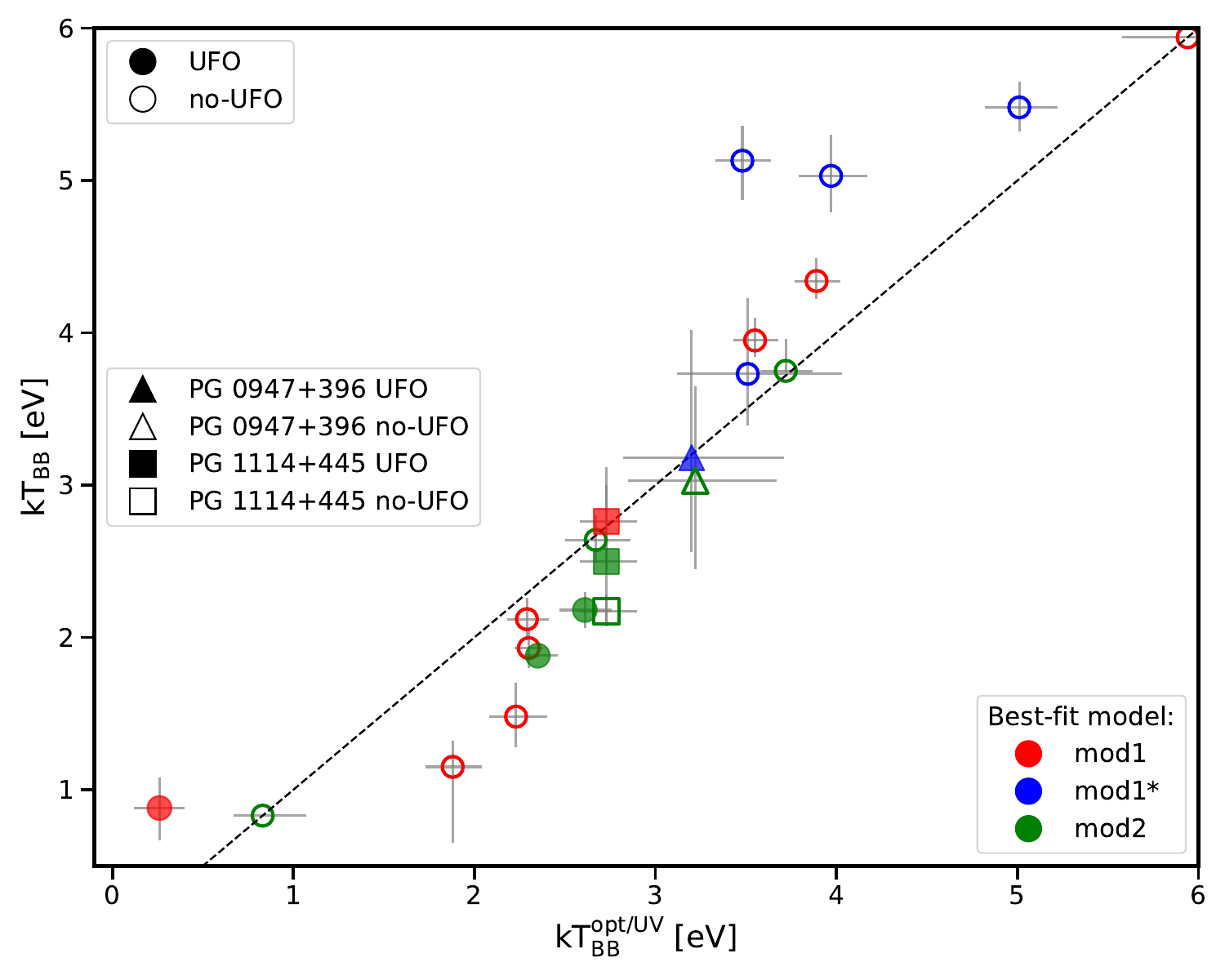}
    \caption[]
    {{Comparison between seed photon temperatures obtained by fitting optical/UV (x-axis) and optical/UV/X-ray data (y-axis). The best-fit models are reported: mod1 in red, mod1* in blue, and mod2 in green. The gray dashed line corresponds to y = x. The two temperature show a significant correlation with Log(NHP) = $-$12.22, indicating that the inclusion of the X-ray data does not introduce a systematic change in the inferred values. For four AGN, kT$_{\rm BB}$ is fixed to the optical/UV only value, as it is either unconstrained or reached unrealistically high values, see Table~\ref{fig:distributions}.}}
    \label{fig:ktbb_bis}
\end{figure}

In Fig.~\ref{fig:ktbb_bis}, we present a comparison between the best-fit values of the seed photon temperature derived from only the optical/UV data (kT$_{\rm bb}^{\rm opt/UV}$) and those from the combined optical/UV/X-ray data (kT$_{\rm bb}$). 
Overall, the two estimates show no statistically significant systematic offset, with a scatter of $\sim$0.6 eV\footnote{The scatter is defined as the root-mean-square of the residuals with respect to the one-to-one relation, i.e., $\sqrt{\langle (\rm kT_{bb} - kT_{bb}^{opt/UV})^2 \rangle}$.}.

\begin{figure}[h!]
\centering
\includegraphics[width=.35\textwidth, trim=10 10 10 10, clip]{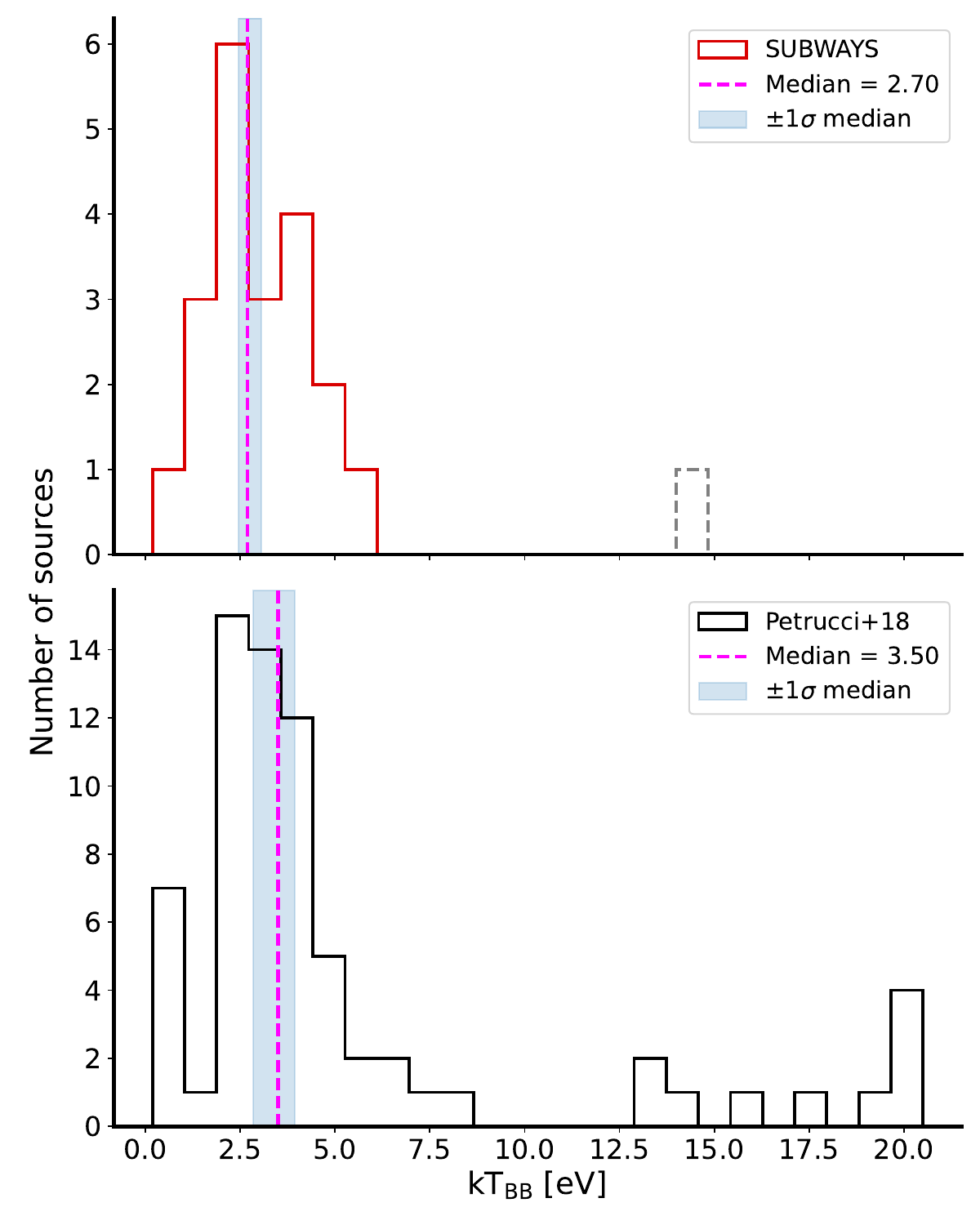}
    \caption[]
    {{Comparison of the best-fit seed photon temperature distributions obtained in this work and by \citetalias{petrucci18}. The median values are shown as dashed magenta lines with 1$\sigma$ uncertainties in gray bars. The KS test indicates no statistically significant difference between the two distributions.}}
    \label{fig:ktbb_comp}
\end{figure}

In Fig.~\ref{fig:ktbb_comp}, we compare the kT$_{\rm bb}$ distribution of our AGN sample with that of \citetalias{petrucci18}.
By applying the KS test on the two distributions, we find them statistically comparable.
The \citetalias{petrucci18} distribution shows a clear separation around kT$_{\rm bb} \sim 10$ eV, with most sources lying below $\sim$5 eV and a smaller sub-set extending to higher temperatures, $\gtrsim$12 eV. This latter sub-group is populated by observations of three AGN: 1H~0419-577 (1/8 spectra), IRASF~12397+3333 (1/2 spectra), and RE~1034+396 (8/8 spectra).
Considering SUBWAYS, PG~1435$-$067 shows kT$_{\rm bb} \sim$14 eV (shown as a dashed gray histogram in the upper part of Fig.~\ref{fig:ktbb_comp}). 
We subsequently fixed kT$_{\rm bb}$ to the optical/UV best-fit value, obtaining $\Delta\chi^2 = +11$ for one d.o.f. more. 
However, if left free, the kT$_{\rm bb}$ value falls within the second peak of the distribution. 
To further investigate this behavior, we examine whether AGN with kT$_{\rm bb} > 10$ eV share similar properties and whether these differ from the rest of the AGN population considered in \citetalias{petrucci18} and in the SUBWAYS sample. For the four AGN, we consider only the observations in which high blackbody temperatures are measured. In particular, we consider the BH mass, X-ray and bolometric luminosities, Eddington ratio, spectral parameters (i.e., photon index and electron temperature of both WC and HC) of the SE, and iron-line broadening. 
We also account for the physical properties of the SE and we consider three additional parameters derived by \citet{palit2024}: the hardness ratio, the dissipation fraction (i.e., the fraction of energy released within the WC), and the model normalization N$_{\rm SE}$, which serves as an indicator of the emitting surface area of the WC\footnote{N$_{\rm SE}$ is available only for \citetalias{petrucci18} sample and is not derived for PG~1435$-$067.}. 
Overall, we do not find a specific range of parameters (in terms of BH mass, luminosity, Eddington ratio, spectral and physical properties of WC, or SE diagnostics) that clearly distinguishes these AGN from the rest of the population.
For the SUBWAYS sample, we derive the reflection fraction R, defined as the ratio of the observed flux of the neutral reflection to that of primary component in the 20--40 keV. The value obtained for PG~1435$-$067 (R = 0.7$\pm$0.3) lies well within the range of the rest of the sample (0.1 $\leq$ R $\leq$ 2.1), indicating that its reflection contribution is not significantly different from that of the other sources.
Although not statistically significant, two trends are worth noting. First, 3/4 AGN populating the second peak show only a narrow Fe K$\alpha$ line component \citep[e.g.,][]{delacalle2010}. A partial broadening of the line is reported for the observation of 1H~0419-577 \citep{digesu2014}. However, despite the high data quality, the authors could not simultaneously constrain the broad component width and narrow component normalization. They also note that the Fe k$\alpha$ emission from cold matter may appear broadened by blending of the line core with the Compton shoulder \citep{yaqoob2010}, although this effect is much weaker than relativistic broadening.
In the case of PG~1435$-$067, when modeling the Fe line with \texttt{gsmooth}, we obtain only an upper limit ($<$40 keV) to its width.
Second, the photon index of the HC is generally soft (with median value $\sim$2.20) for these AGN. The only apparent exception is 1H~0419-577, for which \citetalias{petrucci18} report $\Gamma = 1.77 \pm 0.06$; however, other studies of the same observation find a softer photon index, consistent with the remaining sources in the second peak \citep[e.g.,][]{digesu2014,jiang2019,nandi2023}.

\end{appendix}

\end{document}